\documentclass{article}
\pdfoutput=1
\usepackage{graphicx}
\usepackage{subcaption}
\usepackage[includefoot,margin=1in]{geometry} 

\usepackage{tikz}
\usetikzlibrary{arrows.meta}
\usetikzlibrary{plotmarks}
\usetikzlibrary{shapes}
 
\usepackage{array} 
\usepackage{footnote}
 
\usepackage{amssymb}
\usepackage{amsmath}
\usepackage{amsfonts}
\usepackage{bm}

\usepackage{url}
\usepackage{hyperref}
\hypersetup{pdfstartview={XYZ null null 1.00}}

\newcommand{\beq}{\begin{equation}}
\newcommand{\eeq}{\end{equation}}

\newcommand{\beqr}{\begin{eqnarray}}
\newcommand{\eeqr}{\end{eqnarray}}
\def\bal#1\eal{\begin{align}#1\end{align}}
\def\bat#1#2\eat{\begin{alignat}{#1}#2\end{alignat}}

\begin{document}

\title{Exactly solvable flat-foldable quadrilateral origami tilings}
\author{Michael Assis\footnote{School of Mathematics and Statistics, University of Melbourne, Carlton, VIC, Australia}
}
\date{\today}
\maketitle

\begin{abstract}
We consider several quadrilateral origami tilings, including the Miura-ori crease pattern, allowing for crease-reversal defects above the ground state which maintain local flat-foldability. Using exactly solvable models, we show that these origami tilings can have phase transitions as a function of crease state variables, as a function of the arrangement of creases around vertices, and as a function of local layer orderings of neighboring faces. We use the exactly solved cases of the staggered odd 8-vertex model  as well as Baxter's exactly solved 3-coloring problem on the square lattice to study these origami tilings. By treating the crease-reversal defects as a lattice gas, we find exact analytic expressions for their density, which is directly related to the origami material's elastic modulus. The density and phase transition analysis has implications for the use of these origami tilings as tunable metamaterials; our analysis shows that Miura-ori's density is more tunable than Barreto's Mars, for example. We also find that there is a broader range of tunability as a function of the density of layering defects compared to as a function of the density of crease order defects before the phase transition point is reached; material and mechanical properties that depend on local layer ordering properties will have a greater amount of tunability. The defect density of Barreto's Mars, on the other hand, can be increased until saturation without passing through a phase transition point. We further consider relaxing the requirement of local flat-foldability by mapping to a solvable case of the 16-vertex model, demonstrating a different phase transition point for this case.
\end{abstract}

\section{Introduction}
Recently foldable origami crease patterns (CPs) have seen much direct use as programmable matter~\cite{hawkes2010abtkdrw,an2011bdr}, tunable metamaterials~\cite{silverberg2014emhhsc,dudte2016vtm}, self-deployable systems~\cite{felton2013tsodrw,tolley2014fmarw}, architecture~\cite{tachi20104}, and medical devices~\cite{
shim2012khjy,randall2012gg,fernandes2012g,gracias2013}. In many of these explorations, the origami CP hinges are self-folding after application of heat~\cite{silverberg2015nelhslhc,na2015ebcslhh}, electric current~\cite{hawkes2010abtkdrw,an2011bdr,felton2013tsodrw}, lasers~\cite{ocampo2003vfwkaosihn}, and various liquids~\cite{mulakkal2016swmt}, to name some examples. See~\cite{perazahernandez2014hml} for a recent review. 

In~\cite{silverberg2014emhhsc} the authors show experimentally how the elastic modulus of a Miura-ori origami CP depends on the density of defects present, that is, the number of creases with opposite orientations than the original CP. Defects can also arise in the production of reversible self-folding origami based on hydrogel bilayers~\cite{evans2017_temp}, and in~\cite{felton2013tsodrw} the authors report that initially some creases started self-folding in the opposite orientation before correcting themselves. Presumably, defects naturally arise after several cycles of reversible self-folding and unfolding, and they certainly arise through environmental factors. Along these lines, there has been recent interest in finding the minimum number of crease orientations necessary to force the orientations of the remaining creases in the lattice as the origami CP is being activated, the so-called forcing sets~\cite{ballinger2015defgh,abel2016cdehklt}; due to multistability, however, defects can easily arise from forcing sets~\cite{waitukaitis2015mch,stern2017pm_temp}. 

We are interested in characterizing flat-foldable defects which can arise in origami CPs, and so we here study for the first time origami CPs from the perspective of exactly solvable equilibrium statistical mechanics. Our statistical ensembles are presumed to be either a large collection of manufactured origami CPs, self-folding once to their final shape, or else a single origami undergoing many cycles of reversible self-folding and unfolding, in which case it would approach equilibrium after a number of such cycles. We do not consider localized tuning of creases on the lattice, such as forcing sets, but rather use homogeneous variables throughout the lattice and assume that all creases must be folded. Our exact solutions allow us to derive not only exact phase transition point locations for the first time, but also free-energy expressions which allow the derivation of its thermodynamical properties. 

The nature of this work bridges the theoretical models of exactly solvable lattice statistical mechanics with the more experimental work on origami engineering, condensed matter physics, and materials science. We do not assume a familiarity with all of these areas and attempt to explain a sufficient amount of origami results and theory in order to allow both the statistical mechanics theorists to understand the applications of the models to origami, as well as to explain sufficiently the methods and models of statistical mechanics to those not familiar with the exactly solved models. Those more familiar with origami and more interested in the applications of the theory may wish to skip some details of the models and focus on sections~\ref{sec:latticegas} and \ref{sec:discussion}, while those not familiar with origami and its applications may wish to spend more time studying the introduction and sections~\ref{sec:defects} and \ref{sec:weights}.

In our models, the ground state represents a known origami CP crease configuration, and excitations above the ground state represent defects in the foldable lattice due to reversal of the states of the creases, from mountain to valley or vice-versa. The phase transitions of our models represent points at which the long-range order in the lattice disappears, either with respect to crease states in the lattice or else with respect to relative face layer orderings, so that correlations of either of these variables decay exponentially above the phase transition point. As a consequence of our analysis, we are able to characterize CPs which are more stable against defects than others, and conversely, which are more tunable as a metamaterial. Some CPs do not feature a phase transition point, and so defects can be added to the lattice until saturation without changing its long range order properties. Our models enable us to predict the defect density as a function of the preference of creases for the reversed state. To the extent that mechanical properties of the origami CP depend on the defect density, knowledge of how much to bias the creases in the lattice allows for a tuning of the defect density, and hence, of the mechanical properties. 

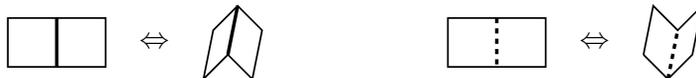
\begin{figure}[htpb]
\begin{center}
\scalebox{0.65}{
\begin{tikzpicture}
\draw[line width = 1pt] (-1,0) rectangle (1,1);
\draw[line width = 2pt] (0,1) -- (0,0);
\node at (2,0.5) {\LARGE$\Leftrightarrow$}; 
\begin{scope}[shift={(0,-0.25)}]
\draw[line width = 1pt] (3,0) -- (3.5,0.5) -- (4,0) -- (4.2,1) -- (3.7,1.5) -- (3.2,1) -- (3,0);
\draw[line width = 2pt] (3.5,0.5) -- (3.7,1.5);
\end{scope}
\begin{scope}[shift={(9,0)}]
\draw[line width = 1pt] (-1,0) rectangle (1,1);
\draw[dashed,line width = 2pt] (0,1) -- (0,0);
\node at (2,0.5) {\LARGE$\Leftrightarrow$}; 
\begin{scope}[shift={(0,-0.25)}]
\draw[line width = 1pt] (3,0.5) -- (3.5,0) -- (4,0.5) -- (4.2,1.5) -- (3.7,1) -- (3.2,1.5) -- (3,0.5);
\draw[dashed,line width = 2pt] (3.5,0) -- (3.7,1);
\end{scope}
\end{scope}
\end{tikzpicture}
}
\end{center}
\caption{The mapping of mountain folds to solid lines (left) and valley folds to dashed lines (right).\label{fig:mountainvalley}}
\end{figure}

Origami CPs readily lend themselves to vertex model interpretations in statistical mechanics. If we represent mountain creases by solid lines and valley creases by dashed lines as in figure~\ref{fig:mountainvalley}, then we have a direct correspondence with vertex models with 2-state edges, having Boltzmann weights $v_i=\exp(-\beta\varepsilon_i)$ defined in terms of the configuration of mountain and valley creases around a vertex in the lattice, where $\varepsilon_i$ represents an interaction energy or chemical potential, and $\beta=k_BT$, where $k_B$ is Boltzmann's constant and $T$ is the temperature. With these definitions, we sum over all of the possible valid vertex configurations in the lattice to arrive at the partition function $Z$,
\beq
Z= \sum_{\scriptscriptstyle\mathrm{configs}}\,\prod_i v_i^{m_i}
\eeq
where $m_j$ are the number of vertex weights $v_i$ in the lattice. In the thermodynamic limit where the number of lattice sites $\mathcal{N}\to\infty$, the free-energy $f$ is defined by
\beq
-\beta\,f = \lim_{\mathcal{N}\to\infty} \ln\left(Z^{1/\mathcal{N}}\right)
\eeq
From the free-energy, all thermodynamic quantities and critical phenomena can be derived.

Vertex models in statistical mechanics are often defined on a regular lattice. The lattice edge lengths and angles are typically disregarded --- only the graph connectivity of vertices is usually of interest. For origami CPs, on the other hand, not only the graph but the angles and edge lengths are important for its foldability properties. In particular, necessary flat-foldability conditions require that the alternating sum of angles around a vertex add to $\pi$ (Kawasaki's theorem), that each vertex be of even degree, and that the difference in the number of mountain and valley creases around each vertex equal two (Maekawa's theorem)~\cite{demaine2007o}. Also, changing the angles of creases around a vertex can change the number of valid mountain-valley crease assignments which are flat-foldable around that vertex, even while still satisfying Kawaski's theorem at that vertex. Therefore, since we are interested in changing crease assignments to allow for defects, we must also incorporate the effect of the CP angles on the number of allowed crease assignments around a vertex into our models. Furthermore, aside from the importance of angles in flat-foldability and determining the number of valid crease assignments, flat-foldable CPs with the same connectivity and the same crease asignments but different angles can have very different folding properties, as seen in~\cite{evans2015lmh}, where the Miura-ori, Barreto's Mars, the quadrilateral mesh, and the dual square twist CPs only differ in their angles, even though their resulting global foldings are quite different. We will not directly consider the resulting folding of the CP.

Sufficient conditions for the global flat-foldability of an origami CP is an NP-hard problem~\cite{bern1996h}, depending on the global layer ordering of the facets. This layer ordering condition can in principle be translated into a statistical mechanics model, by mapping to an SOS model with a partial height ordering around each vertex and summing over those configurations which admit a global height ordering. As far as we are aware this kind of model has not been considered before, and it cannot be reduced to only nearest-neighbor interactions. We therefore do not consider global flat-foldability. Because of the inherent difficulty of dealing with global flat-foldability conditions, we will only consider local flat-foldability requirements.

Ours is not the first attempt to relate origami crease patterns to statistical mechanics. Most other work has been done considering foldings of polymerized membranes or tethered membranes on a lattice, that is, random crumplings of the lattice where not all bonds need to be folded. See~\cite{mori1996k} for such a study on the square lattice, \cite{francesco1994g2,bowick1995fgg,munkel1995h,bowick1996fgg,cirillo1996gp,cirillo1996gp2, mori1996k2,francesco1997gm,bowick1997ggm,mori1997g,popova2007m,popova2009m} for studies on the triangular lattice, and~\cite{francesco1998,francesco19982,cirillo2000gp} for studies on the union-jack lattice. Also, one-dimensional folding as a meander problem was considered in~\cite{francesco1997gg,francesco1997gg2}. In~\cite{francesco1998eg,francesco2001} foldings of all edges of triangulations by regular triangles of arbitrary genus surfaces are considered, but since their only restrictions are even degree vertices which are 3-colorable, in order to allow the mapping of all triangular faces onto each other, they disregard Kawasaki and Maekawa's theorems. See~\cite{francesco2000,bowick2001t,francesco2005g} for reviews on these topics just mentioned. We also note~\cite{shender1993chb}, where a mapping is given from the kagom{\'e} lattice Heisenberg antiferromagnet to a folded triangular sheet. Except for the exact calculation of the folding entropies of random crumpling on triangular lattice~\cite{francesco1994g} and of flat-foldable Miura-ori states~\cite{ginepro2014h}, no other exact results are known, as far as we are aware. Our work here appears to be the first time that exactly solvable models have been used to study origami CPs in general. 

As a first means of studying origami CPs using the methods of solvable equilibrium statistical mechanics, we will confine ourselves in this paper to origami CPs whose graph connectivities are of the form of a square lattice, that is, regular degree four lattices. We consider staggering units of up to four vertices in our models of isohedral quadrilateral tilings of the plane which are flat-foldable. These flat-foldable tilings are the simple square tiling, the parallelogram or rhombus (pmg) tiling commonly known as Miura-ori, the trapezoid tiling, called chicken wire in~\cite{evans2015lmh}, and the kite tiling, called Huffman in~\cite{evans2015lmh}, where we use the IUC short crystallographic notation for the wallpaper groups to distinguish Miura-ori from the parallelogram p2 and rhombus cmm tilings.  We will use the standard naming conventions of the trapezoid and kite tiling but use the Miura-ori name, which is well known in the origami literature, rather than the parallelogram or rhombus pmg tiling naming convention; we will also refer to the square tiling as the simple square CP.

In our translation of the CPs to our square lattice models we impose the flat-foldability restrictions inherent in the origami CP due to the values of its angles. We will generally only consider locally flat-foldable origami CPs, except when looking at CPs which break Maekawa's theorem at vertices. Since we are not considering global flat-foldability, it is possible that some of the configurations of crease assignments which are included in the partition function summation are not globally flat-foldable. Indeed, even the ground state crease assignment of the trapezoid CP can self-intersect globally without taking proper precautions~\cite{lang2017_temp}. 

\begin{figure}[htpb]
\begin{center}
\scalebox{0.65}{
\begin{tikzpicture}
\draw[dashed,line width = 2pt] (0,1) -- (0,0); 
\draw[line width = 2pt] (0,0) -- (0,-1); 
\draw[dashed,line width = 2pt] (-1,0) -- (0,0); 
\draw[dashed,line width = 2pt] (0,0) -- (1,0);
\node at (-.7,.6) {{\Large $v_1$}};
\begin{scope}[shift={(3,0)}]
\draw[line width = 2pt] (0,1) -- (0,0); 
\draw[dashed,line width = 2pt] (0,0) -- (0,-1); 
\draw[line width = 2pt] (-1,0) -- (0,0); 
\draw[line width = 2pt] (0,0) -- (1,0);
\node at (-.7,.6) {{\Large $v_2$}};
\end{scope}
\begin{scope}[shift={(6,0)}]
\draw[line width = 2pt] (0,1) -- (0,0); 
\draw[dashed,line width = 2pt] (0,0) -- (0,-1); 
\draw[dashed,line width = 2pt] (-1,0) -- (0,0); 
\draw[dashed,line width = 2pt] (0,0) -- (1,0);
\node at (-.7,.6) {{\Large $v_3$}};
\end{scope}
\begin{scope}[shift={(9,0)}]
\draw[dashed,line width = 2pt] (0,1) -- (0,0); 
\draw[line width = 2pt] (0,0) -- (0,-1); 
\draw[line width = 2pt] (-1,0) -- (0,0); 
\draw[line width = 2pt] (0,0) -- (1,0); 
\node at (-.7,.6) {{\Large $v_4$}};
\end{scope}
\begin{scope}[shift={(12,0)}]
\draw[dashed,line width = 2pt] (0,1) -- (0,0);
\draw[dashed,line width = 2pt] (0,0) -- (0,-1); 
\draw[dashed,line width = 2pt] (-1,0) -- (0,0); 
\draw[line width = 2pt] (0,0) -- (1,0);
\node at (-.7,.6) {{\Large $v_5$}};
\end{scope}
\begin{scope}[shift={(15,0)}]
\draw[line width = 2pt] (0,1) -- (0,0); 
\draw[line width = 2pt] (0,0) -- (0,-1); 
\draw[line width = 2pt] (-1,0) -- (0,0); 
\draw[dashed,line width = 2pt] (0,0) -- (1,0);
\node at (-.7,.6) {{\Large $v_6$}};
\end{scope}
\begin{scope}[shift={(18,0)}]
\draw[dashed,line width = 2pt] (0,1) -- (0,0); 
\draw[dashed,line width = 2pt] (0,0) -- (0,-1); 
\draw[line width = 2pt] (-1,0) -- (0,0); 
\draw[dashed,line width = 2pt] (0,0) -- (1,0);
\node at (-.7,.6) {{\Large $v_7$}};
\end{scope}
\begin{scope}[shift={(21,0)}]
\draw[line width = 2pt] (0,1) -- (0,0); 
\draw[line width = 2pt] (0,0) -- (0,-1); 
\draw[dashed,line width = 2pt] (-1,0) -- (0,0); 
\draw[line width = 2pt] (0,0) -- (1,0);
\node at (-.7,.6) {{\Large $v_8$}};
\end{scope}
\end{tikzpicture}
}
\end{center}
\caption{The odd 8-vertex model weights, with bond states shown in terms of line type, dashed or solid, representing valley or mountain creases, respectively.\label{fig:oddconfigs}}
\end{figure}
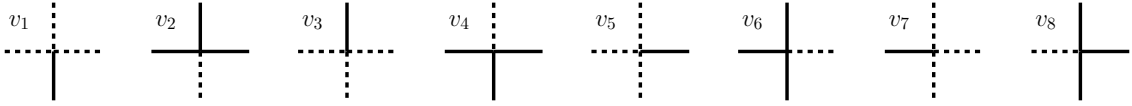

Maekawa's theorem applied to degree 4 vertices demands an odd number of mountain and valley creases. Thus there are eight valid locally flat-foldable vertex configurations at each square lattice vertex, shown in figure~\ref{fig:oddconfigs}, which immediately recalls the odd 8-vertex model~\cite{wu2004k,assis2017temp}. 

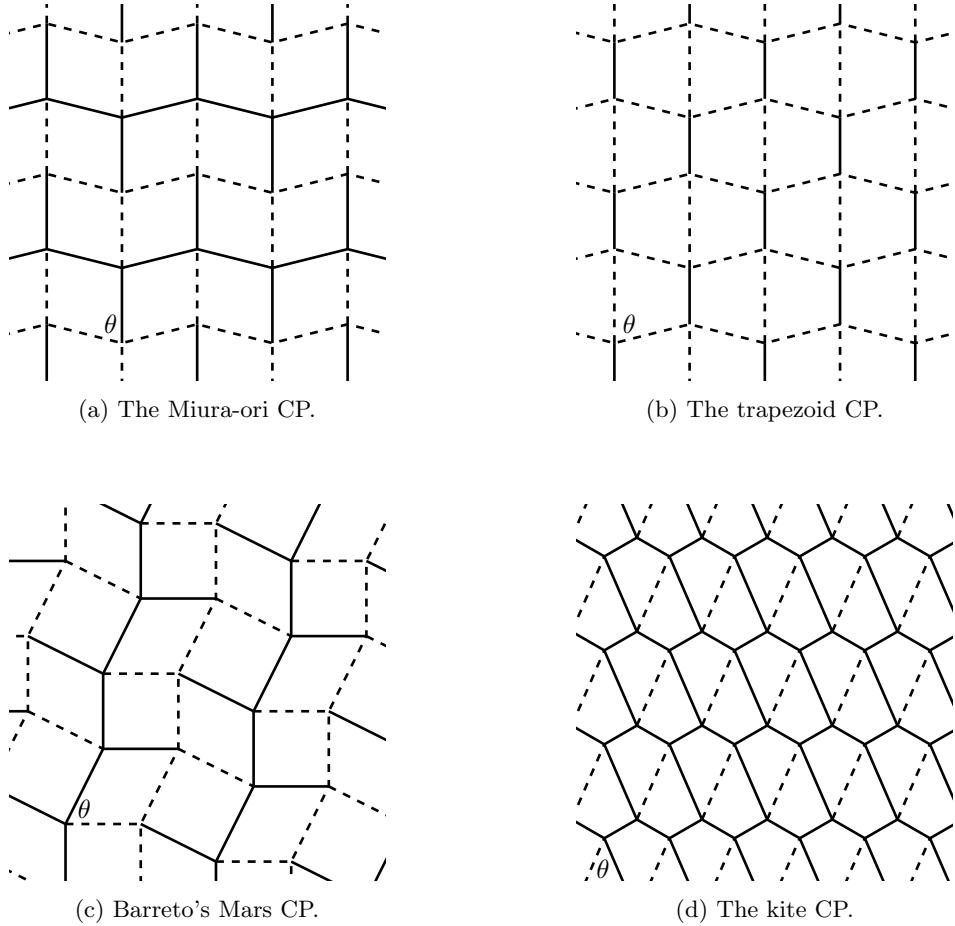
\begin{figure}[htpb]
\begin{center}
\begin{subfigure}{0.45\textwidth}
\centering
\scalebox{1.}{
\begin{tikzpicture}
\clip (0.5,0.5) rectangle (5.5,5.5);
\node at (1.85,1.25) {$\theta$};
\foreach \position in {(0,0),(0,2),(0,4),(0,6)}
{
	\begin{scope}[shift={\position}]
	\draw[line width = 1pt] (0,0) -- (1,0.25) -- (2,0) -- (3,0.25) -- (4,0) -- (5,0.25) -- (6,0);
	\draw[line width = 1pt, dashed] (0,1) -- (1,1.25) -- (2,1) -- (3,1.25) -- (4,1) -- (5,1.25) -- (6,1);
	\draw[line width = 1pt] (1,0.25) -- (1,1.25);
	\draw[line width = 1pt] (3,0.25) -- (3,1.25);
	\draw[line width = 1pt] (5,0.25) -- (5,1.25);
	\draw[line width = 1pt] (0,1) -- (0,2);
	\draw[line width = 1pt] (2,1) -- (2,2);
	\draw[line width = 1pt] (4,1) -- (4,2);
	\draw[line width = 1pt] (6,1) -- (6,2);
	\draw[line width = 1pt,dashed] (0,0) -- (0,1);
	\draw[line width = 1pt,dashed] (2,0) -- (2,1);
	\draw[line width = 1pt,dashed] (4,0) -- (4,1);
	\draw[line width = 1pt,dashed] (6,0) -- (6,1);
	\draw[line width = 1pt,dashed] (1,1.25) -- (1,2.25);
	\draw[line width = 1pt,dashed] (3,1.25) -- (3,2.25);
	\draw[line width = 1pt,dashed] (5,1.25) -- (5,2.25);
	\end{scope}
}
\end{tikzpicture}
}
\caption{The Miura-ori CP.\label{fig:miuraori}}
\end{subfigure}
\begin{subfigure}{0.45\textwidth}
\centering
\scalebox{1.}{
\begin{tikzpicture}
\clip (0.5,0.5) rectangle (5.5,5.5);
\node at (1.2,1.25) {$\theta$};
\foreach \position in {(0,0),(0,2),(0,4),(0,6)}
{
	\begin{scope}[shift={\position}]
	\draw[line width = 1pt,dashed] (0,0) -- (1,0.25) -- (2,0) -- (3,0.25) -- (4,0) -- (5,0.25) -- (6,0);
	\draw[line width = 1pt, dashed] (-1,1) -- (0,1.25) -- (1,1) -- (2,1.25) -- (3,1) -- (4,1.25) -- (5,1) -- (6,1.25);
	\draw[line width = 1pt] (1,0.25) -- (1,1);
	\draw[line width = 1pt] (3,0.25) -- (3,1);
	\draw[line width = 1pt] (5,0.25) -- (5,1);
	\draw[line width = 1pt] (0,1.25) -- (0,2);
	\draw[line width = 1pt] (2,1.25) -- (2,2);
	\draw[line width = 1pt] (4,1.25) -- (4,2);
	\draw[line width = 1pt] (6,1.25) -- (6,2);
	\draw[line width = 1pt,dashed] (0,0) -- (0,1.25);
	\draw[line width = 1pt,dashed] (2,0) -- (2,1.25);
	\draw[line width = 1pt,dashed] (4,0) -- (4,1.25);
	\draw[line width = 1pt,dashed] (6,0) -- (6,1.25);
	\draw[line width = 1pt,dashed] (1,1) -- (1,2.25);
	\draw[line width = 1pt,dashed] (3,1) -- (3,2.25);
	\draw[line width = 1pt,dashed] (5,1) -- (5,2.25);
	\end{scope}
}
\end{tikzpicture}
}
\caption{The trapezoid CP.\label{fig:trapezoid}}
\end{subfigure}
\\[0.4in]
\begin{subfigure}{0.45\textwidth}
\centering
\scalebox{1.}{
\begin{tikzpicture}
\clip (1.75,1.75) rectangle (6.75,6.75);
\node at (2.75,2.7) {$\theta$};
\foreach \position in {(0,0),(2,-0.5),(4,-1),(6,-1.5),(8,-2)}
{
	\begin{scope}[shift={\position}]
	\draw[line width = 1pt] (0,0) -- (0,1) -- (0.5,2) -- (0.5,3) -- (1,4) -- (1,5) -- (1.5,6) -- (1.5,7) -- (2,8);
	\end{scope}
}
\foreach \position in {(1,0),(3,-0.5),(5,-1),(7,-1.5)}
{
	\begin{scope}[shift={\position}]
	\draw[line width = 1pt,dashed] (0,0) -- (0,1) -- (0.5,2) -- (0.5,3) -- (1,4) -- (1,5) -- (1.5,6) -- (1.5,7) -- (2,8);
	\end{scope}
}
\foreach \position in {(0,0),(0.5,2),(1,4),(1.5,6),(2,8)}
{
	\begin{scope}[shift={\position}]
	\draw[line width = 1pt] (0,0) -- (1,0); 
	\draw[line width = 1pt,dashed] (1,0) -- (2,-0.5);
	\draw[line width = 1pt] (2,-0.5) -- (3,-0.5);
	\draw[line width = 1pt,dashed] (3,-0.5) -- (4,-1);
	\draw[line width = 1pt] (4,-1) -- (5,-1);
	\draw[line width = 1pt,dashed] (5,-1) -- (6,-1.5);
	\draw[line width = 1pt] (6,-1.5) -- (7,-1.5);
	\draw[line width = 1pt,dashed] (7,-1.5) -- (8,-2);
	\end{scope}
}
\foreach \position in {(0,1),(0.5,3),(1,5),(1.5,7)}
{
	\begin{scope}[shift={\position}]
	\draw[line width = 1pt,dashed] (0,0) -- (1,0); 
	\draw[line width = 1pt] (1,0) -- (2,-0.5);
	\draw[line width = 1pt,dashed] (2,-0.5) -- (3,-0.5);
	\draw[line width = 1pt] (3,-0.5) -- (4,-1);
	\draw[line width = 1pt,dashed] (4,-1) -- (5,-1);
	\draw[line width = 1pt] (5,-1) -- (6,-1.5);
	\draw[line width = 1pt,dashed] (6,-1.5) -- (7,-1.5);
	\draw[line width = 1pt] (7,-1.5) -- (8,-2);
	\end{scope}
}
\node at (1.8,-0.1) {$\theta$};
\end{tikzpicture}
}
\caption{Barreto's Mars CP.\label{fig:barreto}}
\end{subfigure}
\begin{subfigure}{0.45\textwidth}
\centering
\scalebox{1.}{
\begin{tikzpicture}
\clip (.5,7.44) rectangle (5.5,12.44);
\node at (.85,7.6) {$\theta$};
\foreach \position in {(0,0),(0,1.25),(0,2.5),(0,3.75),(0,5),(0,6.25),(0,7.5),(0,8.75),(0,10),(0,11.25),(0,12.5),(0,13.75)}
{
	\begin{scope}[shift={\position}]
	\draw[line width = 1pt] (0,8) -- (0.433,7) -- (.866,6.75) -- (1.299,5.75) -- (1.732,5.5) -- (2.165,4.5) -- (2.598,4.25) -- (3.031,3.25) -- (3.464,3) -- (3.897,2) -- (4.33,1.75) -- (4.763,0.75) -- (5.196,0.5) -- (5.629,-.5) -- (6.062,-.75) -- (6.495,-1.75);
	\draw[line width = 1pt] (0,6.75) -- (0.433,7);
	\draw[line width = 1pt] (.866,5.5) -- (1.299,5.75);
	\draw[line width = 1pt] (1.732,4.25) -- (2.165,4.5);
	\draw[line width = 1pt] (2.598,3) -- (3.031,3.25);
	\draw[line width = 1pt] (3.464,1.75) -- (3.897,2);
	\draw[line width = 1pt] (4.33,.5) -- (4.763,0.75);
	\draw[line width = 1pt] (5.196,-.75) -- (5.629,-.5);
	\draw[line width = 1pt] (6.062,-2) -- (6.495,-1.75);
	\draw[line width = 1pt,dashed] (0.433,7) -- (.866,8);
	\draw[line width = 1pt,dashed] (1.299,5.75) -- (1.732,6.75);
	\draw[line width = 1pt,dashed] (2.165,4.5) -- (2.598,5.5);
	\draw[line width = 1pt,dashed] (3.031,3.25) -- (3.464,4.25);
	\draw[line width = 1pt,dashed] (3.897,2) -- (4.33,3);
	\draw[line width = 1pt,dashed] (4.763,0.75) -- (5.196,1.75);
	\draw[line width = 1pt,dashed] (5.629,-.5) -- (6.062,0.5);
	\end{scope}
}
\end{tikzpicture}
}
\caption{The kite CP.\label{fig:kite}}
\end{subfigure}
\end{center}\caption{Four of the five origami CPs we consider; the simple square tiling is not shown.\label{fig:cps}}
\end{figure}

In the CPs we consider there is one continuous degree of freedom in their definitions, given in terms of an angle $\theta$ in the staggering quadrilateral unit, as shown in figure~\ref{fig:cps}. For any angle~$\theta< 90^{\circ}$, geometrical folding constraints force two or four vertex weights to be disallowed, that is $v_i=0$, in these CPs, as shown in figure~\ref{fig:angles}. When the angle $\theta=90^{\circ}$, all four CPs become degenerate with the homogeneous square lattice CP where all 8 vertex weights $v_i$ can be non-zero. Therefore we will assume in talking about these CPs that $\theta\neq90^{\circ}$.

\begin{figure}
\begin{center}
\begin{subfigure}{0.15\textwidth}
\centering
\scalebox{.8}{
\begin{tikzpicture}
\draw[line width = 1.5pt] (-1,0) -- (1,0);
\draw[line width = 1.5pt] (0,-1) -- (0,1);
\node at (0.2,0.2) {$\alpha$};
\node at (0.2,-0.25) {$\delta$};
\node at (-0.25,0.2) {$\beta$};
\node at (-0.25,-0.25) {$\gamma$};

\end{tikzpicture}
}
\end{subfigure}
\begin{subfigure}{0.8\textwidth}
\centering
\scalebox{0.95}{
\begin{tabular}{|c|c|c|}
\hline
Angle pattern example & Disallowed weights & Representative CP \\\hline
$\alpha=\beta<90^{\circ}$,~ $\gamma=\delta = \alpha+90^{\circ}$ & $v_1=v_2=0$ & Miura-ori, trapezoid \\\hline
$\alpha<90^{\circ}$,~ $\gamma= \alpha+90^{\circ}$,~ $\beta=\delta=90^{\circ}$ & $v_1=v_2=v_7=v_8=0$ & Barreto's Mars, kite \\\hline
$\alpha=\beta=\gamma=\delta = 90^{\circ}$ & & Square \\\hline
\end{tabular}
}
\end{subfigure}
\end{center}\caption{On the left, a single vertex with angles labeled, and on the right, the dependence of the allowed flat-foldable vertex weights on the angle pattern examples.\label{fig:angles}}
\end{figure}

From the perspective of allowed vertex weights at each vertex of the CP according to the table in figure~\ref{fig:angles}, the Miura-ori is column staggered with units of two vertices, trapezoid and kite are bi-partite staggered with units of two vertices, and Barreto's Mars is column staggered with units of four vertices. However, as we will see below, it is more useful, in order to set the correct ground state, to consider Miura-ori as a column staggered lattice with units of four vertices. Very few exact results are known for staggered vertex models, the majority being free-fermion 8-vertex models; the only solved staggered vertex models which are not free-fermion models have extra interactions which do not have a natural origami interpretation~\cite{assis2017temp}. Therefore, we will focus on free-fermion models, which have simple solution constructions in terms of Pfaffian/dimer methods. Bi-partite staggered even 8-vertex models with units of two vertices were studied in~\cite{hsue1975lw}, column staggered even and odd 8-vertex models with units of two vertices were studied in~\cite{assis2017temp}, and column staggered even 8-vertex models with units of four vertices were considered in~\cite{lin1977w}. Because staggered even 8-vertex models can always be mapped to staggered odd 8-vertex models by re-interpreting bond occupation variables~\cite{wu2004k,assis2017temp}, we can make use of the results in these papers for our purposes. For reference, in appendix~\ref{app:dimer} we give the dimer method of solution for the column staggered odd 8-vertex model with units 
of four vertices, from which all of the remaining vertex model results can be specialized. The  free-energies are given in appendix~\ref{app:freeenergy}. 

We find that the Miura-ori and trapezoid CPs have phase transition points as a function of the vertex weights, points beyond which the long-range order of the ground state disappears. We also find that the Barreto's Mars CP has purely non-interacting defects and that the kite CP can be treated as an effective one-dimensional model. Therefore, neither Barreto's Mars nor kite have a phase transition point. For the simple square CP, the equivalent homogeneous odd 8-vertex free-fermion model does not have a phase transition unless at least two of the vertex weights are disallowed at each vertex~\cite{wu2004k}, but if we consider a staggering of two or four units, then there exist phase transitions for all positive weights, which we give in section~\ref{sec:square}.

The Miura-ori and trapezoid CPs can also be mapped to the exactly solved three-coloring problem on the square lattice~\cite{baxter19702,pegg1982,pearce1989s,ginepro2014h}, where colors represent relative layering orders of neighboring faces. The model does not allow two neighboring faces to have the same color, and the ground state is chosen to only have two colors in a checkerboard fashion. The introduction of the third color maps to the appearance of defects on the lattice. The three-coloring problem has a known phase transition which these two CPs also exhibit as a function of these face layer defects. Because the three-coloring problem does not satisfy the free-fermion condition, this is a phase transition point outside of those seen in the mapping to the staggered free-fermion odd 8-vertex model solutions, and it also happens for a different order parameter in the problem, namely the relative local face layer orderings.

In section~\ref{sec:latticegas} we treat the models as a lattice gas of defects, and using the free-energy results, re-interpreted as pressure, we find exact analytic expressions for the density of defects in these models. We consider crease-reversal defects as ``particles" as well as layer ordering defects. Our analysis of the densities allow us to conclude that Miura-ori and trapezoid are less stable against defects, and hence more tunable, than Barreto's Mars. We also give analytic expressions for the isothermal compressibility, and compute the equations of state, of these lattice gas models.

Finally, we also consider relaxing the local flat-foldability requirement by breaking Maekawa's theorem on the simple square CP in section~\ref{sec:maekawa}. There are then a total of 16 valid vertex weights at each vertex and by imposing a crease reversal symmetry to the vertex weights, we make use of a weak-graph transformation to map this model to an even 8-vertex model with the known solvable subcase of the free-fermion model~\cite{wu1972,assis2017temp}. We find for this model another phase transition point. We discuss in section~\ref{sec:discussion} rigid foldability, finite lattice results, higher degree CPs and free-fermion models, and lattice versions of random crumpling. We finish the paper in section~\ref{sec:conclusions} with some conclusions.

\section{Flat-foldable crease-reversal defects}\label{sec:defects}
\begin{figure}[htpb]
\begin{center}
\scalebox{0.65}{
\begin{tikzpicture}
\draw[color=red,line width = 2pt] (0,1) -- (0,0); 
\draw[dotted,line width = 1pt] (0,0) -- (0,-1); 
\draw[dotted,line width = 1pt] (-1,0) -- (0,0); 
\draw[color=red,line width = 2pt] (0,0) -- (1,0);
\begin{scope}[shift={(3,0)}]
\draw[dotted,line width = 1pt] (0,1) -- (0,0); 
\draw[color=red,line width = 2pt] (0,0) -- (0,-1); 
\draw[color=red,line width = 2pt] (-1,0) -- (0,0); 
\draw[dotted,line width = 1pt] (0,0) -- (1,0); 
\end{scope}
\begin{scope}[shift={(6,0)}]
\draw[dotted,line width = 1pt] (0,1) -- (0,0); 
\draw[color=red,line width = 2pt] (0,0) -- (0,-1); 
\draw[dotted,line width = 1pt] (-1,0) -- (0,0); 
\draw[color=red,line width = 2pt] (0,0) -- (1,0);
\end{scope}
\begin{scope}[shift={(9,0)}]
\draw[color=red,line width = 2pt] (0,1) -- (0,0); 
\draw[dotted,line width = 1pt] (0,0) -- (0,-1); 
\draw[color=red,line width = 2pt] (-1,0) -- (0,0); 
\draw[dotted,line width = 1pt] (0,0) -- (1,0);
\end{scope}
\begin{scope}[shift={(12,0)}]
\draw[color=red,line width = 2pt] (0,1) -- (0,0); 
\draw[color=red,line width = 2pt] (0,0) -- (0,-1); 
\draw[color=red,line width = 2pt] (-1,0) -- (0,0); 
\draw[color=red,line width = 2pt] (0,0) -- (1,0);
\end{scope}
\begin{scope}[shift={(15,0)}]
\draw[color=red,line width = 2pt] (0,1) -- (0,0); 
\draw[color=red,line width = 2pt] (0,0) -- (0,-1); 
\draw[dotted,line width = 1pt] (-1,0) -- (0,0); 
\draw[dotted,line width = 1pt] (0,0) -- (1,0);
\end{scope}
\begin{scope}[shift={(18,0)}]
\draw[dotted,line width = 1pt] (0,1) -- (0,0); 
\draw[dotted,line width = 1pt] (0,0) -- (0,-1); 
\draw[color=red,line width = 2pt] (-1,0) -- (0,0); 
\draw[color=red,line width = 2pt] (0,0) -- (1,0);
\end{scope}
\end{tikzpicture}
}
\end{center}
\caption{Schematic representation of the valid crease reversals around vertices, with crease reversals indicated by solid red lines. Kite only admits the first two types, Miura-ori, trapezoid, and Barreto's Mars admit the first five types, and the simple square CP admits all seven types. Barreto's Mars admits the first two only on one sublattice and the next two on the other sublattice only.\label{fig:creasereversals}}
\end{figure}
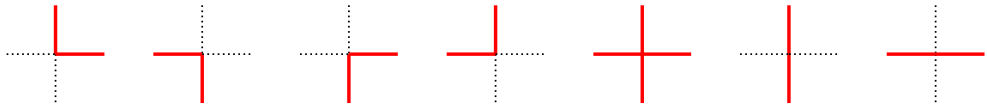

The flat-foldability requirement about each vertex means that crease assignment defects cannot occur in isolation. In order to satisfy Maekawa's theorem, an even number of crease reversals from the ground state must occur at each vertex, either two or four, and since a crease joins two vertices, neighboring vertices are affected. The arrangement of angles in the CP further limits the number of possible crease reversals around a vertex. We show the valid types of crease reversals around vertices in figure~\ref{fig:creasereversals}, where we also indicate which types are valid for each CP.

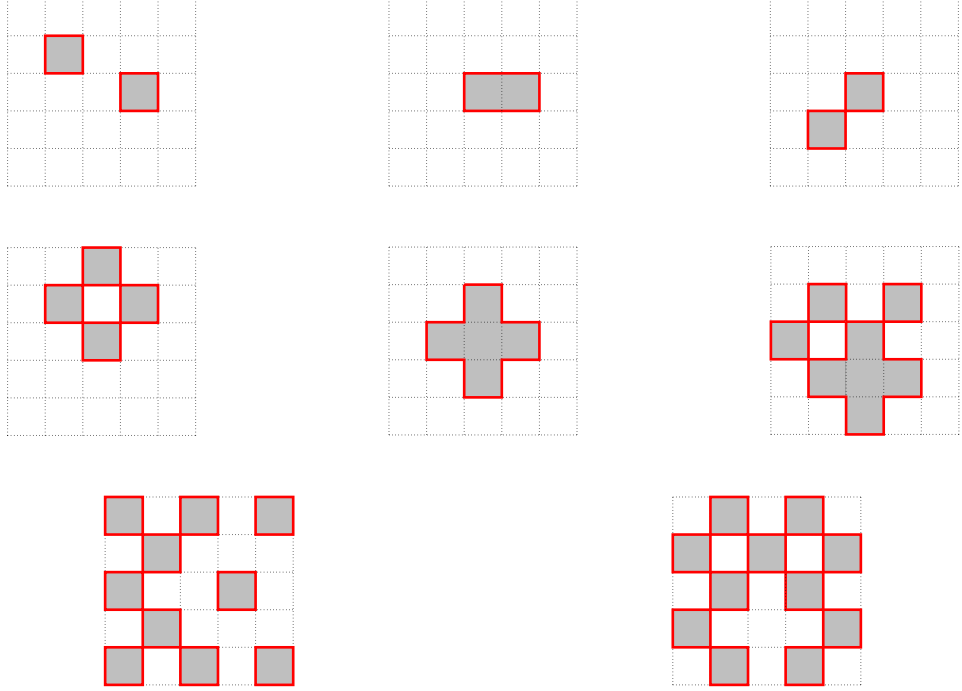
\begin{figure}[htpb]
\begin{center}
\begin{subfigure}{0.3\textwidth}
\centering
\scalebox{0.5}{
\begin{tikzpicture}
\draw[dotted] (0,0) grid (5,5);
\filldraw[color=red,fill=gray!50!white,line width = 2pt] (1,4) rectangle (2,3);
\filldraw[color=red,fill=gray!50!white,line width = 2pt] (3,3) rectangle (4,2);
\end{tikzpicture}
}
\end{subfigure}
\begin{subfigure}{0.3\textwidth}
\centering
\scalebox{0.5}{
\begin{tikzpicture}
\draw[dotted] (0,0) grid (5,5);
\filldraw[color=red,fill=gray!50!white,line width = 2pt] (2,3) rectangle (4,2);
\draw[dotted] (3,2)--(3,3);
\end{tikzpicture}
}
\end{subfigure}
\begin{subfigure}{0.3\textwidth}
\centering
\scalebox{0.5}{
\begin{tikzpicture}
\draw[dotted] (0,0) grid (5,5);
\filldraw[color=red,fill=gray!50!white,line width = 2pt] (2,3) rectangle (3,2);
\filldraw[color=red,fill=gray!50!white,line width = 2pt] (1,1) rectangle (2,2);
\end{tikzpicture}
}
\end{subfigure}
\\[0.3in]
\begin{subfigure}{0.3\textwidth}
\centering
\scalebox{0.5}{
\begin{tikzpicture}
\draw[dotted] (0,0) grid (5,5);
\filldraw[color=red,fill=gray!50!white,line width = 2pt] (2,5) rectangle (3,4);
\filldraw[color=red,fill=gray!50!white,line width = 2pt] (2,3) rectangle (3,2);
\filldraw[color=red,fill=gray!50!white,line width = 2pt] (1,4) rectangle (2,3);
\filldraw[color=red,fill=gray!50!white,line width = 2pt] (3,4) rectangle (4,3);
\end{tikzpicture}
}
\end{subfigure}
\begin{subfigure}{0.3\textwidth}
\centering
\scalebox{0.5}{
\begin{tikzpicture}
\draw[dotted] (1,1) grid (6,6);
\filldraw[fill=gray,color=gray!50!white] (3,5) rectangle (4,4);
\filldraw[fill=gray,color=gray!50!white] (3,3) rectangle (4,2);
\filldraw[fill=gray,color=gray!50!white] (2,4) rectangle (3,3);
\filldraw[fill=gray,color=gray!50!white] (4,4) rectangle (5,3);
\filldraw[fill=gray,color=gray!50!white] (3,4) rectangle (4,3); 
\draw[dotted] (3,4) rectangle (4,3);
\draw[color=red,line width = 2pt] (3,5) -- (4,5) -- (4,4) -- (5,4) -- (5,3) -- (4,3) -- (4,2) -- (3,2) -- (3,3) -- (2,3) -- (2,4) -- (3,4) -- (3,5);
\end{tikzpicture}
}
\end{subfigure}
\begin{subfigure}{0.3\textwidth}
\centering
\scalebox{0.5}{
\begin{tikzpicture}
\draw[dotted] (1,1) grid (6,6);
\filldraw[color=red,fill=gray!50!white,line width = 2pt] (1,3) rectangle (2,4);
\filldraw[color=red,fill=gray!50!white,line width = 2pt] (2,4) rectangle (3,5);
\filldraw[fill=gray!50!white] (2,3) rectangle (3,2);
\filldraw[fill=gray,color=gray!50!white] (3,1) rectangle (4,4);
\filldraw[fill=gray,color=gray!50!white] (3,2) rectangle (5,3);
\filldraw[color=red,fill=gray!50!white,line width = 2pt] (4,4) rectangle (5,5);
\draw[dotted] (3,3) rectangle (4,2);
\draw[color=red,line width = 2pt] (3,4) -- (4,4) -- (4,3) -- (5,3) -- (5,2) -- (4,2) -- (4,1) -- (3,1) -- (3,2) -- (2,2) -- (2,3) -- (3,3) -- (3,4);
\end{tikzpicture}
}
\end{subfigure}
\\[0.3in]
\begin{subfigure}{0.45\textwidth}
\centering
\scalebox{0.5}{
\begin{tikzpicture}
\draw[dotted] (0,0) grid (5,5);
\filldraw[color=red,fill=gray!50!white,line width = 2pt] (1,4) rectangle (2,3);
\filldraw[color=red,fill=gray!50!white,line width = 2pt] (3,3) rectangle (4,2);
\filldraw[color=red,fill=gray!50!white,line width = 2pt] (0,2) rectangle (1,3);
\filldraw[color=red,fill=gray!50!white,line width = 2pt] (0,0) rectangle (1,1);
\filldraw[color=red,fill=gray!50!white,line width = 2pt] (1,1) rectangle (2,2);
\filldraw[color=red,fill=gray!50!white,line width = 2pt] (2,0) rectangle (3,1);
\filldraw[color=red,fill=gray!50!white,line width = 2pt] (4,0) rectangle (5,1);
\filldraw[color=red,fill=gray!50!white,line width = 2pt] (0,4) rectangle (1,5);
\filldraw[color=red,fill=gray!50!white,line width = 2pt] (2,4) rectangle (3,5);
\filldraw[color=red,fill=gray!50!white,line width = 2pt] (4,4) rectangle (5,5);
\end{tikzpicture}
}
\end{subfigure}
\begin{subfigure}{0.45\textwidth}
\centering
\scalebox{0.5}{
\begin{tikzpicture}
\draw[dotted] (0,0) grid (5,5);
\filldraw[color=red,fill=gray!50!white,line width = 2pt] (0,3) rectangle (1,4);
\filldraw[color=red,fill=gray!50!white,line width = 2pt] (0,1) rectangle (1,2);
\filldraw[color=red,fill=gray!50!white,line width = 2pt] (1,0) rectangle (2,1);
\filldraw[color=red,fill=gray!50!white,line width = 2pt] (3,0) rectangle (4,1);
\filldraw[color=red,fill=gray!50!white,line width = 2pt] (2,3) rectangle (3,4);
\filldraw[color=red,fill=gray!50!white,line width = 2pt] (3,2) rectangle (4,3);
\filldraw[color=red,fill=gray!50!white,line width = 2pt] (1,2) rectangle (2,3);
\filldraw[color=red,fill=gray!50!white,line width = 2pt] (1,4) rectangle (2,5);
\filldraw[color=red,fill=gray!50!white,line width = 2pt] (3,4) rectangle (4,5);
\filldraw[color=red,fill=gray!50!white,line width = 2pt] (4,1) rectangle (5,2);
\filldraw[color=red,fill=gray!50!white,line width = 2pt] (4,3) rectangle (5,4);
\draw[dotted] (3,2)--(3,3);
\end{tikzpicture}
}
\end{subfigure}
\end{center}
\caption{Schematic representation of face-flip defect configurations, where solid red lines represent crease reversals for Miura-ori, trapezoid, Barreto's Mars, and the simple square CPs. The first two rows are at low density and the bottom row is at high density. The top middle configuration cannot occur except for the simple square CP, and Barreto's Mars face-flip defects only occur for its diamond faces. \label{fig:defectexamples}}
\end{figure}

From the valid crease-reversals around each vertex, we can see that kite's crease defects form lines which traverse the entire lattice and Barreto's Mars forms only small crease defect loops around the diamond faces of the CP which are independent of each other. The types of crease defects which occur in Miura-ori, trapezoid, and the simple square CPs are more interesting and can interact with each other. For these three CPs, all finite terminating crease defects form loops enclosing polygonal areas which can be viewed as being composed of a number of individual face defects that have been joined together. For these three CPs, then, the finite sized defects have as their basic units what we call ``face-flips", where all four creases around a face are reversed. All of the defects in Barreto's Mars have face-flips as their building blocks, but the other CPs can also have isolated lines of defects which traverse the entire lattice which cannot be built up using only face-flip defects. In figure~\ref{fig:defectexamples} we show a sample of possible face-flip defects for these models at low and high densities of crease reversals.

Though face-flips are useful for understanding defects, except for Barreto's Mars it is important that the crease-reversal models use crease reversals as a defect variable and not face-flips. At low density, the two cases will agree, but at high face-flip density the lattice will be covered by face-flips, which is equivalent to the ground state and not the fully crease-reversed lattice. Since we are interested in characterizing the effect of defects in the lattice and want the high density to correspond to the lattice being fully crease-reversed, we see that we cannot use face-flips as a useful Boltzmann weight in these models. For Barreto's Mars, since the only allowed defects occur on the diamond faces and the defects are non-interacting, face-flips are an equally valid Boltzmann weight, and the model can be viewed as a non-interacting lattice gas of diamonds.

\section{Origami Boltzmann weights}\label{sec:weights}
Origami CPs fold in a manner which depends on the crease assignments, the arrangement of those creases around individual vertices, and the ordering of the layers of faces. We consider Boltzmann weights for each of these three cases, weights that depend on the arrangement of crease assignments around a vertex using the standard odd 8-vertex model weights shown in figure~\ref{fig:oddconfigs}, individual crease Boltzmann weights for mountain and valley crease assignments separately, and relative layer ordering weights for faces with respect to their neighboring faces. We consider each type of weight in turn below.

\subsection{Vertex Boltzmann weights and free-fermion models}
If treated purely as a vertex model, the origami models then have Boltzmann weights which depend only on the pattern of mountain and valley creases around a vertex, as shown in figure~\ref{fig:oddconfigs}. From the perspective of flat-foldability, this is a natural set of variables, since flat-foldability depends crucially on the arrangement of the creases around a vertex~\cite{demaine2007o}. Since vertex models are well studied in statistical mechanics, we can make use of exact solutions to study flat-foldable origami CPs, in particular we make use of ``free-fermion" models. Since we generally consider that crease assignment defects must also respect local flat-foldability requirements, we limit the vertex weights to only the set of 8 weights showin in figure~\ref{fig:oddconfigs}, except in section~\ref{sec:maekawa} where we break Maekawa's theorem.

So-called free-fermion models are 8-vertex models which satisfy a ``free-fermion condition", which for the odd 8-vertex model is~\cite{assis2017temp}
\beq
v_1v_2+v_3v_4=v_5v_6+v_7v_8
\eeq
This condition allows the model to be solved via Pfaffian techniques, such as via dimer methods~\cite{mccoy1973w,mccoy2014w}, which we demonstrate explicitely in appendix~\ref{app:dimer}. On the homogeneous square lattice the free-fermion model is equivalent to the Ising model on the union-jack or checkerboard lattices~\cite{assis2017temp}. Except for the simple square CP, the other origami CPs require staggering units, and it is unknown whether there are similar interpretations for staggered free-fermion models in terms of Ising models on a different lattice.

\begin{figure}[htpb]
\begin{center}
\scalebox{.65}{
\begin{tikzpicture}
\draw[color=red,line width = 2pt] (0,-1) -- (0,1); 
\draw[color=red,line width = 2pt] (-1,0) -- (1,0); 
\node at (2,0) {\large\textbf{=}};
\begin{scope}[shift={(4,0)}]
\draw[dotted,line width = 1pt] (-1,0) -- (1,0); 
\draw[dotted,line width = 1pt] (0,-1) -- (0,1); 
\draw[color=red,line width = 2pt] (0,1) to[out=90,in=45] (-.2,.2) to[out=-135,in=0] (-1,0); 
\draw[color=red,line width = 2pt] (1,0) to[out=180,in=45] (.2,-.2) to[out=-135,in=90] (0,-1); 
\node at (2,0) {\large\textbf{+}};
\end{scope}
\begin{scope}[shift={(8,0)}]
\draw[dotted,line width = 1pt] (-1,0) -- (1,0); 
\draw[dotted,line width = 1pt] (0,-1) -- (0,1); 
\draw[color=red,line width = 2pt] (0,1) to[out=90,in=135] (.2,.2) to[out=-45,in=0] (1,0); 
\draw[color=red,line width = 2pt] (-1,0) to[out=0,in=135] (-.2,-.2) to[out=-45,in=-90] (0,-1); 
\end{scope}
\end{tikzpicture}
}
\\[0.1in]
\scalebox{.65}{
\begin{tikzpicture}
\draw[dotted] (-2,-2) grid (2,2);
\filldraw[color=red,fill=gray!50!white,line width = 2pt] (-1,0) rectangle (0,1);
\filldraw[color=red,fill=gray!50!white,line width = 2pt] (0,-1) rectangle (1,0);
\node at (3,0) {\large\textbf{=}};
\begin{scope}[shift={(6,0)}]
\draw[dotted] (-2,-2) grid (2,2);
\filldraw[color=red,fill=gray!50!white,line width = 2pt] (-1,1) -- (0,1) -- (0,0.5) to[out=90,in=45] (-.1,.1) to[out=-135,in=0] (-0.5,0) -- (-1,0) -- (-1,1); 
\draw[color=red,fill=gray!50!white,line width = 2pt] (1,-1) -- (1,0) -- (0.5,0) to[out=180,in=45] (.1,-.1) to[out=-135,in=90] (0,-0.5) -- (0,-1) -- (1,-1); 
\node at (3,0) {\large\textbf{+}};
\end{scope}
\begin{scope}[shift={(12,0)}]
\draw[dotted] (-2,-2) grid (2,2);
\filldraw[color=red,fill=gray!50!white,line width = 2pt] (-1,0) -- (-1,1) -- (0,1) -- (0,0.5) to[out=90,in=135] (.1,.1) to[out=-45,in=0] (0.5,0) -- (1,0) -- (1,-1) -- (0,-1) -- (0,-0.5) to[out=90,in=-45] (-.1,-.1) to[out=135,in=180] (-0.5,0) -- (-1,0); 
\end{scope}
\end{tikzpicture}
}
\end{center}
\caption{Graphical interpretation of the factor of 2 in the free-fermion condition of the Miura-ori and trapezoid models. On top, the interpretation around each vertex with four crease reversals, below an example with two face flip defects.\label{fig:factoroftwo}}
\end{figure}
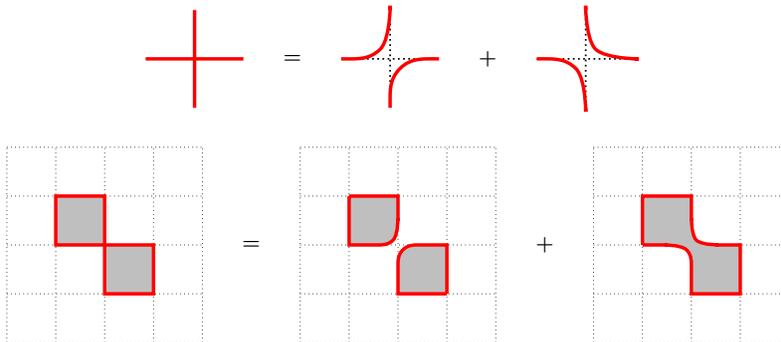

Since we are primarily interested in staggered lattices, we choose independent sets of vertex weights for each vertex of the two or four vertices appearing in each unit, each of which satisfies its own free-fermion condition. Except for the simple square CP, the pattern of angles around a CP's vertices causes certain vertex weights to be disallowed, with examples shown in figure~\ref{fig:angles}. For the trapezoid CP we use two independent sets of weights $v_i$ and $w_i$ which satisfy
\bal
v_1v_2 &= v_5v_6+v_7v_8 \\
w_3w_4 &= w_5w_6+w_7w_8
\eal
We are interested in the ground state being given by alternations of $v_1$ and $w_3$ with the remaining allowed vertex weights being considered defects. If we choose $v_1=w_3=1$ as the ground state weights, and if we choose $v_5=v_6=v_7=v_8=y$ and $w_5=w_6=w_7=w_8=y$ for the defects which involve two crease reversals (each crease is shared among two vertices, so the factor of $y^2$ is split among them), then we must have $v_2=2y^2$ and $w_4=2y^2$ in order to respect the free-fermion condition. The defect corresponding to reversing all four creases at a vertex, then, is twice as large as the naive factor of $y^2$ one might presume for each crease reversal. We can interpret this extra factor of 2 occuring in the fully crease-reversed vertex weights in the following graphical sense, shown in figure~\ref{fig:factoroftwo}, that each of the vertices with four crease reversals corresponds to two different arrangements of the defect loops. Therefore, the effect of imposing the free-fermion condition is to favor defects clustering together in the lattice. As long as $y<\tfrac{1}{2}$, defect weights $v_2$ and $w_4$ will be smaller than the other defect weights. However, as we will see below, the model has a phase transition point at $y=\sqrt{2}/2$, at which point the vertex weights which are crease reversals of the ground state weights are equally favored, that is, $v_2=w_4=v_1=w_3=1$. For Miura-ori, similar considerations and results hold.

For Barretos's Mars, there are four vertex weights disallowed at each vertex and we have the following free-fermion conditions for the four sets of vertex weights
\beq
v_1v_2 = v_7v_8,\qquad w_3w_4 = w_7w_8,\qquad t_1t_2 = t_5t_6,\qquad u_3u_4 = u_5u_6 
\eeq
with the ground state given by $v_1$, $w_3$ $t_2$, $u_4$ according to figure~\ref{fig:barretonotation}. Again setting the ground state weights to equal 1, we can choose defect weights given by a factor $y$ that corresponds to the number of crease reversals at each vertex, that is, the fully crease-reversed defect weights $v_2$, $w_4$, $t_1$, $u_3$ all equal $y^2$ and the remaining weights equal $y$. The extra factor of 2 in the fully crease-reversed weights of Miura-ori and trapezoid is not present in this model. 

\subsection{Crease Boltzmann weights}
The partition function of any vertex model can be modified to show explicitely the dependence of the partition function on the edge states, that is, the crease assignments; this is done by using a transformation of the vertex weights, which we now show. We define crease Boltzmann weights $a_m,~a_v,\ldots h_v$ as shown in figure~\ref{fig:creaseweights} for a general unit of four vertices, with individual weights for mountain and valley assignments of the same crease. 
\begin{figure}[htpb]
\begin{center}
\scalebox{1.15}{
\begin{tikzpicture}
\draw[line width = 1.5pt] (0,-1) -- (0,2);
\draw[line width = 1.5pt] (1,-1) -- (1,2);
\draw[line width = 1.5pt] (-1,0) -- (2,0);
\draw[line width = 1.5pt] (-1,1) -- (2,1);
\node at (-0.2,-0.2) {$v_i$};
\node at (1.25,-0.2) {$w_i$};
\node at (-0.2,1.2) {$t_i$};
\node at (1.25,1.15) {$u_i$};
\node at (0.15,-0.5) {$a$}; \node at (0.15,1.5) {$a$};
\node at (0.85,-0.5) {$c$}; \node at (0.85,1.5) {$c$};
\node at (-0.5,0.2) {$d$}; \node at (1.5,.2) {$d$};
\node at (-0.5,0.8) {$h$}; \node at (1.5,.8) {$h$};
\node at (0.5,0.2) {$b$};
\node at (0.5,0.8) {$f$};
\node at (0.15,0.5) {$e$};
\node at (0.85,0.5) {$g$};
\end{tikzpicture}
}
\end{center}
\caption{The crease assignment weight notation convention on a unit of four vertices for crease weights $a\ldots h$, each of which represents two weights for the mountain and valley assignment separately, e.g. $a_m$, $a_v$ for the $a$ crease notation, respectively. The notation for the four vertex weights $t_i$, $u_i$, $v_i$, $w_i$ are also shown. \label{fig:creaseweights}}
\end{figure}

In order to avoid either double-counting or using square roots, we use the following asymmetric transformation of the vertex weights, where for each vertex weight we add in its lower and right crease assignment dependence
\bat{4}
v_1 &~\to~ a_m b_v v_1,&\qquad v_3 &~\to~ a_v b_v v_3,&\qquad v_5 &~\to~ a_v b_m v_5,&\qquad v_7 &~\to~ a_v b_v v_7, \nonumber \\
v_2 &~\to~ a_v b_m v_2,&\qquad v_4 &~\to~ a_m b_m v_4,&\qquad v_6 &~\to~ a_m b_v v_6,&\qquad v_8 &~\to~ a_m b_m v_8, \label{creasetrans1}\\[0.1in]
w_1 &~\to~ c_m d_v w_1,&\qquad w_3 &~\to~ c_v d_v w_3,&\qquad w_5 &~\to~ c_v d_m w_5,&\qquad w_7 &~\to~ c_v d_v w_7, \nonumber \\
w_2 &~\to~ c_v d_m w_2,&\qquad w_4 &~\to~ c_m d_m w_4,&\qquad w_6 &~\to~ c_m d_v w_6,&\qquad w_8 &~\to~ c_m d_m w_8, \\[0.1in]
t_1 &~\to~ e_m f_v t_1,&\qquad t_3 &~\to~ e_v f_v t_3,&\qquad t_5 &~\to~ e_v f_m t_5,&\qquad t_7 &~\to~ e_v f_v t_7, \nonumber \\
t_2 &~\to~ e_v f_m t_2,&\qquad t_4 &~\to~ e_m f_m t_4,&\qquad t_6 &~\to~ e_m f_v t_6,&\qquad t_8 &~\to~ e_m f_m t_8, \\[0.1in]
u_1 &~\to~ g_m h_v u_1,&\qquad u_3 &~\to~ g_v h_v u_3,&\qquad u_5 &~\to~ g_v h_m u_5,&\qquad u_7 &~\to~ g_v h_v u_7, \nonumber \\
u_2 &~\to~ g_v h_m u_2,&\qquad u_4 &~\to~ g_m h_m u_4,&\qquad u_6 &~\to~ g_m h_v u_6,&\qquad u_8 &~\to~ g_m h_m u_8 \label{creasetrans4}
\eat
All of the odd 8-vertex results below can incorporate the extra dependence on the crease assignments after performing these transformations. We can choose the crease assignment variables corresponding to the ground state crease configuration to all have value 1. Then as long as the remaining crease assignment variables are $<1$, the ground state will continue to be favored. 

For free-fermion models, the free-fermion condition is valid identically with or without such a crease assignment dependence. For the simple square CP, Barreto's Mars, and the kite CP, the free-energy can be written purely in terms of crease assignment weights, since in the free-fermion condition each of the vertex weights can be chosen to have the value $v_i=w_i=t_i=u_i=1$. For the Miura-ori and trapezoid CPs, though, the free-fermion condition cannot be satisfied in this manner, so that the vertex weight dependence must continue to be explicitely shown. 


We show explicitely below for the simple square CP that a phase transition point can be found which depends solely on these crease assignment weights.

\subsection{Relative local face layering Boltzmann weights}
The third type of origami Boltzmann weight we consider captures the relative layer ordering of neighboring faces. We show in section~\ref{sec:threecolor} below how to map from the vertex weights of the Miura-ori and the trapezoid models to the 3-coloring problem on the square lattice, where each face on the square lattice is assigned one of 3 colors such that no two neighboring faces have the same color. The convention we define in figure~\ref{fig:coloringmap} for both models shows how to assign face colors so that mountain and valley creases between two neighboring faces describes the relative layering of the two faces consistently in the lattice. We then consider color Boltzmann weights $z_i$ for the states of each face. The exactly solvable 3-coloring problem has a known phase transition point, which happens whenever all three color fugacities are equal $z_0=z_1=z_2$, so that another phase transition point depending on the local layer ordering of faces can be established analytically, falling outside of the free-fermion model's free-fermion condition constraints.

\section{Flat-foldable vertex models}\label{sec:vertexmodels}
We now consider the origami CPs as staggered free-fermion odd 8-vertex models which have flat-foldable crease assignment defects. Using vertex weights as well as crease assignment weights we consider their free-energies as well as their critical phenomena. 

\subsection{Miura-ori}\label{sec:miura}
\begin{figure}[htpb]
\begin{center}
\scalebox{1.25}{
\begin{tikzpicture}
\draw[line width = 1pt] (0,2) -- (1,2.25) -- (2,2) -- (3,2.25);
\draw[line width = 1pt, dashed] (0,1) -- (1,1.25) -- (2,1) -- (3,1.25);
\draw[line width = 1pt] (1,0.25) -- (1,1.25);
\draw[line width = 1pt] (1,2.25) -- (1,3.25);
\draw[line width = 1pt] (2,1) -- (2,2);
\draw[line width = 1pt,dashed] (2,0) -- (2,1);
\draw[line width = 1pt,dashed] (2,2) -- (2,3);
\draw[line width = 1pt,dashed] (1,1.25) -- (1,2.25);
\node at (0.8,0.97) {$v_i$};
\node at (1.8,0.82) {$w_i$};
\node at (1.16,2.4) {$t_i$};
\node at (2.2,2.22) {$u_i$};
\end{tikzpicture}
}
\end{center}
\caption{The column staggered Miura-ori four vertex unit and vertex weight notation convention.\label{fig:miuranotation}}
\end{figure}
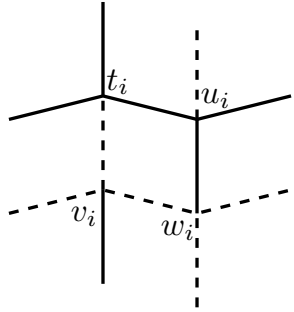

For the Miura-ori four vertex unit we use the vertex weight notation shown in figure~\ref{fig:miuranotation}.

At each vertex, the pattern of the angles causes two vertex weights to be disallowed, so that we have
\beq
v_3 = v_4 = t_3 = t_4 = w_1 = w_2 = u_1= u_2=0
\eeq
We can use the free-fermion conditions for each independent set of vertex weights to set the ground state to be given only in terms of the weights $v_1$, $w_3$, $t_2$, $u_4$ as discussed above. We give the fully asymmetric free-energy in appendix~\ref{app:freeenergy}. 

The fully asymmetric model has phase transition points given when the following conditions hold
\bal
&t_1v_2w_4u_3+v_1w_3t_2u_4+(v_5w_6+v_7w_8)(t_5u_6+t_7u_8)+(v_6w_5+v_8w_7)(t_6u_5+t_8u_7)\nonumber\\
&\qquad\qquad\qquad\qquad\qquad\qquad\qquad \pm(v_5w_7-v_7w_5)(t_5u_7-t_7u_5)\pm(v_6w_8-v_8w_6)(t_6u_8-t_8u_6)=0 \\
&t_1v_2w_4u_3+v_1w_3t_2u_4\pm(v_5w_7+v_7w_5)(t_5u_7+t_7u_5)\pm(v_6w_8+v_8w_6)(t_6u_8+t_8u_6)\nonumber\\
&\qquad\qquad\qquad\qquad\qquad\qquad\qquad -(v_5w_6-v_7w_8)(t_5u_6-t_7u_8)-(v_6w_5-v_8w_7)(t_6u_5-t_8u_7)=0 
\eal
There can be up to 5 phase transitions in this model, which are generically logarithmic of second order, although in special cases it can have two first-order phase transitions or up to three second order phase transitions with exponent $\alpha=\tfrac{1}{2}$~\cite{lin1977w}.

If we assume that the vertex weights $t_i$ and $u_i$ are given by the corresponding crease-inverted $v_i$ and $w_i$ weights respectively, the free-energy factorizes as an exact square and the phase transition points become more simply
\beqr
& v_1w_3\pm v_2 w_4 =0 \label{miuraphasetranssym1}\\
& v_1^2w_3^2+v_2^2w_4^2+2(v_5w_6+v_7w_8)(v_6w_5+v_8w_7)-2(v_5w_7-v_7w_5)(v_6w_8-v_8w_6) = 0 \\
& v_1^2w_3^2+v_2^2w_4^2-2(v_5w_6-v_7w_8)(v_6w_5-v_8w_7)+2(v_5w_7+v_7w_5)(v_6w_8+v_8w_6) = 0
\eeqr
If we further assume full symmetry of the weights, so that the $w_i$ weights are equal to the $v_i$ weights rotated by $180^{\circ}$, the phase transition conditions become
\beqr
& v_1^2\pm v_2^2 =0 \label{miuraphasetranssym2}\\
& v_1^4+v_2^4+2(v_5v_8+v_6v_7)(v_6v_7+v_5v_8)-2(v_5^2-v_7^2)(v_6^2-v_8^2) = 0 \\
& v_1^4+v_2^4-2(v_5v_8-v_6v_7)(v_6v_7-v_5v_8)+2(v_5^2+v_7^2)(v_6^2+v_8^2) = 0
\eeqr
If we choose $v_1=1$ to set the ground state and $v_5=v_6=v_7=v_8=y$ with $y<1$, the free-fermion condition gives $v_2=2y^2$ and the free-energy is given as follows
\bal
-\beta f_{Mi} =  \frac{1}{16\pi^2}\int_0^{2\pi}\int_0^{2\pi} \ln\big[1+4y^4+16y^8+4y^4\left(\cos\theta_1+\cos\theta_2-\cos\theta_1\cos\theta_2 \right) \big] d\theta_1d\theta_2 \label{miurafreey}
\eal
The phase transition points then become
\beqr
& 1\pm 4y^4 =0 \\
& 1+16y^8+ 8y^4 =0
\eeqr
and we see that there's a physical phase transition point at a positive $y$ value
\beq
y_c = \sqrt{2}/2
\eeq
This point $y_c$ corresponds to the point where the fully crease-reversed vertex weights equal the ground state weights, $v_2=w_4=t_1=u_3=1$, although the remaining valid vertex weight defects are still less favored, $<1$, at this point.

\subsection{Trapezoid}\label{sec:trapezoid}
\begin{figure}[htpb]
\begin{center}
\scalebox{1.25}{
\begin{tikzpicture}
\draw[line width = 1pt, dashed] (0,1.25) -- (1,1) -- (2,1.25) -- (3,1);
\draw[line width = 1pt] (1,0.25) -- (1,1);
\draw[line width = 1pt] (2,1.25) -- (2,2);
\draw[line width = 1pt,dashed] (2,0) -- (2,1.25);
\draw[line width = 1pt,dashed] (1,1) -- (1,2.25);
\node at (0.8,0.85) {$v_i$};
\node at (1.8,1.4) {$w_i$};
\end{tikzpicture}
}
\end{center}
\caption{The bi-partite staggered trapezoid two vertex unit and vertex weight notation convention.\label{fig:trapezoidnotation}}
\end{figure}
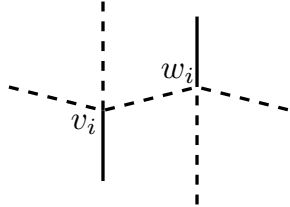

For the trapezoid two vertex unit we use the vertex weight notation shown in figure~\ref{fig:trapezoidnotation}.

At each vertex, the pattern of the angles causes two vertex weights to be disallowed, so that we have
\beq
v_3 = v_4 = w_1 = w_2=0
\eeq
We can use the free-fermion conditions for each independent set of vertex weights to set the ground state to be given only in terms of the weights $v_1$ and $w_3$, as discussed above. We give the free-energy in appendix~\ref{app:freeenergy}. 

This model has phase transition points when the following conditions hold
\bal
v_1w_3+v_2w_4\pm v_5w_7\pm v_6w_8\pm v_7w_5\pm v_8w_6 &=0 \\
v_1w_3+v_2w_4\pm v_5w_7\pm v_6w_8\mp v_7w_5\mp v_8w_6 &=0 \\
\eal
There can be up to three phase transitions of the model which are in general logarithmic of second order, although in special cases there is only one phase transition which has exponent $\alpha=\tfrac{1}{2}$~\cite{hsue1975lw}. 

If we assume the symmetry such that the $w_i$ vertex weights are equal to the $180^{\circ}$ rotated $v_i$ weights, then the free-energy factorizes into two parts and the phase transition points are given by
\bal
v_1^2+v_2^2\pm v_5^2\pm v_6^2\pm v_7^2\pm v_8^2 &=0 \\
v_1^2+v_2^2\pm v_5^2\pm v_6^2\mp v_7^2\mp v_8^2 &=0 \\
\eal
If we choose $v_1=1$ to set the ground state and $v_5=v_6=v_7=v_8=y$ with $y<1$, the free-fermion condition gives $v_2=2y^2$ and the free-energy becomes 
\bal
-\beta f_{C} =  \frac{1}{8\pi^2}\int_0^{2\pi}\int_0^{2\pi} \ln\big(1+4y^4+2y^2\left[\cos\theta_2-\cos(\theta_1+\theta_2) \right]\big) d\theta_1d\theta_2 \label{trapfreey}
\eal
The phase transition points then become
\bal
1+4y^4\pm 4y^2 &=0 \\
1+ 4y^4 &=0
\eal
and we see that there's a physical phase transition point at positive $y$ value
\beq
y_c = \sqrt{2}/2
\eeq
It can be shown that the free-energy of~(\ref{trapfreey}) is equal to the free-energy~(\ref{miurafreey}) of Miura-ori. In this symmetric defect case, but not in general, the two models are therefore equal. As with the Miura-ori case, $y_c$ corresponds to the point where the fully crease-reversed vertex weights equal the ground state weights, $v_2=w_4=1$, although the remaining valid vertex weight defects are still less favored, $<1$, at this point.

\subsection{Barreto's Mars}\label{sec:barreto}
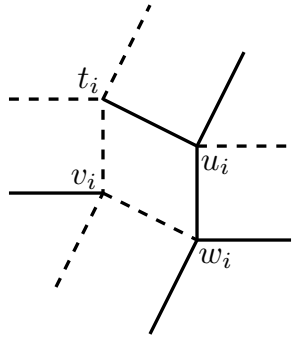
\begin{figure}[htpb]
\begin{center}
\scalebox{1.25}{
\begin{tikzpicture}
\draw[line width = 1pt] (1.5,-1.5) -- (2,-.5) -- (2,0.5) -- (2.5,1.5);
\draw[line width = 1pt,dashed] (.5,-1) -- (1,0) -- (1,1) -- (1.5,2);
\draw[line width = 1pt] (0,0) -- (1,0); 
\draw[line width = 1pt,dashed] (1,0) -- (2,-0.5);
\draw[line width = 1pt] (2,-0.5) -- (3,-0.5);
\draw[line width = 1pt,dashed] (0,1) -- (1,1); 
\draw[line width = 1pt] (1,1) -- (2,0.5);
\draw[line width = 1pt,dashed] (2,0.5) -- (3,0.5);
\node at (.8,0.15) {$v_i$};
\node at (2.2,-0.7) {$w_i$};
\node at (.85,1.2) {$t_i$};
\node at (2.2,0.3) {$u_i$};
\end{tikzpicture}
}
\end{center}
\caption{The column staggered Barreto's Mars four vertex unit and vertex weight notation convention.\label{fig:barretonotation}}
\end{figure}

For the Barreto's Mars four vertex unit we use the vertex weight notation shown in figure~\ref{fig:barretonotation}.

At each vertex, the pattern of the angles causes four vertex weights to be disallowed, so that we have
\beq
t_3 = t_4 = t_7=t_8 = u_1= u_2=u_7=u_8 = v_3 = v_4 = v_5=v_6 = w_1 = w_2=w_5=w_6=0
\eeq
We can use the free-fermion conditions for each independent set of vertex weights to set the ground state to be given only in terms of the weights $v_7$, $w_8$, $t_5$, $u_6$ as discussed above. As shown in appendix~\ref{app:freeenergy}, the free-energy is given simply by the expression
\bal
-\beta f_{Ma} = \frac{1}{8}\, \ln\big[(t_1u_3v_2w_4+t_2u_4v_1w_3+t_5u_6v_7w_8+t_6u_5v_8w_7)^2-2u_3u_4w_3w_4(t_1t_2v_1v_2-t_5t_6v_7v_8)\big]
\eal
Upon expanding, the argument of the logarithm contains only positive terms, so that there is no physical phase transition point for this model. We can understand this as follows. All of the possible flat-foldable crease-reversal defects correspond to the reversal of all creases around only the diamond faces of the lattice. Since two diamond faces never share a mutual crease, all of the defects occur independently of all others --- there is no interaction among the defects. Therefore, a decimation procedure could be performed for each diamond face, and the resulting model would be in a frozen state. 

If we re-write the free-energy with defect fugacities as follows
\beqr
& t_5 = u_6 = v_7 = w_8 = 1 \\
& t_6=u_5=v_8=w_7=y^2 \\
& t_1=t_2=u_3=u_4=v_1=v_2=w_3=w_4=y 
\eeqr
we have the very simple expression
\beq
-\beta f_{Ma} = \frac{1}{2}\,\ln(1+y^4)
\eeq
This agrees with our earlier interpretation of the face-flip defects in Barreto's Mars as a non-interacting lattice gas of diamond faces. A non-interacting lattice gas with fugacity $z$ on a lattice of size $\mathcal{N}$ will have a partition function give by
\beq
Z = 1+\mathcal{N}z+\binom{\mathcal{N}}{2}z^2+\ldots +\binom{\mathcal{N}}{\mathcal{N}}z^\mathcal{N} = (1+z)^{\mathcal{N}}
\eeq
with a free-energy given by
\beq
-\beta f = \ln(1+z)
\eeq
The defect free-energy of Barreto's Mars now follows, since each diamond in Barreto's Mars requires four $y$ creases, $z=y^4$ and there are $\mathcal{N}/2$ diamond faces in the lattice.

\subsection{Kite}\label{sec:kite}
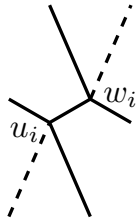
\begin{figure}[htpb]
\begin{center}
\scalebox{1.25}{
\begin{tikzpicture}
\draw[line width = 1pt,dashed] (0,0) -- (0.433,1);
\draw[line width = 1pt,dashed] (.866,1.25) -- (1.299,2.25);
\draw[line width = 1pt] (0,1.25) -- (0.433,1) -- (.866,0);
\draw[line width = 1pt] (.433,1) -- (.866,1.25);
\draw[line width = 1pt] (.433,2.25) -- (.866,1.25) -- (1.299,1);
\node at (.17,.9) {$u_i$};
\node at (1.19,1.27) {$w_i$};
\end{tikzpicture}
}
\end{center}
\caption{The bi-partite staggered kite two vertex unit and vertex weight notation convention.\label{fig:kitenotation}}
\end{figure}

For the kite two vertex unit we use the vertex weight notation shown in figure~\ref{fig:kitenotation}.

At each vertex, the pattern of the angles causes four vertex weights to be disallowed, so that we have
\beq
v_3 = v_4 = v_5=v_6 = w_1 = w_2 = w_7=w_8=0
\eeq
We can use the free-fermion conditions for each independent set of vertex weights to set the ground state to be given only in terms of the weights $v_8$ and $w_6$, as discussed above. 

We give the free-energy as a specialization of the four unit column staggered odd 8-vertex free-fermion model in appendix~\ref{app:freeenergy}, but this is unnecessary. It can be shown by considering the possible valid neighbors of each vertex weight that a given vertex weight must always be repeated on the lower-right diagonal of the weight. For a toroidal boundary condition lattice, the model is effectively a one-dimensional model. We can therefore solve it directly by a standard transfer matrix procedure and without needing to impose the free-fermion condition on the weights.

A given unit of two vertices can only have the following 8 valid vertex weight combinations
\beq
v_1w_3,\quad v_1w_5,\quad v_8w_4,\quad v_8w_6,\quad v_2w_4,\quad v_2w_6,\quad v_7w_3,\quad v_7w_5 \label{huffvalid}
\eeq
We then define the following asymmetric one-dimensional transfer matrix, with rows and columns indexed by the order shown in~(\ref{huffvalid})
\beq
T = \begin{pmatrix}
v_1w_3 & 0 & 0 & v_8w_6 & & v_2w_6 & v_7w_3 & 0 \\
v_1w_3 & 0 & 0 & v_8w_6 & & v_2w_6 & v_7w_3 & 0 \\
v_1w_3 & 0 & 0 & v_8w_6 & & v_2w_6 & v_7w_3 & 0 \\
v_1w_3 & 0 & 0 & v_8w_6 & & v_2w_6 & v_7w_3 & 0 \\
0 & v_1w_5 & v_8w_4 & 0 & v_2w_4 & 0 & 0 & v_7w_5 \\
0 & v_1w_5 & v_8w_4 & 0 & v_2w_4 & 0 & 0 & v_7w_5 \\
0 & v_1w_5 & v_8w_4 & 0 & v_2w_4 & 0 & 0 & v_7w_5 \\
0 & v_1w_5 & v_8w_4 & 0 & v_2w_4 & 0 & 0 & v_7w_5 
\end{pmatrix}
\eeq
The matrix $T$ has rank 2 with non-zero eigenvalues
\beq
\lambda_{\pm} = \frac{1}{2}\,\left[v_1w_3+v_2w_4+v_7w_5+v_8w_6 \pm \sqrt{D}\right]
\eeq
where
\bal
D &= (v_1w_3-v_2w_4)^2+(v_7w_5-v_8w_6)^2+2(v_1w_3+v_2w_4)(v_7w_5+v_8w_6)+4v_1v_2w_5w_6+4v_7v_8w_3w_4
\eal

The partition of the one-dimensional model with $N$ sites and periodic boundary conditions is given by
\beq
Z_H = \mathrm{Tr}\left(T^N\right)
\eeq
so that it can be written as
\beq
Z_{H} = \lambda_+^N+\lambda_-^N
\eeq 

In the case of the two dimensional lattice with toroidal boundary conditions, since each row is repeated above and below, only shifted diagonally, the partition function of the lattice with $M$ rows is found by taking the $M$-th power of each of the weights which appear in the one-dimensional partition function.

The quantity $D$ cannot vanish for positive weights, so that the kite model, like Barreto's Mars, does not have a phase transition.

\subsection{Simple square}\label{sec:square}
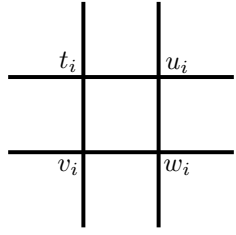
\begin{figure}[htpb]
\begin{center}
\scalebox{1.}{
\begin{tikzpicture}
\draw[line width = 1.5pt] (0,-1) -- (0,2);
\draw[line width = 1.5pt] (1,-1) -- (1,2);
\draw[line width = 1.5pt] (-1,0) -- (2,0);
\draw[line width = 1.5pt] (-1,1) -- (2,1);
\node at (-0.2,-0.2) {$v_i$};
\node at (1.25,-0.2) {$w_i$};
\node at (-0.2,1.2) {$t_i$};
\node at (1.25,1.15) {$u_i$};
\end{tikzpicture}
}
\end{center}
\caption{The simple square four vertex unit and vertex weight notation convention. \label{fig:squarenotation}}
\end{figure}

The simple square CP admits all eight of the odd 8-vertex weights at each vertex of the lattice. We look at the square lattice with homogeneous vertex weights, with units of two vertices as either column or bi-partite staggered, and with units of four vertices, column staggered. We show the vertex weight notation convention we use in figure~\ref{fig:squarenotation}.

In appendix~\ref{app:freeenergy} we list the free-energies of each of these cases except for the units of four vertices case, whose expression is quite large but can be found, alternatively, via a mapping of~\cite{assis2017temp} from the even to the odd staggered 8-vertex model of the expressions found in~\cite{lin1977w}. All of the previous results except for the one-dimensional derivation of the kite model can be found by suitable specializations of this general free energy, although in some cases the specialization is not immediately obvious; see~\cite{hsue1975lw} and~\cite{lin1977w} for examples of specializations for the staggered even 8-vertex model. Therefore, the free-energies of the staggered odd 8-vertex models with units of two vertices can be found more simply from the derivations in~\cite{assis2017temp}.

The phase transition points of the homogeneous lattice are given by the following conditions~\cite{wu2004k,assis2017temp}
\beqr
& (v_1v_2+v_3v_4)=(v_5v_6+v_7v_8) = 0 \\
& (v_1v_3+v_2v_4) = 0 \\
& (v_5v_7+v_6v_8) = 0 \\
& (v_1v_2+v_3v_4)(v_5v_6+v_7v_8)+(v_1v_3-v_2v_4)^2+(v_5v_7-v_6v_8)^2 = 0 
\eeqr
From the phase transition point conditions, we can see that at least two of the weights must vanish for a phase transition to occur, for example $v_1=v_2=0$. The phase transitions will in general be logarithmic of second order, except in special cases where it can have an exponent of $\alpha=\tfrac{1}{2}$~\cite{fan1970w}.

The column staggered lattice with units of two vertices has phase transitions at the following points~\cite{assis2017temp}
\bal
(v_5+v_8)(w_6+w_7)\pm(v_6-v_7)(w_5-w_8)&=0 \\
(v_6+v_7)(w_5+w_8)\pm(v_5-v_8)(w_6-w_7)&=0
\eal
and the bi-partite staggered lattice with units of two vertices has phase transitions at the following points~\cite{assis2017temp}
\bal
-v_1w_3-v_2w_4+v_3w_1+v_4w_2+v_5w_7+v_6w_8+v_7w_5+v_8w_6 &= 0 \\
v_1w_3+v_2w_4-v_3w_1-v_4w_2+v_5w_7+v_6w_8+v_7w_5+v_8w_6 &= 0 \\
v_1w_3+v_2w_4+v_3w_1+v_4w_2-v_5w_7-v_6w_8+v_7w_5+v_8w_6 &= 0 \\
v_1w_3+v_2w_4+v_3w_1+v_4w_2+v_5w_7+v_6w_8-v_7w_5-v_8w_6 &= 0
\eal
There can be up to three phase transitions of these models which are in general logarithmic of second order, although in special cases there is only one phase transition which has exponent $\alpha=\tfrac{1}{2}$~\cite{hsue1975lw}. 

The expressions for the phase transition points of the general four unit column staggered model are given by the following new expressions
\bal
-\Omega_1+\Omega_2+\Omega_3+\Omega_4 &=0 \\
\Omega_1-\Omega_2+\Omega_3+\Omega_4 &=0 \\
\Omega_1+\Omega_2-\Omega_3+\Omega_4 &=0 \\
\Omega_1+\Omega_2+\Omega_3-\Omega_4 &=0
\eal
where
\bal
\Omega_1 &= t_1u_1v_2w_2+t_2u_2v_1w_1+t_3u_3v_4w_4+t_4u_4v_3w_3+t_5u_7v_7w_5+t_6u_8v_8w_6+  t_7u_5v_5w_7+t_8u_6v_6w_8 \\
\Omega_2 &=t_1u_1v_3w_3+t_2u_2v_4w_4+t_3u_3v_1w_1+t_4u_4v_2w_2+t_5u_7v_5w_7+t_6u_8v_6w_8
+ t_7u_5v_7w_5+t_8u_6v_8w_6 \\
\Omega_3 &=t_1u_3v_2w_4+t_2v_1u_4w_3+t_3u_1v_4w_2+t_4u_2v_3w_1+t_5u_6v_7w_8+t_6u_5v_8w_7
+ t_7u_8v_5w_6+t_8u_7v_6w_5 \\
\Omega_4 &=t_1u_3v_3w_1+t_2u_4v_4w_2+t_3u_1v_1w_3 +t_4u_2v_2w_4+t_5u_6v_5w_6+t_6u_5v_6w_5
+ t_7u_8v_7w_8+t_8u_7v_8w_7
\eal
There can be up to 5 phase transitions in this model, which are generically logarithmic of second order, although in special cases it can have two first-order phase transitions or up to three second order phase transitions with exponent $\alpha=\tfrac{1}{2}$~\cite{lin1977w}.

We can also consider the general free-energy as a function only of crease assignment weights instead of vertex weights using the transformations~(\ref{creasetrans1})--(\ref{creasetrans4}), and setting all vertex weights to unity, $v_i=w_i=t_i=u_i=1$. The phase transition points are then given by the following conditions
\bal
&\pm(f_mh_m-f_vh_v)(b_md_m-b_vd_v)(a_mc_me_vg_v+a_vc_ve_mg_m) \nonumber\\
&\qquad-(f_mh_m+f_vh_v)(b_md_m+b_vd_v)(a_mc_ve_vg_m+a_vc_me_mg_v) \nonumber\\
&\qquad\qquad\pm(f_mh_v-f_vh_m)(b_md_v-b_vd_m)(a_mc_me_mg_m+a_vc_ve_vg_v) \nonumber\\
&\qquad\qquad\qquad-(f_mh_v+f_vh_m)(b_md_v+b_vd_m)( a_mc_ve_m g_v+a_vc_me_vg_m) =0 \\
&\pm(f_mh_m-f_vh_v)(b_md_m-b_vd_v)(a_mc_ve_vg_m+a_vc_me_mg_v) \nonumber\\
&\qquad+(f_mh_m+f_vh_v)(b_md_m+b_vd_v)(a_mc_me_vg_v+a_vc_ve_mg_m) \nonumber\\
&\qquad\qquad\pm(f_mh_v-f_vh_m)(b_md_v-b_vd_m)(a_mc_ve_mg_v+a_vc_me_vg_m) \nonumber\\
&\qquad\qquad\qquad+(f_mh_v+f_vh_m)(b_md_v+b_vd_m)( a_mc_me_m g_m+a_vc_ve_vg_v)=0
\eal
A sufficient but not necessary condition for a phase transition in these expressions is that at least four crease assignment weights are identically zero, such as for example $f_m=h_v=b_m=d_m=0$, or $a_m=c_v=e_m=g_m=0$, or etc.

We see that in general for the square lattice, phase transitions do not occur unless the symmetry of the lattice is broken.

\section{Flat-foldable 3-coloring models}\label{sec:threecolor}
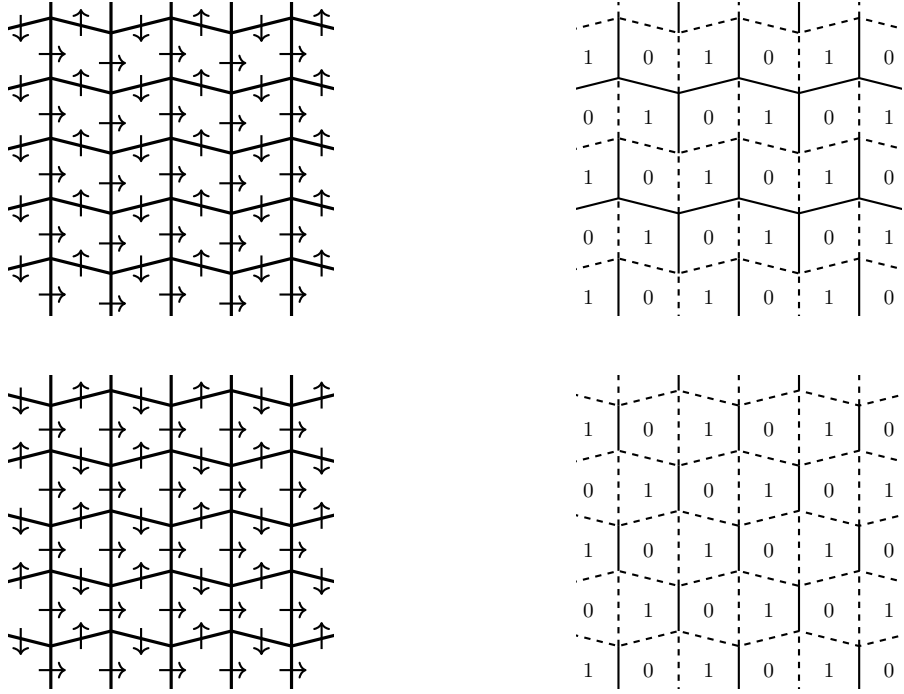
\begin{figure}[htpb]
\begin{center}
\begin{subfigure}{0.45\textwidth}
\centering
\scalebox{.8}{
\begin{tikzpicture}
\clip (0.3,0.3) rectangle (5.7,5.5);
\foreach \position in {(0,0),(0,2),(0,4),(0,6)}
{
	\begin{scope}[shift={\position}]
	\draw[line width = 1.5pt] (0,0) -- (1,0.25) -- (2,0) -- (3,0.25) -- (4,0) -- (5,0.25) -- (6,0);
	\draw[line width = 1.5pt] (0,1) -- (1,1.25) -- (2,1) -- (3,1.25) -- (4,1) -- (5,1.25) -- (6,1);
	\draw[line width = 1.5pt] (1,0.25) -- (1,1.25);
	\draw[line width = 1.5pt] (3,0.25) -- (3,1.25);
	\draw[line width = 1.5pt] (5,0.25) -- (5,1.25);
	\draw[line width = 1.5pt] (0,1) -- (0,2);
	\draw[line width = 1.5pt] (2,1) -- (2,2);
	\draw[line width = 1.5pt] (4,1) -- (4,2);
	\draw[line width = 1.5pt] (6,1) -- (6,2);
	\draw[line width = 1.5pt] (0,0) -- (0,1);
	\draw[line width = 1.5pt] (2,0) -- (2,1);
	\draw[line width = 1.5pt] (4,0) -- (4,1);
	\draw[line width = 1.5pt] (6,0) -- (6,1);
	\draw[line width = 1.5pt] (1,1.25) -- (1,2.25);
	\draw[line width = 1.5pt] (3,1.25) -- (3,2.25);
	\draw[line width = 1.5pt] (5,1.25) -- (5,2.25);
	\end{scope}
}
\foreach \position in {(0,0),(0,1),(0,2),(0,3),(0,4),(2,0),(2,1),(2,2),(2,3),(2,4),(4,0),(4,1),(4,2),(4,3),(4,4)}
{
	\begin{scope}[shift={\position}]
	\draw[->,line width = 1pt] (0.5,1.3) -- (0.5,0.85);	
	\draw[->,line width = 1pt] (1.5,0.95) -- (1.5,1.4);
	\draw[->,line width = 1pt] (0.8,0.65) -- (1.25,0.65);
	\draw[->,line width = 1pt] (1.8,0.5) -- (2.25,0.5);
	\end{scope}
}
\end{tikzpicture}
}
\end{subfigure}
\begin{subfigure}{0.45\textwidth}
\centering
\scalebox{.8}{
\begin{tikzpicture}
\clip (0.3,0.3) rectangle (5.7,5.5);
\foreach \position in {(0,0),(0,2),(0,4),(0,6)}
{
	\begin{scope}[shift={\position}]
	\draw[line width = 1pt] (0,0) -- (1,0.25) -- (2,0) -- (3,0.25) -- (4,0) -- (5,0.25) -- (6,0);
	\draw[line width = 1pt, dashed] (0,1) -- (1,1.25) -- (2,1) -- (3,1.25) -- (4,1) -- (5,1.25) -- (6,1);
	\draw[line width = 1pt] (1,0.25) -- (1,1.25);
	\draw[line width = 1pt] (3,0.25) -- (3,1.25);
	\draw[line width = 1pt] (5,0.25) -- (5,1.25);
	\draw[line width = 1pt] (0,1) -- (0,2);
	\draw[line width = 1pt] (2,1) -- (2,2);
	\draw[line width = 1pt] (4,1) -- (4,2);
	\draw[line width = 1pt] (6,1) -- (6,2);
	\draw[line width = 1pt,dashed] (0,0) -- (0,1);
	\draw[line width = 1pt,dashed] (2,0) -- (2,1);
	\draw[line width = 1pt,dashed] (4,0) -- (4,1);
	\draw[line width = 1pt,dashed] (6,0) -- (6,1);
	\draw[line width = 1pt,dashed] (1,1.25) -- (1,2.25);
	\draw[line width = 1pt,dashed] (3,1.25) -- (3,2.25);
	\draw[line width = 1pt,dashed] (5,1.25) -- (5,2.25);
	\end{scope}
}
\foreach \position in {(0,0),(2,0),(4,0),(6,0),(1,1),(3,1),(5,1),(0,2),(2,2),(4,2),(6,2),(1,3),(3,3),(5,3),(0,4),(2,4),(4,4),(6,4)}
{
	\begin{scope}[shift={\position}]
	\node at (-.5,0.6) {0};
	\node at (0.5,0.6) {1};
	\end{scope}
}
\end{tikzpicture}
}
\end{subfigure}
\\[0.3in]
\begin{subfigure}{0.45\textwidth}
\centering
\scalebox{.8}{
\begin{tikzpicture}
\clip (0.3,0.3) rectangle (5.7,5.5);
\foreach \position in {(0,0),(0,2),(0,4),(0,6)}
{
	\begin{scope}[shift={\position}]
	\draw[line width = 1.5pt] (0,0) -- (1,0.25) -- (2,0) -- (3,0.25) -- (4,0) -- (5,0.25) -- (6,0);
	\draw[line width = 1.5pt] (-1,1) -- (0,1.25) -- (1,1) -- (2,1.25) -- (3,1) -- (4,1.25) -- (5,1) -- (6,1.25);
	\draw[line width = 1.5pt] (1,0.25) -- (1,1);
	\draw[line width = 1.5pt] (3,0.25) -- (3,1);
	\draw[line width = 1.5pt] (5,0.25) -- (5,1);
	\draw[line width = 1.5pt] (0,1.25) -- (0,2);
	\draw[line width = 1.5pt] (2,1.25) -- (2,2);
	\draw[line width = 1.5pt] (4,1.25) -- (4,2);
	\draw[line width = 1.5pt] (6,1.25) -- (6,2);
	\draw[line width = 1.5pt] (0,0) -- (0,1.25);
	\draw[line width = 1.5pt] (2,0) -- (2,1.25);
	\draw[line width = 1.5pt] (4,0) -- (4,1.25);
	\draw[line width = 1.5pt] (6,0) -- (6,1.25);
	\draw[line width = 1.5pt] (1,1) -- (1,2.25);
	\draw[line width = 1.5pt] (3,1) -- (3,2.25);
	\draw[line width = 1.5pt] (5,1) -- (5,2.25);
	\end{scope}
}
\foreach \position in {(0,0),(2,0),(4,0),(6,0),(1,1),(3,1),(5,1),(0,2),(2,2),(4,2),(6,2),(1,3),(3,3),(5,3),(0,4),(2,4),(4,4),(6,4)}
{
	\begin{scope}[shift={\position}]
	\draw[->,line width = 1pt] (0.5,1.3) -- (0.5,0.85);
	\draw[->,line width = 1pt] (-.5,0.95) -- (-.5,1.4);
	\draw[->,line width = 1pt] (-.2,0.6) -- (.25,0.6);	
	\draw[->,line width = 1pt] (0.8,0.6) -- (1.25,0.6);
	\end{scope}
}
\end{tikzpicture}
}
\end{subfigure}
\begin{subfigure}{0.45\textwidth}
\centering
\scalebox{.8}{
\begin{tikzpicture}
\clip (0.3,0.3) rectangle (5.7,5.5);
\foreach \position in {(0,0),(0,2),(0,4),(0,6)}
{
	\begin{scope}[shift={\position}]
	\draw[line width = 1pt,dashed] (0,0) -- (1,0.25) -- (2,0) -- (3,0.25) -- (4,0) -- (5,0.25) -- (6,0);
	\draw[line width = 1pt, dashed] (-1,1) -- (0,1.25) -- (1,1) -- (2,1.25) -- (3,1) -- (4,1.25) -- (5,1) -- (6,1.25);
	\draw[line width = 1pt] (1,0.25) -- (1,1);
	\draw[line width = 1pt] (3,0.25) -- (3,1);
	\draw[line width = 1pt] (5,0.25) -- (5,1);
	\draw[line width = 1pt] (0,1.25) -- (0,2);
	\draw[line width = 1pt] (2,1.25) -- (2,2);
	\draw[line width = 1pt] (4,1.25) -- (4,2);
	\draw[line width = 1pt] (6,1.25) -- (6,2);
	\draw[line width = 1pt,dashed] (0,0) -- (0,1.25);
	\draw[line width = 1pt,dashed] (2,0) -- (2,1.25);
	\draw[line width = 1pt,dashed] (4,0) -- (4,1.25);
	\draw[line width = 1pt,dashed] (6,0) -- (6,1.25);
	\draw[line width = 1pt,dashed] (1,1) -- (1,2.25);
	\draw[line width = 1pt,dashed] (3,1) -- (3,2.25);
	\draw[line width = 1pt,dashed] (5,1) -- (5,2.25);
	\end{scope}
}
\foreach \position in {(0,0),(2,0),(4,0),(6,0),(1,1),(3,1),(5,1),(0,2),(2,2),(4,2),(6,2),(1,3),(3,3),(5,3),(0,4),(2,4),(4,4),(6,4)}
{
	\begin{scope}[shift={\position}]
	\node at (-.5,0.6) {0};
	\node at (0.5,0.6) {1};
	\end{scope}
}
\end{tikzpicture}
}
\end{subfigure}
\end{center}\caption{On the left, conventions for changing the face color across creases for the Miura-ori (top row) and trapezoid (bottom row). Following the arrow direction, a valley crease increases the face color number while mountain creases decrease the face color number (mod 3). On the right, one of the three possible color ground states the Miura-ori (top row) and trapezoid (bottom row) map to. The other two possible ground states are found by globally increasing or decreasing all face color values by one (mod 3). \label{fig:coloringmap}}
\end{figure}

The Miura-ori and trapezoid CPs can be put into a 3-to-1 correspondence with the 3-coloring model on the square lattice, as shown in figure~\ref{fig:coloringmap}. This was first pointed out in~\cite{ginepro2014h} for the Miura-ori, although they only consider the total enumeration of colorings and not the generalization to Baxter's 3-coloring problem with three color fugacity weights $z_i$~\cite{baxter19702}. 

Baxter's 3-coloring problem on the square lattice is an exactly solved model where each face of the square lattice is allowed one of three colors with the condition that no two neighbors can share the same color. This model has a 3-to-1 mapping from the even 6-vertex model~\cite{baxter1982} and can also be mapped to Baxter's symmetric even 8-vertex model~\cite{pegg1982}. It is always possible to map a staggered odd 8-vertex model to a staggered even 8-vertex model and then specialize to a homogeneous even 8-vertex model~\cite{assis2017temp}, for example
\bat{6}
v_1 &= w_4 &= \omega^{(\mathrm{e})}_5, \quad\quad\quad & v_2 &= w_3 &= \omega^{(\mathrm{e})}_6, \quad\quad\quad & v_3 &= w_2 &= \omega^{(\mathrm{e})}_8, \quad\quad\quad & v_4 &= w_1 &= \omega^{(\mathrm{e})}_7, \nonumber\\
v_5 &= w_7 &= \omega^{(\mathrm{e})}_1, \quad\quad\quad & v_6 &= w_8 &= \omega^{(\mathrm{e})}_2, \quad\quad\quad & v_7 &= w_5 &= \omega^{(\mathrm{e})}_4, \quad\quad\quad & v_8 &= w_6 &= 
\omega^{(\mathrm{e})}_3
\eat
where the weights $\omega^{(\mathrm{e})}_i$ refer to the even 8-vertex model weights; see figure~\ref{fig:evenconfigsbonds} for the $\omega^{(\mathrm{e})}_i$ notation convention. When this mapping is specialized to the even 6-vertex model, so that $\omega^{(\mathrm{e})}_7=\omega^{(\mathrm{e})}_8=0$, we see that both the Miura-ori and trapezoid models have a mapping to the even 6-vertex model, and hence to the 3-coloring problem. Alternatively, we show a direct mapping to the 3-coloring problem in figure~\ref{fig:coloringmap}, giving a convention for how to change face colors across crease assignments. We represent the colors by numbers~0,~1,~2, and we increase the color across a valley crease or decrease the color across a mountain crease in the direction of the arrows (mod 3). The mapping is unique except for the initial color chosen for a face somewhere in the lattice. The Miura-ori mapping we show is different than the one given in~\cite{ginepro2014h}. In figure~\ref{fig:coloringmap} we also show color ground states for the Miura-ori and trapezoid CPs, although two other different ground states are possible by globally increasing or decreasing all face colors by 1 (mod 3).

Mapping from the 3-coloring problem to the vertex models, we see that it is necessary to consider that the vertex weights come in three colors $w_{i,j}$. We then have the following mapping from color fugacity variables $z_k$ to colored 6-vertex weights $w_{i,j}$~\cite{pegg1982}
\bat{3}
\left.\omega{}^{(\mathrm{e})}_{1,j}\right.^4 &= ~~\left.\omega{}^{(\mathrm{e})}_{2,j}\right.^4& &= z_j{}^2 z_{j-1}z_{j+1}, &\qquad \left.\omega{}^{(\mathrm{e})}_{3,j}\right.^4 &= z_j z_{j-1}{}^2z_{j+1} \nonumber\\
\left.\omega{}^{(\mathrm{e})}_{5,j}\right.^2 &= \left.\omega{}^{(\mathrm{e})}_{6,j-1}\right.^2& &= z_j{}^2 z_{j-1}{}^2, &\qquad \left.\omega{}^{(\mathrm{e})}_{4,j}\right.^4 &= z_j z_{j-1}z_{j+1}{}^2 \label{6VMmap}
\eat
The 3-coloring problem can also be mapped to Baxter's symmetric even 8-vertex model~\cite{pegg1982},
\bat{2}
(a^2-c^2)(b^2-d^2) &= 0,\qquad & (b^2c^2-a^2d^2)(-a^2+b^2+c^2-d^2) &=(z_0z_1z_2)^2 \nonumber\\
(a^2-b^2)(c^2-d^2) &= 0,\qquad & (a^2+b^2+c^2+d^2) &= z_0z_1+z_1z_2+z_2z_0
\eat
where
\beqr
\omega{}^{(\mathrm{e})}_1=\omega{}^{(\mathrm{e})}_2=a,\qquad \omega{}^{(\mathrm{e})}_3=\omega{}^{(\mathrm{e})}_4=b,\qquad \omega{}^{(\mathrm{e})}_5=\omega{}^{(\mathrm{e})}_6=c,\qquad \omega{}^{(\mathrm{e})}_7=\omega{}^{(\mathrm{e})}_8=d
\eeqr

Baxter has given the free-energy of the 3-coloring problem in the following form~\cite{baxter19702}
\beq
-\beta\, f = \frac{1}{3}\,\ln(z_0z_1z_2)+\frac{1}{2}\,\ln\left[\frac{64\,(1-9t^2)^{2/3}}{27\,(1+t)^3(1-3t)}\right]
\eeq
where $t$ is found from
\beq
\frac{(1-3t^2)^3}{(1-9t^2)}=\frac{(z_0z_1+z_1z_2+z_2z_0)^3}{27(z_0z_1z_2)^2}
\eeq
and where $t$ is the root in the range $0\leq t<\tfrac{1}{3}$.

The face colors have a convenient origami interpretation: they give the local ordering of the layers of each face with respect to its neighbors. As discussed above, determining the global layer ordering of an origami CP, even one which is locally flat-foldable, is an NP-hard problem. We see from this interpretation of the 3-coloring problem, however, how to map the global flat-foldability problem to a suitable SOS model. We require, rather than 3 colors around a vertex, a partial ordering of the layers around each vertex, and we then sum over those configurations which admit a global layer ordering, counting multiplicity. As far as we aware, this type of SOS model has not been considered before, and further consideration is beyond the scope of this work. 

We can see from figure~\ref{fig:coloringmap} that the ground states for these models have equal numbers of two colors, so that we assume without loss of generality that $z_0=z_1=1$. It can easily be seen, then, that individual face-flip defects have the effect of introducing the third color into the lattice, although in combination they can add extra~0 or~1 colors, for example in the middle case of the middle row of figure~\ref{fig:defectexamples}. The third color faces can be considered as particles with fugacity $z=z_2$, so that we can treat the model as a kind of hard square lattice gas, as discussed by Baxter~\cite{baxter19702}. In this case, we have
\beq
P = \frac{1}{3}\,\ln(z)+\frac{1}{2}\,\ln\left[\frac{64\,(1-9t^2)^{2/3}}{27\,(1+t)^3(1-3t)}\right]\label{threecolorP}
\eeq
where
\beq
t = \begin{cases}
\displaystyle\frac{\sqrt{2z\left[1+8z-\sqrt{1+12z+36z^2+32z^3}\right]}}{6z},\quad &z<1\\
\displaystyle \frac{\sqrt{z(z-1)}}{3z},\quad &z>1
\end{cases} \label{threecolort}
\eeq

The 3-coloring problem has a second order phase transition with critical exponent $\alpha=\tfrac{1}{2}$ when $z_0=z_1=z_2$, or in our defect case, $z=1$~\cite{baxter19702,baxter1982,baxter2007}. We see from the colored 6-vertex mapping in~\ref{6VMmap} that this corresponds to 
\beq
\omega^{(\mathrm{e})}_1=\omega^{(\mathrm{e})}_2=\omega^{(\mathrm{e})}_3=\omega^{(\mathrm{e})}_4=\omega^{(\mathrm{e})}_5=\omega^{(\mathrm{e})}_6
\eeq
and correspondingly
\beq
v_1=v_2=v_5=v_6=v_7=v_8=w_3=w_4=w_5=w_6=w_7=w_8
\eeq
which falls outside of the free-fermion constraint for the Miura-ori and trapezoid models. Therefore, we have found an extra phase transition point of these models, which characterizes the state where all vertex weights have equal strength, or equivalently, all local layer orderings are equlally probable. Above we have found phase transition points dependent on vertex weights, that is, the arrangement of creases around a vertex, as well as dependent on the crease assignments. We now also have found a phase transition dependent on the local layering of neighboring faces. 

\begin{figure}[htpb]
\begin{center}
\begin{subfigure}{0.45\textwidth}
\centering
\scalebox{.8}{
\begin{tikzpicture}
\clip (0.3,7.3) rectangle (5.7,12.5);
\foreach \position in {(0,0),(0,1.25),(0,2.5),(0,3.75),(0,5),(0,6.25),(0,7.5),(0,8.75),(0,10),(0,11.25),(0,12.5),(0,13.75)}
{
	\begin{scope}[shift={\position}]
	\draw[line width = 1.5pt] (0,8) -- (0.433,7) -- (.866,6.75) -- (1.299,5.75) -- (1.732,5.5) -- (2.165,4.5) -- (2.598,4.25) -- (3.031,3.25) -- (3.464,3) -- (3.897,2) -- (4.33,1.75) -- (4.763,0.75) -- (5.196,0.5) -- (5.629,-.5) -- (6.062,-.75) -- (6.495,-1.75);
	\draw[line width = 1.5pt] (0,6.75) -- (0.433,7);
	\draw[line width = 1.5pt] (.866,5.5) -- (1.299,5.75);
	\draw[line width = 1.5pt] (1.732,4.25) -- (2.165,4.5);
	\draw[line width = 1.5pt] (2.598,3) -- (3.031,3.25);
	\draw[line width = 1.5pt] (3.464,1.75) -- (3.897,2);
	\draw[line width = 1.5pt] (4.33,.5) -- (4.763,0.75);
	\draw[line width = 1.5pt] (5.196,-.75) -- (5.629,-.5);
	\draw[line width = 1.5pt] (6.062,-2) -- (6.495,-1.75);
	\draw[line width = 1.5pt] (0.433,7) -- (.866,8);
	\draw[line width = 1.5pt] (1.299,5.75) -- (1.732,6.75);
	\draw[line width = 1.5pt] (2.165,4.5) -- (2.598,5.5);
	\draw[line width = 1.5pt] (3.031,3.25) -- (3.464,4.25);
	\draw[line width = 1.5pt] (3.897,2) -- (4.33,3);
	\draw[line width = 1.5pt] (4.763,0.75) -- (5.196,1.75);
	\draw[line width = 1.5pt] (5.629,-.5) -- (6.062,0.5);
	
	\draw[->,line width = 1pt] (0.1,7.05) -- (0.383,6.7);
	\draw[->,line width = 1pt] (.966,5.8) -- (1.249,5.45);
	\draw[->,line width = 1pt] (1.832,4.55) -- (2.115,4.2);
	\draw[->,line width = 1pt] (2.698,3.3) -- (2.981,2.95);
	\draw[->,line width = 1pt] (3.564,2.05) -- (3.847,1.7);
	\draw[->,line width = 1pt] (4.43,.8) -- (4.713,0.45);
	\draw[->,line width = 1pt] (5.296,-.45) -- (5.579,-.8);
	\draw[->,line width = 1pt] (6.162,-1.7) -- (6.445,-2.05);
	\draw[->,line width = 1pt] (0.533,7.6) -- (.866,7.3);
	\draw[->,line width = 1pt] (1.399,6.35) -- (1.732,6.05);
	\draw[->,line width = 1pt] (2.265,5.1) -- (2.598,4.8);
	\draw[->,line width = 1pt] (3.131,3.85) -- (3.464,3.55);
	\draw[->,line width = 1pt] (3.997,2.6) -- (4.33,2.3);
	\draw[->,line width = 1pt] (4.863,1.35) -- (5.196,1.05);
	\draw[->,line width = 1pt] (5.729,.1) -- (6.062,-0.2);
	
	\draw[->,line width = 1pt] (.966,7.25) -- (1.299,7.7);
	\draw[->,line width = 1pt] (1.832,6.0) -- (2.165,6.45);
	\draw[->,line width = 1pt] (2.698,4.75) -- (3.031,5.2);
	\draw[->,line width = 1pt] (3.564,3.5) -- (3.897,3.95);
	\draw[->,line width = 1pt] (4.43,2.25) -- (4.763,2.7);
	\draw[->,line width = 1pt] (5.296,1.0) -- (5.629,1.45);
	\draw[->,line width = 1pt] (6.162,-.25) -- (6.495,.2);
	
	\draw[->,line width = 1pt] (0.683,7.05) -- (0.533,6.7);
	\draw[->,line width = 1pt] (1.549,5.8) -- (1.399,5.45);
	\draw[->,line width = 1pt] (2.415,4.55) -- (2.265,4.2);
	\draw[->,line width = 1pt] (3.281,3.3) -- (3.131,2.95);
	\draw[->,line width = 1pt] (4.147,2.05) -- (3.997,1.7);
	\draw[->,line width = 1pt] (5.013,.8) -- (4.863,0.45);
	\draw[->,line width = 1pt] (5.879,-.45) -- (5.729,-.8);
	\end{scope}
}
\end{tikzpicture}
}
\end{subfigure}
\begin{subfigure}{0.45\textwidth}
\centering
\scalebox{.8}{
\begin{tikzpicture}
\clip (0.3,7.3) rectangle (5.7,12.5);
\foreach \position in {(0,0),(0,1.25),(0,2.5),(0,3.75),(0,5),(0,6.25),(0,7.5),(0,8.75),(0,10),(0,11.25),(0,12.5),(0,13.75)}
{
	\begin{scope}[shift={\position}]
	\draw[line width = 1pt] (0,8) -- (0.433,7) -- (.866,6.75) -- (1.299,5.75) -- (1.732,5.5) -- (2.165,4.5) -- (2.598,4.25) -- (3.031,3.25) -- (3.464,3) -- (3.897,2) -- (4.33,1.75) -- (4.763,0.75) -- (5.196,0.5) -- (5.629,-.5) -- (6.062,-.75) -- (6.495,-1.75);
	\draw[line width = 1pt] (0,6.75) -- (0.433,7);
	\draw[line width = 1pt] (.866,5.5) -- (1.299,5.75);
	\draw[line width = 1pt] (1.732,4.25) -- (2.165,4.5);
	\draw[line width = 1pt] (2.598,3) -- (3.031,3.25);
	\draw[line width = 1pt] (3.464,1.75) -- (3.897,2);
	\draw[line width = 1pt] (4.33,.5) -- (4.763,0.75);
	\draw[line width = 1pt] (5.196,-.75) -- (5.629,-.5);
	\draw[line width = 1pt] (6.062,-2) -- (6.495,-1.75);
	\draw[line width = 1pt,dashed] (0.433,7) -- (.866,8);
	\draw[line width = 1pt,dashed] (1.299,5.75) -- (1.732,6.75);
	\draw[line width = 1pt,dashed] (2.165,4.5) -- (2.598,5.5);
	\draw[line width = 1pt,dashed] (3.031,3.25) -- (3.464,4.25);
	\draw[line width = 1pt,dashed] (3.897,2) -- (4.33,3);
	\draw[line width = 1pt,dashed] (4.763,0.75) -- (5.196,1.75);
	\draw[line width = 1pt,dashed] (5.629,-.5) -- (6.062,0.5);
	\node at (0.866,7.25) {1};
	\node at (1.732,6.) {1};
	\node at (2.598,4.75) {1};
	\node at (3.464,3.5) {1};
	\node at (4.33,2.25) {1};
	\node at (5.196,1.0) {1};
	\node at (6.062,-.25) {1};
	\node at (0.433,6.5) {0};
	\node at (1.299,5.25) {0};
	\node at (2.165,4.) {0};
	\node at (3.031,2.75) {0};
	\node at (3.897,1.5) {0};
	\node at (4.763,0.25) {0};
	\node at (5.629,-1.) {0};
	\end{scope}
}
\end{tikzpicture}
}
\end{subfigure}
\end{center}\caption{On the left the convention for changing the face color across creases for the kite CP. Following the arrow direction, a valley crease increases the face color number while mountain creases decrease the face color number (mod 3). On the right, one of the three possible color ground states the kite model maps to. The other two possible ground states are found by globally increasing or decreasing all face color values by one (mod 3). \label{fig:coloringmapkite}}
\end{figure}
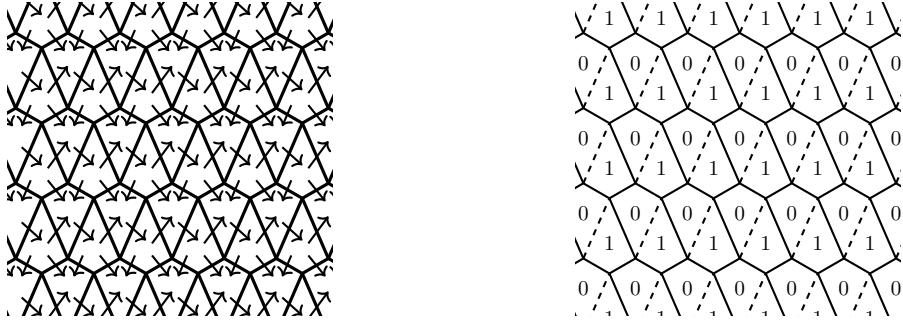
We show in figure~\ref{fig:coloringmapkite} a mapping from the kite CP to the 3-coloring problem, but since the only defects in this model are lines which traverse the entire lattice, the mapping is not surjective. It can be seen from this mapping that each line of defects causes a shift by one (mod 3) of all faces below the defect line. This agrees with the demonstration above that the kite model is effectively a one-dimensional model.

We note that the ground state configuration of Barreto's Mars has a mapping to the 3-coloring problem but its defects do not. 

\section{Lattice gas of defects}\label{sec:latticegas}
We would like to understand analytically the density of defects in these models, since the material properties of the origami tiling, such as their elastic modulus, depend on the density of defects~\cite{silverberg2014emhhsc}. As such, we can re-interpret these models as lattice gas models where the particles are the defects. If we choose our defect ``particles" to be the creases themselves, we will use the variable $y$, in agreement with the analysis in section~\ref{sec:vertexmodels}. At low densities, each defect requires 4 creases, so it is reasonable to choose as another defect variable $z=y^{1/4}$, which at low densities also corresponds to face-flip defects. For Barreto's Mars, this correspondence at all densities is one-to-one, since the defects never interact with each other. But for the other models, the correspondence is only approximate except at low densities, as can be seen in figure~\ref{fig:defectexamples} and discussed in section~\ref{sec:defects}. For Miura-ori and trapezoid, these face-flip defects behave similarly to the hard squares lattice gas, since at low densities they cannot be adjacent to each other and at high densities they form a checkerboard pattern~\cite{gaunt1965f,runnels1966c,chan2012,chan2013}. For intermediate densities, though, four particles can enclose a fifth one, as seen in figure~\ref{fig:defectexamples}, so the two models are not equal. The 3-coloring defect color can also be considered as a hard particle with properties similar but not equal to hard squares~\cite{baxter19702}, so that we will use the notation $z$ for both $y^{1/4}$, as well as the fugacity of the third color $z_2$ in the 3-coloring problem.

As a lattice gas of defects, the partition function can then be re-interpreted as a grand canonical partition function $\mathcal{Z}$, and in the thermodynamic limit, we have the pressure $P$
\beq
\beta\, P = \lim_{\mathcal{N}\to\infty} \ln\left(\mathcal{Z}^{1/\mathcal{N}}\right)
\eeq
Once we rewrite the pressure in terms of the defect fugacity variables $y$ or $z$, we can then find exact expressions for the density $\rho$ of defects in the lattice, that is, the average number of defects $n$ per site,
\beq
\rho = \frac{\langle n\rangle}{\mathcal{N}} = \beta z\,\frac{\partial P}{\partial z} \label{defectdensity}
\eeq
The isothermal compressibility $k_T$ is further given by
\beq
k_T = \frac{1}{\rho}\,\frac{\partial\rho}{\partial P}  =  \frac{\beta z}{\rho^2}\,\frac{\partial\rho}{\partial z}
\eeq
The exact free-energy results above can all be used to calculate the defect density and other thermodynamic quantities exactly. We only make use here of the symmetric defect model results from section~\ref{sec:vertexmodels}.

The crease defect density for Miura-ori and trapezoid are equal and are given by
\bat{2}
\rho_{\mathrm{Mi,Tr}}(y)  &= 1-\frac{2}{\pi}\frac{\left(\tfrac{1}{4}-y^4\right)}{\left(\tfrac{1}{4}+y^4\right)} \, K\left(\frac{y^2}{(\tfrac{1}{4}+y^4)}\right) &&= 4y^4-4y^8+16y^{12}-36y^{16}+\ldots \\
\rho_{\mathrm{Mi,Tr}}(z)  &= \frac{1}{4}\left[ 1-\frac{2}{\pi}\frac{\left(\tfrac{1}{4}-z\right)}{\left(\tfrac{1}{4}+z\right)}   \, K\left(\frac{z^{1/2}}{\left(\tfrac{1}{4}+z\right)}\right) \right] &&= z-z^2+4 z^3-9 z^4+\ldots = \frac{\rho_{\mathrm{Mi,Tr}}\left(y^{1/4}\right)}{4}
\eat
where $K$ is the complete elliptic integral of the first kind, and the crease density of Barreto's Mars defects are given as follows
\bal
\rho_{\mathrm{Ma}}(y) = \frac{2y^4}{(1+y^4)},\qquad \rho_{\mathrm{Ma}}(z) = \frac{z}{2(1+z)}
\eal
The layer order defect density for Miura-ori and trapezoid are large algebraic expressions but can be found through a straightforward application of~(\ref{defectdensity}) to the expressions~(\ref{threecolorP}) and (\ref{threecolort}) in section~\ref{sec:threecolor}.

\begin{figure}[htpb]
\begin{center}
\begin{subfigure}{0.45\textwidth}
\centering
\includegraphics[scale=0.35]{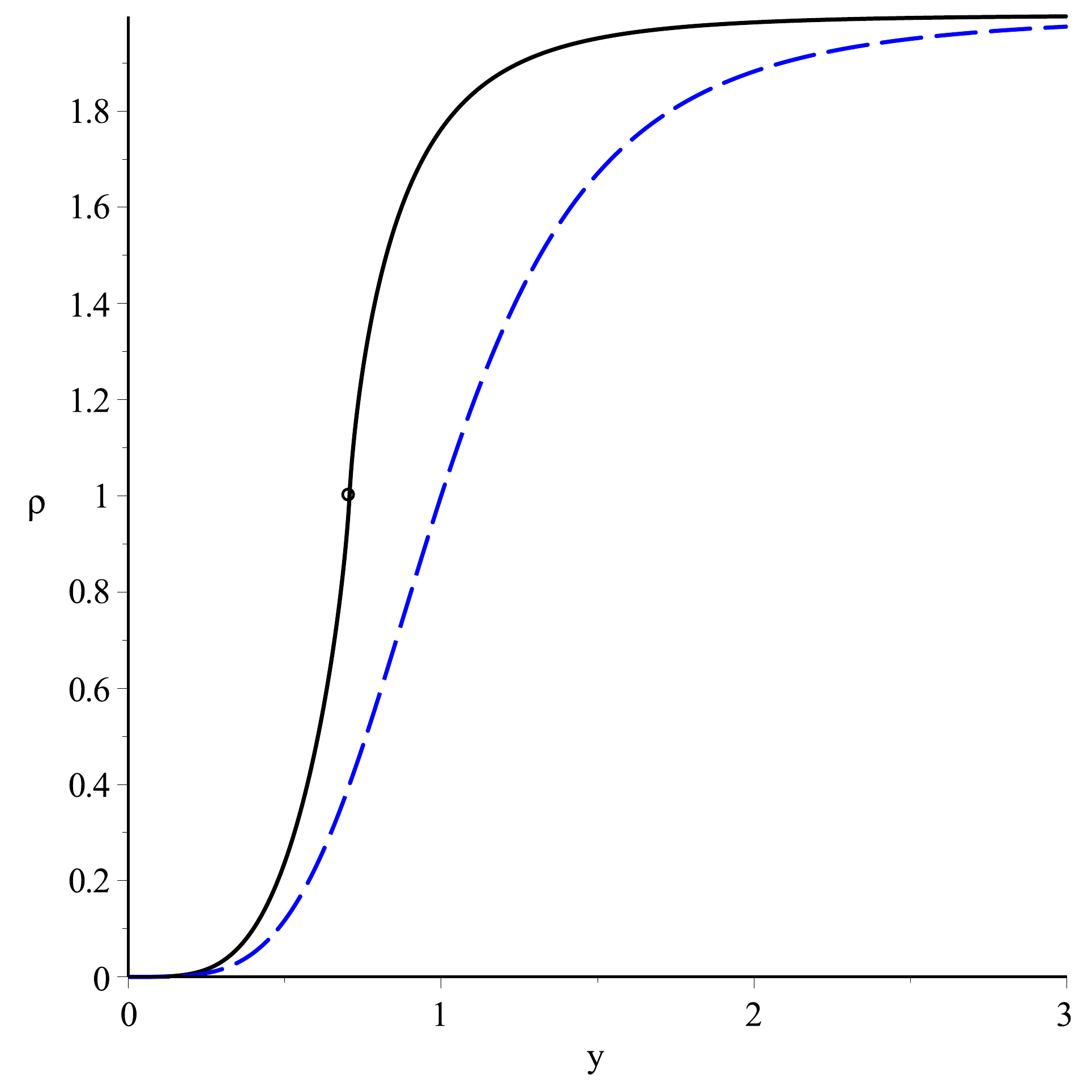}
\end{subfigure}
\begin{subfigure}{0.45\textwidth}
\centering
\includegraphics[scale=0.35]{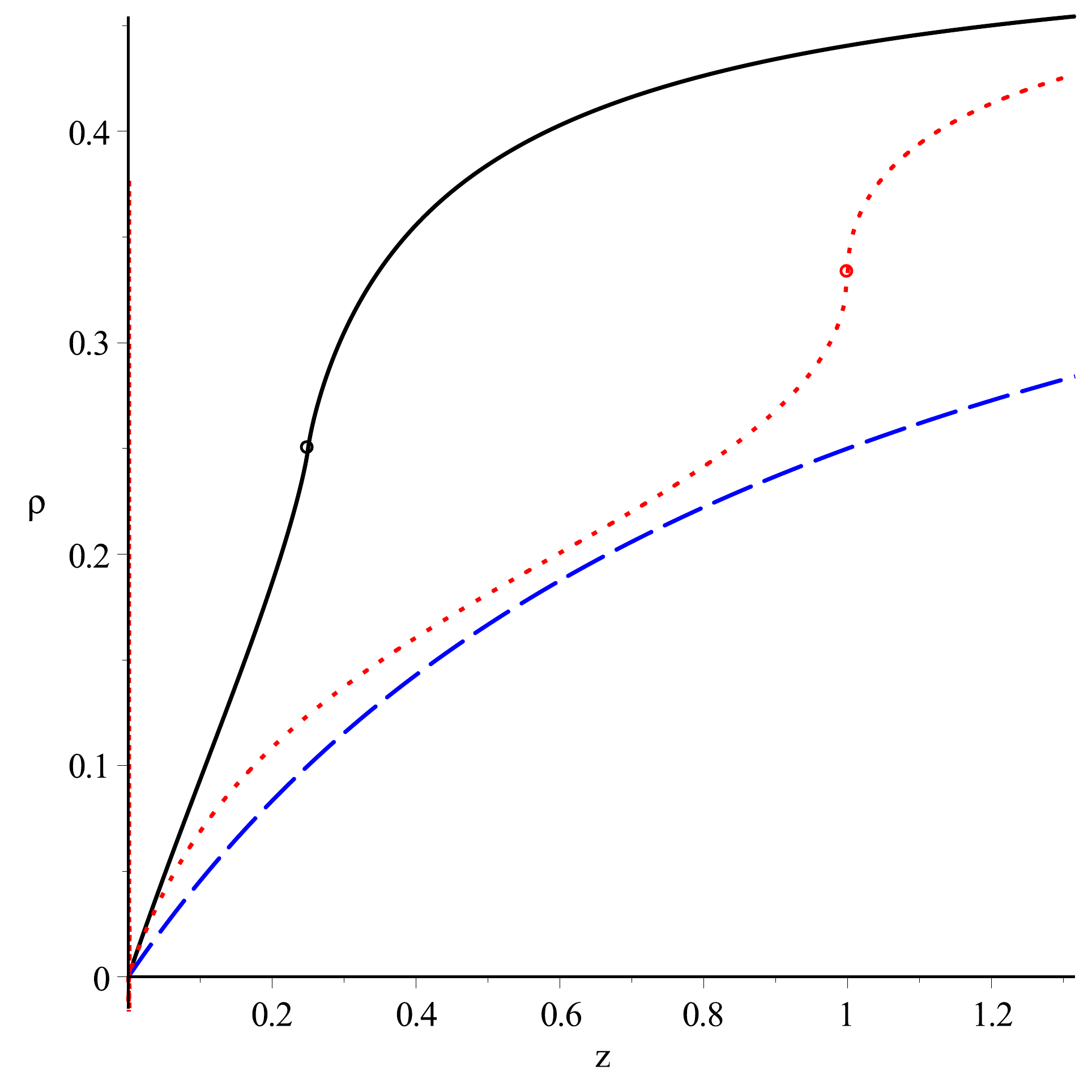}
\end{subfigure}
\end{center}
\caption{Comparison of the crease-reverse defect densities as a function of $y$ on the left and $z$ on the right for Miura-ori and trapezoid (solid black) and Barreto's Mars (blue dashed). On the right also, Miura-ori and trapezoid layer defect density as a function of layer defects $z$ (dotted red). The phase transition points are indicated by circles. \label{fig:densityplot}}
\end{figure}

Miura-ori and trapezoid have a phase transition point at $y_c=\sqrt{2}/2$, $z_c=\tfrac{1}{4}$, where $\rho(y_c)=1$ and $\rho(z_c)=\tfrac{1}{4}$, respectively; Barreto's Mars does not have a phase transition. In all cases the densities approach $\rho\to2$ as $y\to\infty$ and $\rho\to\tfrac{1}{2}$ as $z\to\infty$. We plot in figure~\ref{fig:densityplot} the densities of each case.

For low crease-reversal defect fugacities of the Miura-ori and trapezoid models, the density behaves as a quartic up to near the phase transition point, as can be seen in the plot on the right in figure~\ref{fig:densityplot} where the density is almost linear in $z$ up to $z_c$. The Barreto's Mars defect density is always smaller than the corresponding density for Miura-ori and trapezoid for the same fugacity variable, showing that it is more stable against defects. Conversely, the defect density of Miura-ori and trapezoid is more easily tunable. As a function of layer ordering defects, Miura-ori and trapezoid's density does not agree with their density as a function of crease defects except only at very low and very high densities, and otherwise has a more complex behavior, with a phase transition at $z_c=1$ where $\rho(z_c)=1/3$. Since the phase transition in the 3-coloring problem happens for a larger value of $z$ than the phase transition as a function of creases, we see that the long range crease order disappears before the long range layering order in the lattice. Therefore, to the extent that mechanical properties depend on the layering order versus the crease order, these origami CPs are more stable and tunable for a larger range of defect densities.

We note also that at defect saturation, $y,z\to\infty$, the CP is totally crease reversed and folds in an orderly fashion like the ground state folding, though reversed. Therefore, the points $\rho(y)=1$ and $\rho(z)=\tfrac{1}{4}$ seem to represent states of maximal folding disorder, and we see that the crease defect phase transition points at $y_c=\sqrt{2}/2$ and $z_c=\tfrac{1}{4}$ exactly correspond to these points. The layer defect phase transition point at $z_c=1$, interestingly, is at a higher defect density, corresponding to 2/3 of the defect saturation.

\begin{figure}[htpb]
\begin{center}
\begin{subfigure}{0.45\textwidth}
\centering
\includegraphics[scale=0.35]{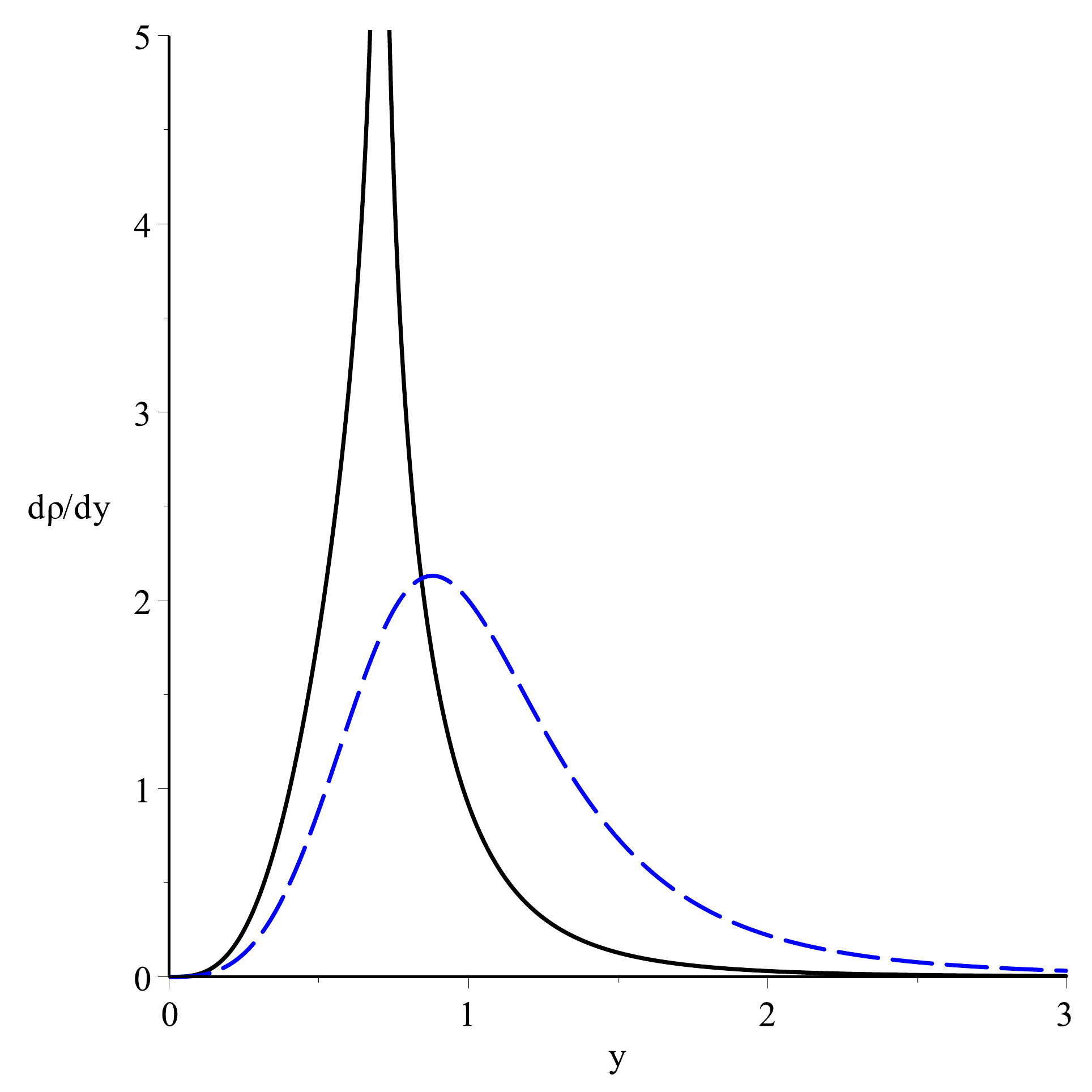}
\end{subfigure}
\begin{subfigure}{0.45\textwidth}
\centering
\includegraphics[scale=0.35]{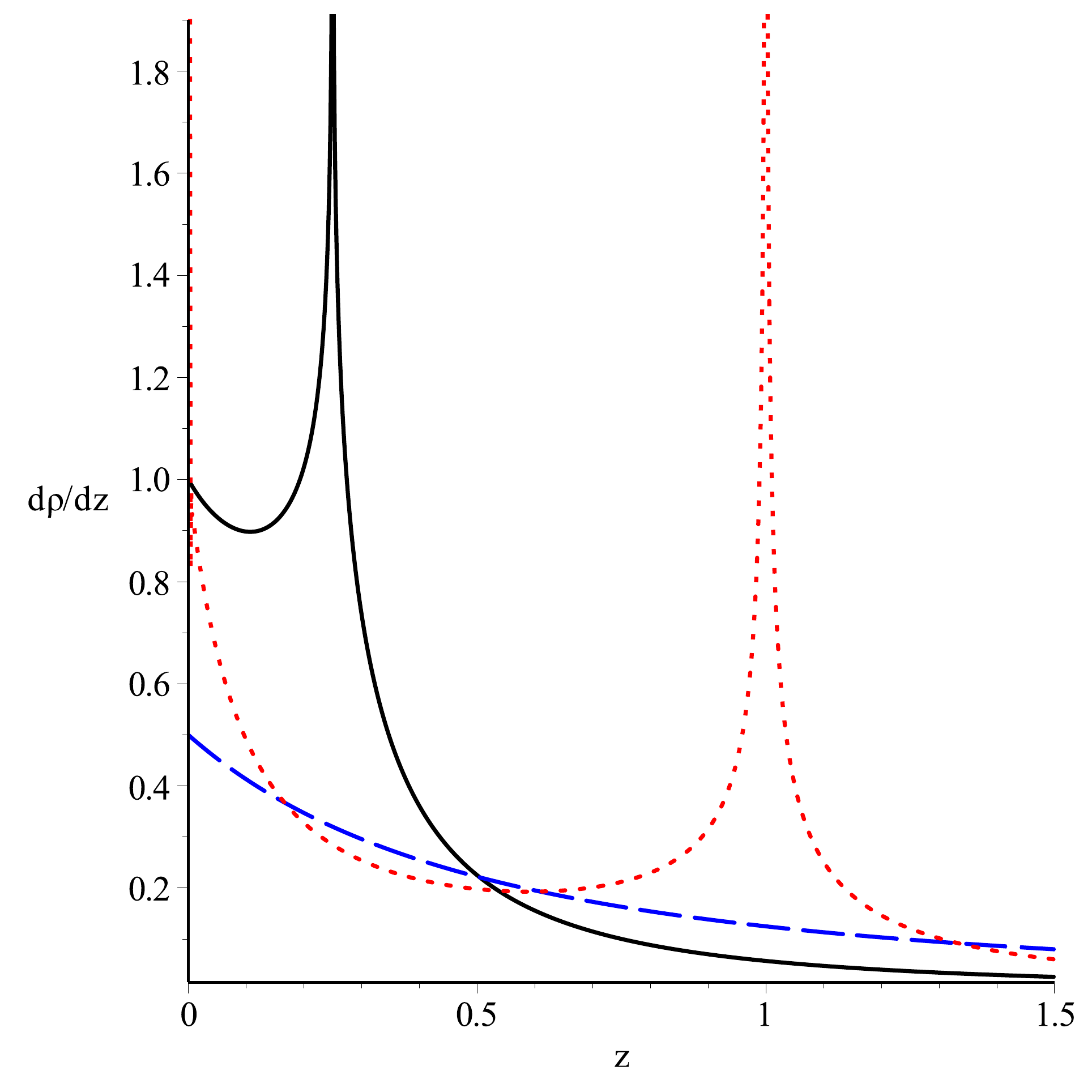}
\end{subfigure}
\end{center}
\caption{Comparison of the derivative of the crease-reverse defect density as a function of $y$ on the left and $z$ on the right for Miura-ori and trapezoid (solid black) and Barreto's Mars (blue dashed). On the right also, Miura-ori and trapezoid layer defect density derivative as a function of layer defects $z$ (dotted red).\label{fig:densitydiffplot}}
\end{figure}

The isothermal compressibility $k_{\mathrm{T}}$ is proportional to the derivative of the density, 
\bal
\frac{d\rho_{\mathrm{Mi,Tr}}(y)}{dy} &= \frac{4}{\pi y}\left[K\left(\frac{y^2}{(\tfrac{1}{4}+y^4)}\right)-E\left(\frac{y^2}{(\tfrac{1}{4}+y^4)}\right) \right] \\
\frac{d\rho_{\mathrm{Mi,Tr}}(z)}{dz} &= \frac{1}{4\pi z}\left[K\left(\frac{z^{1/2}}{\left(\tfrac{1}{4}+z\right)}\right)-E\left(\frac{z^{1/2}}{\left(\tfrac{1}{4}+z\right)}\right) \right]
\eal
where $E$ is the complete elliptic integral of the second kind, and
\beq
\frac{d\rho_{\mathrm{Ma}}(y)}{dy} = \frac{8y^3}{(1+y^4)^2},\qquad \frac{d\rho_{\mathrm{Ma}}(z)}{dy} = \frac{1}{2(1+z)^2}
\eeq
Again, expressions for Miura-ori and trapezoid as a function of the layer defect fugacity $z$ can be found in a straightforward fashion. We plot in figure~\ref{fig:densitydiffplot} the derivative of the densities as a function of both crease-reversal defect fugacities $y$ and $z$ as well as layer defect fugacity $z$. 

\begin{figure}[htpb]
\begin{center}
\begin{subfigure}{0.45\textwidth}
\centering
\includegraphics[scale=0.4]{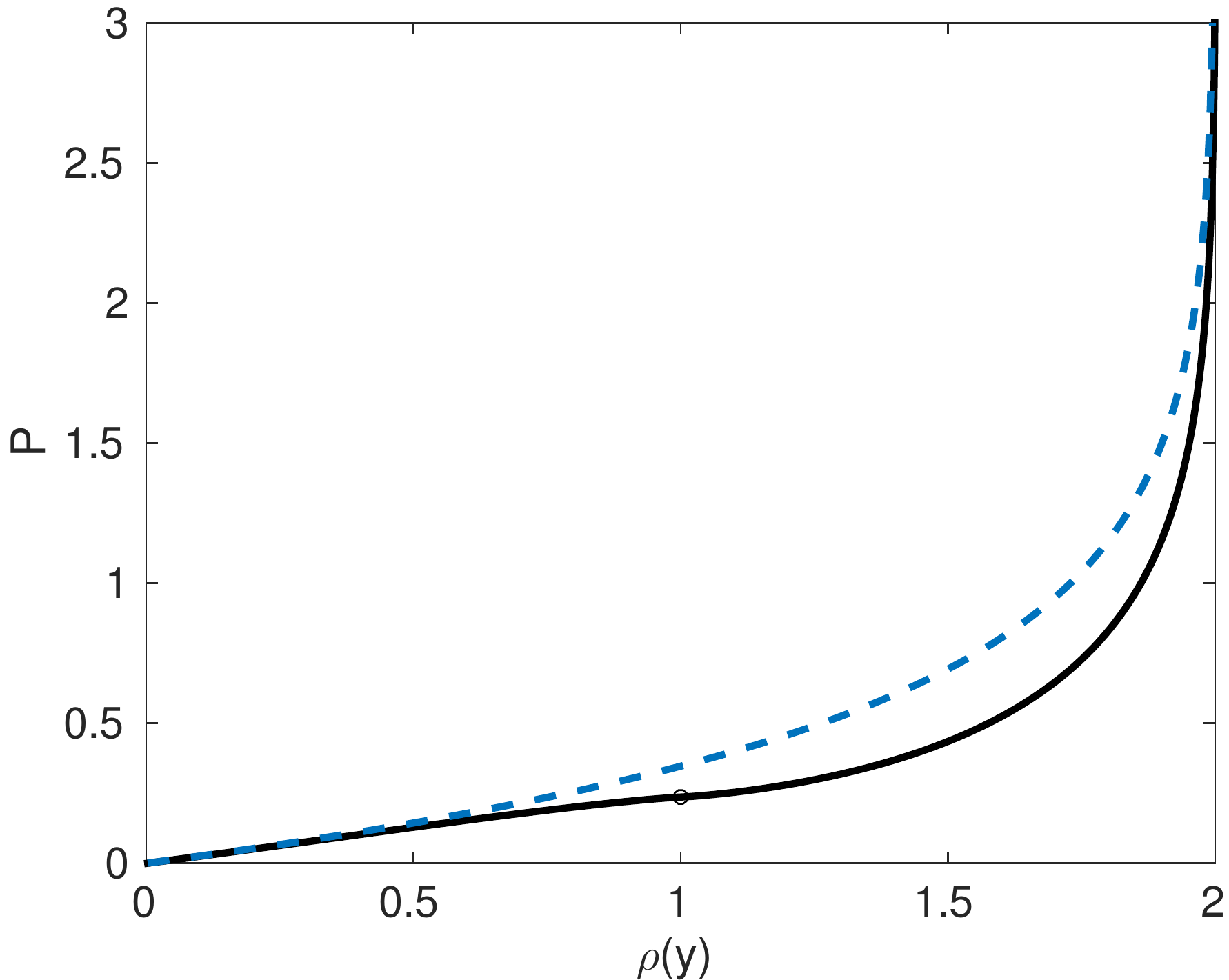}
\end{subfigure}
\hspace{0.2in}
\begin{subfigure}{0.45\textwidth}
\centering
\includegraphics[scale=0.4]{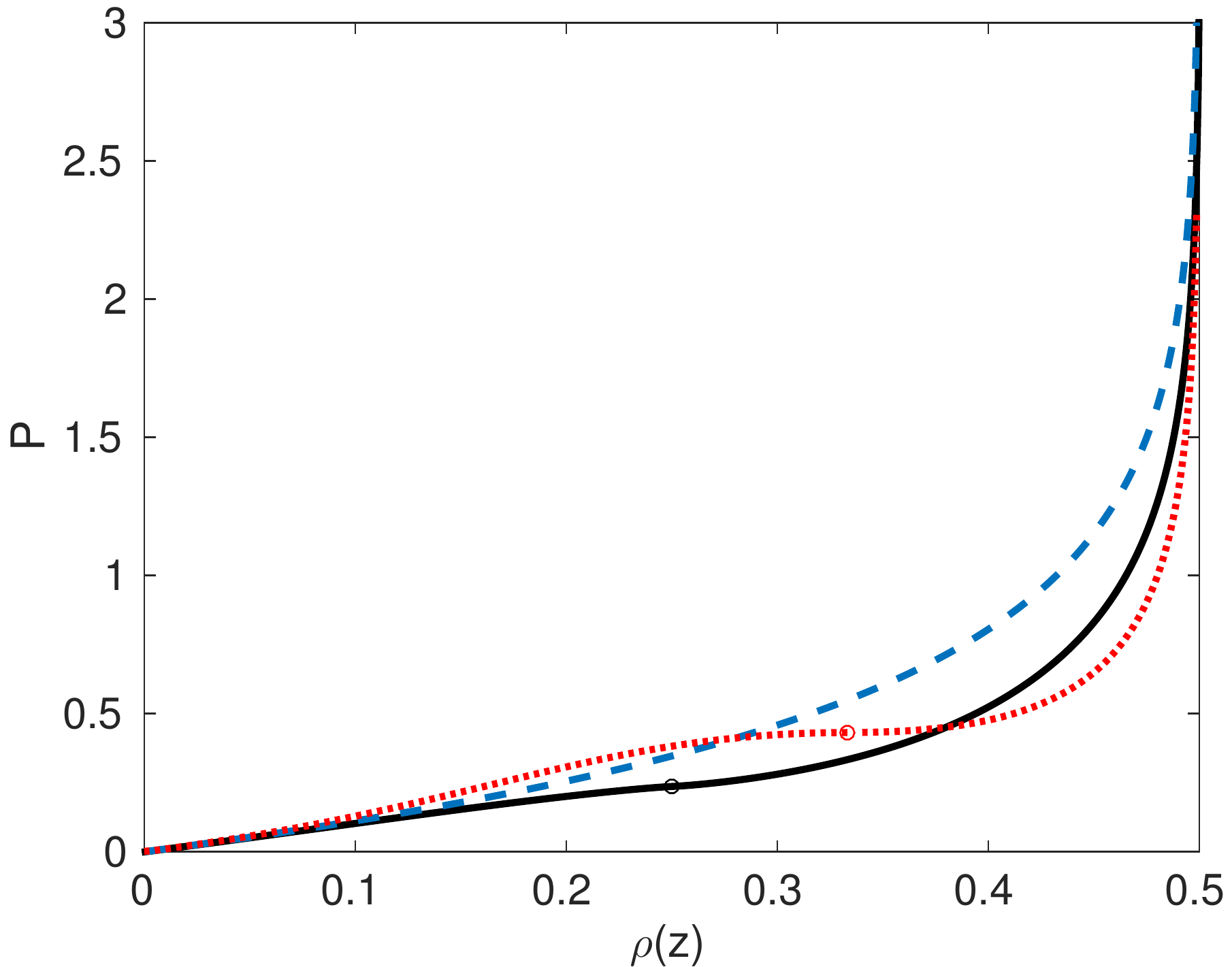}
\end{subfigure}
\end{center}
\caption{Equations of state as a function of crease-reverse density $\rho(y)$ on the left and $\rho(z)$ on the right for Miura-ori and trapezoid (solid black) and Barreto's Mars (blue dashed). On the right also, Miura-ori and trapezoid equation of state as a function of layer defect density $\rho(z)$ (dotted red). The phase transition points are indicated with a circle. \label{fig:eqofstate}}
\end{figure}

In figure~\ref{fig:eqofstate} we further compute the equation of state of these models.

\section{Simple square CP with Maekawa defects}\label{sec:maekawa}
\begin{figure}[htpb]
\begin{center}
\scalebox{0.65}{
\begin{tikzpicture}
\draw[dashed,line width = 2pt] (0,1) -- (0,0); 
\draw[dashed,line width = 2pt] (0,0) -- (0,-1); 
\draw[dashed,line width = 2pt] (-1,0) -- (0,0); 
\draw[dashed,line width = 2pt] (0,0) -- (1,0); 
\node at (-.7,.6) {{\Large $\omega^{(\mathrm{e})}_1$}};
\begin{scope}[shift={(3,0)}]
\draw[line width = 2pt] (0,1) -- (0,0); 
\draw[line width = 2pt] (0,0) -- (0,-1); 
\draw[line width = 2pt] (-1,0) -- (0,0); 
\draw[line width = 2pt] (0,0) -- (1,0);
\node at (-.7,.6) {{\Large $\omega^{(\mathrm{e})}_2$}};
\end{scope}
\begin{scope}[shift={(6,0)}]
\draw[line width = 2pt] (0,1) -- (0,0); 
\draw[line width = 2pt] (0,0) -- (0,-1); 
\draw[dashed,line width = 2pt] (-1,0) -- (0,0); 
\draw[dashed,line width = 2pt] (0,0) -- (1,0);
\node at (-.7,.6) {{\Large $\omega^{(\mathrm{e})}_3$}};
\end{scope}
\begin{scope}[shift={(9,0)}]
\draw[dashed,line width = 2pt] (0,1) -- (0,0); 
\draw[dashed,line width = 2pt] (0,0) -- (0,-1); 
\draw[line width = 2pt] (-1,0) -- (0,0); 
\draw[line width = 2pt] (0,0) -- (1,0);
\node at (-.7,.6) {{\Large $\omega^{(\mathrm{e})}_4$}};
\end{scope}
\begin{scope}[shift={(12,0)}]
\draw[dashed,line width = 2pt] (0,1) -- (0,0); 
\draw[line width = 2pt] (0,0) -- (0,-1); 
\draw[dashed,line width = 2pt] (-1,0) -- (0,0); 
\draw[line width = 2pt] (0,0) -- (1,0);
\node at (-.7,.6) {{\Large $\omega^{(\mathrm{e})}_5$}};
\end{scope}
\begin{scope}[shift={(15,0)}]
\draw[line width = 2pt] (0,1) -- (0,0); 
\draw[dashed,line width = 2pt] (0,0) -- (0,-1); 
\draw[line width = 2pt] (-1,0) -- (0,0); 
\draw[dashed,line width = 2pt] (0,0) -- (1,0);
\node at (-.7,.6) {{\Large $\omega^{(\mathrm{e})}_6$}};
\end{scope}
\begin{scope}[shift={(18,0)}]
\draw[dashed,line width = 2pt] (0,1) -- (0,0); 
\draw[line width = 2pt] (0,0) -- (0,-1); 
\draw[line width = 2pt] (-1,0) -- (0,0); 
\draw[dashed,line width = 2pt] (0,0) -- (1,0);
\node at (-.7,.6) {{\Large $\omega^{(\mathrm{e})}_7$}};
\end{scope}
\begin{scope}[shift={(21,0)}]
\draw[line width = 2pt] (0,1) -- (0,0); 
\draw[dashed,line width = 2pt] (0,0) -- (0,-1); 
\draw[dashed,line width = 2pt] (-1,0) -- (0,0); 
\draw[line width = 2pt] (0,0) -- (1,0);
\node at (-.7,.6) {{\Large $\omega^{(\mathrm{e})}_8$}};
\end{scope}
\end{tikzpicture}
}
\end{center}
\caption{The even 8-vertex model weights, with bond states shown in terms of line type, dashed or solid.\label{fig:evenconfigsbonds}}
\end{figure}
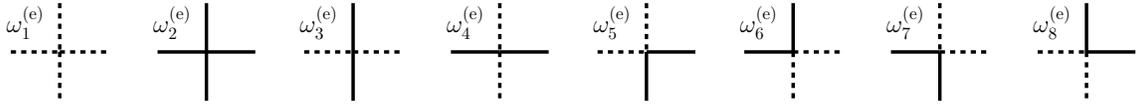

As a final investigation we can consider breaking Maekawa's theorem at each vertex of the homogeneous simple square CP; the model will no longer be locally flat-foldable. This opens up 8 more possible vertex configurations, the even 8-vertex weights $\omega^{(\mathrm{e})}_i$ shown in figure~\ref{fig:evenconfigsbonds}, for a total of 16 weights in the model. This is the 16-vertex model~\cite{assis2017temp}. 

There are few exactly solvable known cases of the general 16-vertex model. The only known cases with all non-zero weights are Baxter's symmetric even 8-vertex model~\cite{wu1972,baxter1982,baxter2007}, the even and odd free-fermion models mapped via the weak-graph expansion transformation~\cite{wu1972,wu2004k,assis2017temp}, and Deguchi's model~\cite{deguchi1991}. Of these, the only solvable case with all positive weights is the 16-vertex representation of the even free-fermion 8-vertex model~\cite{wu1972}, which has weights in the following symmetric form
\beqr
&& \omega^{(\mathrm{e})}_{2i}=\omega^{(\mathrm{e})}_{2i-1},\quad v_{2i}=v_{2i-1},\quad i=1\ldots8 \\
&& \omega^{(\mathrm{e})}_1\omega^{(\mathrm{e})}_3+\omega^{(\mathrm{e})}_5\omega^{(\mathrm{e})}_7 = v_1v_3+v_5v_7 \label{16vfreef}
\eeqr
and where this second condition is the 16-vertex model's representation of the free-fermion condition. This model has a mapping via a weak-graph transformation~\cite{assis2017temp} to the even free-fermion 8-vertex model with weights $\tilde{w}_i$, given below
\beqr
\tilde{w}_1 &=& \frac{1}{2}(\omega^{(\mathrm{e})}_1+\omega^{(\mathrm{e})}_3+\omega^{(\mathrm{e})}_5+\omega^{(\mathrm{e})}_7+v_1+v_3+v_5+v_7) \\
\tilde{w}_2 &=& \frac{1}{2}(\omega^{(\mathrm{e})}_1+\omega^{(\mathrm{e})}_3+\omega^{(\mathrm{e})}_5+\omega^{(\mathrm{e})}_7-v_1-v_3-v_5-v_7) \\
\tilde{w}_3 &=& \frac{1}{2}(\omega^{(\mathrm{e})}_1+\omega^{(\mathrm{e})}_3-\omega^{(\mathrm{e})}_5-\omega^{(\mathrm{e})}_7-v_1-v_3+v_5+v_7) \\
\tilde{w}_4 &=& \frac{1}{2}(\omega^{(\mathrm{e})}_1+\omega^{(\mathrm{e})}_3-\omega^{(\mathrm{e})}_5-\omega^{(\mathrm{e})}_7+v_1+v_3-v_5-v_7) \\
\tilde{w}_5 &=& \frac{1}{2}(\omega^{(\mathrm{e})}_1-\omega^{(\mathrm{e})}_3+\omega^{(\mathrm{e})}_5-\omega^{(\mathrm{e})}_7-v_1+v_3+v_5-v_7) \\
\tilde{w}_6 &=& \frac{1}{2}(\omega^{(\mathrm{e})}_1-\omega^{(\mathrm{e})}_3+\omega^{(\mathrm{e})}_5-\omega^{(\mathrm{e})}_7+v_1-v_3-v_5+v_7) \\
\tilde{w}_7 &=& \frac{1}{2}(\omega^{(\mathrm{e})}_1-\omega^{(\mathrm{e})}_3-\omega^{(\mathrm{e})}_5+\omega^{(\mathrm{e})}_7-v_1+v_3-v_5+v_7) \\
\tilde{w}_8 &=& \frac{1}{2}(\omega^{(\mathrm{e})}_1-\omega^{(\mathrm{e})}_3-\omega^{(\mathrm{e})}_5+\omega^{(\mathrm{e})}_7+v_1-v_3+v_5-v_7)
\eeqr
The exact solution of the even free-fermion 8-vertex model has the following free-energy~\cite{fan1970w}, 
\beq
-\beta\,f = \frac{1}{8\pi^2}\int_0^{2\pi}\int_0^{2\pi}d\theta_1 d\theta_2 \ln(A+2B\cos(\theta_1)+2C\cos(\theta_2)+2D\cos(\theta_1-\theta_2)+2E\cos(\theta_1+\theta_2) 
\eeq
where
\beqr
A &=& \tilde{w}_1^2+\tilde{w}_2^2+\tilde{w}_3^2+\tilde{w}_4^2 \\
B &=& \tilde{w}_1\tilde{w}_3-\tilde{w}_2\tilde{w}_4 \\
C &=& \tilde{w}_1\tilde{w}_4-\tilde{w}_2\tilde{w}_3 \\
D &=& \tilde{w}_3\tilde{w}_4-\tilde{w}_7\tilde{w}_8 = \tilde{w}_5\tilde{w}_6-\tilde{w}_1\tilde{w}_2 \\
E &=& \tilde{w}_3\tilde{w}_4-\tilde{w}_5\tilde{w}_6 = \tilde{w}_7\tilde{w}_8-\tilde{w}_1\tilde{w}_2
\eeqr
which has phase transitions at the points given by each of the following conditions, expressed in terms of the 16-vertex model weights $\omega^{(\mathrm{e})}_i$, $v_i$
\beqr
\omega^{(\mathrm{e})}_1+\omega^{(\mathrm{e})}_3 &=& \omega^{(\mathrm{e})}_5+\omega^{(\mathrm{e})}_7+v_1+v_3+v_5+v_7 \\
\omega^{(\mathrm{e})}_5+\omega^{(\mathrm{e})}_7 &=& \omega^{(\mathrm{e})}_1+\omega^{(\mathrm{e})}_3+v_1+v_3+v_5+v_7 \\
v_1+v_3 &=& \omega^{(\mathrm{e})}_1+\omega^{(\mathrm{e})}_3+\omega^{(\mathrm{e})}_5+\omega^{(\mathrm{e})}_7+v_5+v_7 \\
v_5+v_7 &=& \omega^{(\mathrm{e})}_1+\omega^{(\mathrm{e})}_3+\omega^{(\mathrm{e})}_5+\omega^{(\mathrm{e})}_7+v_1+v_3 
\eeqr
These phase transitions are second order logarithmic phase transitions in general, except in certain subcases where they're second order with critical exponent $\alpha=\tfrac{1}{2}$~\cite{fan1970w}.

If we choose all Maekawa defects to have equal weights so that $\omega^{(\mathrm{e})}_i=\omega$, then the free-fermion condition becomes
\beq
2\omega^2 = v_1v_3+v_5v_7 \label{maekawaff}
\eeq
The free-fermion condition~(\ref{maekawaff}) shows that this solvable case requires Maekawa defects to have Boltzmann weights $\omega^{(\mathrm{e})}_i$ of the same order as the flat-foldable weights $v_i$, which may not be a suitable assumption for certain applications, and it is not a simple extension of the flat-foldable model considered earlier in section~\ref{sec:square}.  Its phase transitions occur at
\beqr
v_1+v_3 &=& 4\omega+v_5+v_7 \\
v_5+v_7 &=& 4\omega+v_1+v_3
\eeqr

\section{Discussion}\label{sec:discussion}
\subsection{Rigid foldability}
In this paper we have concentrated on local flat-foldability properties of origami CPs. A further interest in the study of origami CPs is the determination of their rigid foldability~\cite{tachi2009,tachi20103,tachi20104,tachi2012,evans2015lmh,abel2016cdehklt,tachi2017h}, where as long as the hinges move, the faces are not require to bend for the lattice to fold. The Miura-ori, Barreto's Mars, trapezoid, and kite CPs are rigidly foldable in their ground states~\cite{evans2015lmh}. Based on rigid folding simulations of these CPs in Rigid Origami Simulator version 0.09~\cite{tachi20092,tachi2017}, it appears that the flat-foldable defects in these CPs destroy rigid foldability, so that the faces are required to bend in order to achieve the final flat-folded state (if the crease configuration can achieve global flat-foldability). In practice this may not present an issue. Indeed, an origami triangular lattice ``tessellation" in~\cite{na2015ebcslhh} with 198 creases of size $333\mu$m was able to self-fold correctly while requiring face bending to achieve the final state. Therefore, it would be beneficial to also model the face bending properties of these models~\cite{silverberg2015nelhslhc}. Using a triangular or union-jack lattice from which to construct staggered free-fermion odd 8-vertex models, it would be possible to model some face bending properties. In~\cite{silverberg2014emhhsc} the authors show how defects create subtle diagonal creases along neighboring faces, which a triangular lattice can easily capture. Otherwise, the two states of the diagonal bond can represent a concave or a convex face curvature. Using a union-jack lattice, each face of the origami CP is now covered with four creases which meet at a vertex. The two states of each bond in the face can then be used to model various degrees of curvature of the face.

\subsection{Free-fermion models}
Most of the exactly solvable models used in this work were free-fermion models. The homogeneous free-fermion model, used to model the homogeneous simple square CP as well as the Maekawa defects case, is equivalent to the union-jack and checkerboard lattice Ising models, and its phase transitions can all be mapped to the triangular lattice Ising model~\cite{assis2017temp}. It is unknown whether staggered free-fermion 8-vertex models also have similar interpretations in terms of Ising models on larger or modified lattices. In principle, the correlation functions of all free-fermion models can be studied in a straightforward manner by generalizing the methods in~\cite{mccoy1967w}, but as far as we are aware this has not yet been carried out for any of the cases of interest in this work. 

\subsection{Finite lattice results}
We have focused in this work on the thermodynamic limit, since it is only in this limit that phase transitions arise and can be analyzed. However, finite lattice partition functions are of interest in any experimental implementation of these origami CPs. Free-fermion model partition functions can be solved exactly on the finite lattice, in particular on the finite torus, using the methods in~\cite{mccoy1973w,mccoy2014w}. This may be of interest in practical applications, since toroidal boundary conditions become a better approximation to free boundary conditions as the size of the lattice increases. In fact, for large enough lattice sizes the thermodynamic limit results may be a very good approximation of the properties of finite free boundary origami lattices.  

\subsection{Random crumpling}
Various previous studies have considered random crumplings on lattices, see~\cite{francesco2000,bowick2001t,francesco2005g} for reviews, but they have generally neglected to distinguish between mountain and valley folds and did not incorporate Maekawa's theorem at each vertex; they also did not use exactly solvable models. It would be interesting to seek for an exact solution of these problems. In order to model random lattice crumpling problems, three-state bond variables are necessary, one for the absence of a crease and two for mountain and valley creases, and it is necessary to disallow sudden corner bends of creases. Also, higher degree vertices are preferrable, since they can model more crease angles in the lattice. Three-state vertex models, such as the 19-vertex Izergin-Korepin model~\cite{izergin1981k}, are well known, although the only known exactly solvable models are for the square lattice, and their symmetries do not allow the exclusion of corner bends of creases, which locally break Kawasaki's theorem and hence are not locally flat-foldable, without destroying the integrability of the model.

\section{Conclusions}\label{sec:conclusions}
We have made extensive use of exactly solvable models in order to study phase transition points of origami CP models. We have found phase transitions which depend on the arrangement of crease assignments around a vertex, on individual crease assignments, as well as on the local layer ordering of neighboring faces in the lattice. We also have found a phase transition in the case of breaking Maekawa's theorem at each vertex, so that the model is no longer locally flat-foldable.

Using our exactly solvable models, we have interpreted the flat-foldable crease-reversal and local layering order defects in terms of a lattice gas, which allowed us to find exact analytic expressions for their density as a function of the defect fugacity variables. This has allowed us to characterize how stable different CPs are against flat-foldable defects, or conversely, how tunable the origami CP is for the setting of the defect density. Miura-ori and trapezoid have the same defect density dependence which is less stable, and therefore more tunable, than Barreto's Mars. 

We have found that Miura-ori and trapezoid have phase transition points which determine either the loss of long range crease order in the lattice or else long range layer ordering in the lattice. Since the phase transition point depending on layer ordering is at a greater fugacity location compared to crease order, we see that any mechanical properties depending on layering order has a broader range of tunability compared to mechanical properties depending on crease ordering in the lattice. Barreto's Mars does not have a phase transition point, and though it is more stable against defects, defects can continue to be added until saturation without destroying the long range crease order properties of the CP; Barreto's Mars does not feature a 3-coloring problem interpretation, and so we were not able to analyze its layer ordering properties. The kite CP is effectively a one-dimensional model whose defects are lines which traverse the entire lattice, making it extremely stable; it does not have a phase transition point either. It remains to confirm experimentally the conclusions of this paper.

\section*{Acnowledgements}
We gratefully acknowledge numerous helpful conversations with Arthur A. Evans, as well as with Nathan Clisby and Iwan Jensen, during the preparation of this work. We would like to thank the Australian Research Council for supporting this work under the Discovery Project scheme (project number DP140101110).

\newpage
\appendix
\section{Dimer model solution of four-staggered odd 8-vertex model}\label{app:dimer}
In this appendix we outline the dimer solution method of~\cite{mccoy1973w,mccoy2014w}, generalizing the construction in~\cite{assis2017temp} to construct the free energy of the odd 8-vertex four-unit staggered free-fermion models. 

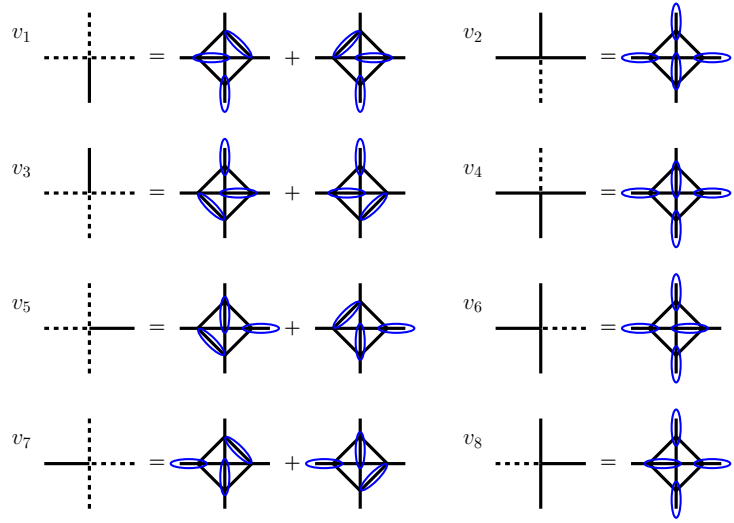
\begin{figure}[htpb]
\begin{center}
\scalebox{.6}{
\begin{tikzpicture}
\begin{scope}[shift={(10,-3)}]
\node at (-1.5,9.5) {{\Large $v_4$}};
\draw[line width = 2pt] (0,8) -- (0,9);
\draw[line width = 2pt,dashed] (0,10) -- (0,9);
\draw[line width = 2pt] (-1,9) -- (0,9);
\draw[line width = 2pt] (1,9) -- (0,9);
\node at (1.5,9) {\large $\bm{=}$};
\draw[line width = 2pt] (3,8) -- (3,9);
\draw[line width = 2pt] (3,10) -- (3,9);
\draw[line width = 2pt] (2,9) -- (3,9);
\draw[line width = 2pt] (4,9) -- (3,9);
\draw[line width = 2pt] (2.4,9) -- (3,9.6) -- (3.6,9) -- (3,8.4) -- (2.4,9);
\draw[line width = 1.3pt, color=blue] (2.2,9) ellipse (0.4 and 0.1);
\draw[line width = 1.3pt, color=blue] (3.8,9) ellipse (0.4 and 0.1);
\draw[line width = 1.3pt, color=blue,rotate around={90:(3,8.2)}] (3,8.2) ellipse (0.4 and 0.1);
\draw[line width = 1.3pt, color=blue,rotate around={90:(3,9.3)}] (3,9.3) ellipse (0.4 and 0.1);
\end{scope}
\begin{scope}[shift={(-7,-3)}]
\node at (5.5,9.5) {{\Large $v_3$}};
\draw[line width = 2pt,dashed] (7,8) -- (7,9);
\draw[line width = 2pt] (7,10) -- (7,9);
\draw[line width = 2pt,dashed] (6,9) -- (7,9);
\draw[line width = 2pt,dashed] (8,9) -- (7,9);
\node at (8.5,9) {\large $\bm{=}$};
\draw[line width = 2pt] (10,8) -- (10,9);
\draw[line width = 2pt] (10,10) -- (10,9);
\draw[line width = 2pt] (9,9) -- (10,9);
\draw[line width = 2pt] (11,9) -- (10,9);
\draw[line width = 2pt] (9.4,9) -- (10,9.6) -- (10.6,9) -- (10,8.4) -- (9.4,9);
\draw[line width = 1.3pt, color=blue] (10.3,9) ellipse (0.4 and 0.1);
\draw[line width = 1.3pt, color=blue,rotate around={135:(9.7,8.7)}] (9.7,8.7) ellipse (0.4 and 0.1);
\draw[line width = 1.3pt, color=blue,rotate around={90:(10,9.8)}] (10,9.8) ellipse (0.4 and 0.1);
\node at (11.5,9) {\large $\bm{+}$};
\draw[line width = 2pt] (13,8) -- (13,9);
\draw[line width = 2pt] (13,10) -- (13,9);
\draw[line width = 2pt] (12,9) -- (13,9);
\draw[line width = 2pt] (14,9) -- (13,9);
\draw[line width = 2pt] (12.4,9) -- (13,9.6) -- (13.6,9) -- (13,8.4) -- (12.4,9);
\draw[line width = 1.3pt, color=blue] (12.7,9) ellipse (0.4 and 0.1);
\draw[line width = 1.3pt, color=blue,rotate around={45:(13.3,8.7)}] (13.3,8.7) ellipse (0.4 and 0.1);
\draw[line width = 1.3pt, color=blue,rotate around={90:(13,9.8)}] (13,9.8) ellipse (0.4 and 0.1);
\end{scope}

\begin{scope}[shift={(10,3)}]
\node at (-1.5,6.5) {{\Large $v_2$}};
\draw[line width = 2pt,dashed] (0,5) -- (0,6);
\draw[line width = 2pt] (0,7) -- (0,6);
\draw[line width = 2pt] (-1,6) -- (0,6);
\draw[line width = 2pt] (1,6) -- (0,6);
\node at (1.5,6) {\large $\bm{=}$};
\draw[line width = 2pt] (3,5) -- (3,6);
\draw[line width = 2pt] (3,7) -- (3,6);
\draw[line width = 2pt] (2,6) -- (3,6);
\draw[line width = 2pt] (4,6) -- (3,6);
\draw[line width = 2pt] (2.4,6) -- (3,6.6) -- (3.6,6) -- (3,5.4) -- (2.4,6);
\draw[line width = 1.3pt, color=blue] (2.2,6) ellipse (0.4 and 0.1);
\draw[line width = 1.3pt, color=blue] (3.8,6) ellipse (0.4 and 0.1);
\draw[line width = 1.3pt, color=blue,rotate around={90:(3,5.7)}] (3,5.7) ellipse (0.4 and 0.1);
\draw[line width = 1.3pt, color=blue,rotate around={90:(3,6.8)}] (3,6.8) ellipse (0.4 and 0.1);
\end{scope}
\begin{scope}[shift={(-7,3)}]
\node at (5.5,6.5) {{\Large $v_1$}};
\draw[line width = 2pt] (7,5) -- (7,6);
\draw[line width = 2pt,dashed] (7,7) -- (7,6);
\draw[line width = 2pt,dashed] (6,6) -- (7,6);
\draw[line width = 2pt,dashed] (8,6) -- (7,6);
\node at (8.5,6) {\large $\bm{=}$};
\draw[line width = 2pt] (10,5) -- (10,6);
\draw[line width = 2pt] (10,7) -- (10,6);
\draw[line width = 2pt] (9,6) -- (10,6);
\draw[line width = 2pt] (11,6) -- (10,6);
\draw[line width = 2pt] (9.4,6) -- (10,6.6) -- (10.6,6) -- (10,5.4) -- (9.4,6);
\draw[line width = 1.3pt, color=blue] (9.7,6) ellipse (0.4 and 0.1);
\draw[line width = 1.3pt, color=blue,rotate around={90:(10,5.2)}] (10,5.2) ellipse (0.4 and 0.1);
\draw[line width = 1.3pt, color=blue,rotate around={135:(10.3,6.3)}] (10.3,6.3) ellipse (0.4 and 0.1);
\node at (11.5,6) {\large $\bm{+}$};
\draw[line width = 2pt] (13,5) -- (13,6);
\draw[line width = 2pt] (13,7) -- (13,6);
\draw[line width = 2pt] (12,6) -- (13,6);
\draw[line width = 2pt] (14,6) -- (13,6);
\draw[line width = 2pt] (12.4,6) -- (13,6.6) -- (13.6,6) -- (13,5.4) -- (12.4,6);
\draw[line width = 1.3pt, color=blue] (13.3,6) ellipse (0.4 and 0.1);
\draw[line width = 1.3pt, color=blue,rotate around={90:(13,5.2)}] (13,5.2) ellipse (0.4 and 0.1);
\draw[line width = 1.3pt, color=blue,rotate around={45:(12.7,6.3)}] (12.7,6.3) ellipse (0.4 and 0.1);
\end{scope}

\begin{scope}[shift={(10,-3)}]
\node at (-1.5,3.5) {{\Large $v_8$}};
\draw[line width = 2pt] (0,2) -- (0,3);
\draw[line width = 2pt] (0,4) -- (0,3);
\draw[line width = 2pt,dashed] (-1,3) -- (0,3);
\draw[line width = 2pt] (1,3) -- (0,3);
\node at (1.5,3) {\large $\bm{=}$};
\draw[line width = 2pt] (3,2) -- (3,3);
\draw[line width = 2pt] (3,4) -- (3,3);
\draw[line width = 2pt] (2,3) -- (3,3);
\draw[line width = 2pt] (4,3) -- (3,3);
\draw[line width = 2pt] (2.4,3) -- (3,3.6) -- (3.6,3) -- (3,2.4) -- (2.4,3);
\draw[line width = 1.3pt, color=blue] (2.7,3) ellipse (0.4 and 0.1);
\draw[line width = 1.3pt, color=blue] (3.8,3) ellipse (0.4 and 0.1);
\draw[line width = 1.3pt, color=blue,rotate around={90:(3,2.2)}] (3,2.2) ellipse (0.4 and 0.1);
\draw[line width = 1.3pt, color=blue,rotate around={90:(3,3.8)}] (3,3.8) ellipse (0.4 and 0.1);
\end{scope}
\begin{scope}[shift={(-7,-3)}]
\node at (5.5,3.5) {{\Large $v_7$}};
\draw[line width = 2pt,dashed] (7,2) -- (7,3);
\draw[line width = 2pt,dashed] (7,4) -- (7,3);
\draw[line width = 2pt] (6,3) -- (7,3);
\draw[line width = 2pt,dashed] (8,3) -- (7,3);
\node at (8.5,3) {\large $\bm{=}$};
\draw[line width = 2pt] (10,2) -- (10,3);
\draw[line width = 2pt] (10,4) -- (10,3);
\draw[line width = 2pt] (9,3) -- (10,3);
\draw[line width = 2pt] (11,3) -- (10,3);
\draw[line width = 2pt] (9.4,3) -- (10,3.6) -- (10.6,3) -- (10,2.4) -- (9.4,3);
\draw[line width = 1.3pt, color=blue] (9.2,3) ellipse (0.4 and 0.1);
\draw[line width = 1.3pt, color=blue,rotate around={135:(10.3,3.3)}] (10.3,3.3) ellipse (0.4 and 0.1);
\draw[line width = 1.3pt, color=blue,rotate around={90:(10,2.7)}] (10,2.7) ellipse (0.4 and 0.1);
\node at (11.5,3) {\large $\bm{+}$};
\draw[line width = 2pt] (13,2) -- (13,3);
\draw[line width = 2pt] (13,4) -- (13,3);
\draw[line width = 2pt] (12,3) -- (13,3);
\draw[line width = 2pt] (14,3) -- (13,3);
\draw[line width = 2pt] (12.4,3) -- (13,3.6) -- (13.6,3) -- (13,2.4) -- (12.4,3);
\draw[line width = 1.3pt, color=blue] (12.2,3) ellipse (0.4 and 0.1);
\draw[line width = 1.3pt, color=blue,rotate around={45:(13.3,2.7)}] (13.3,2.7) ellipse (0.4 and 0.1);
\draw[line width = 1.3pt, color=blue,rotate around={90:(13,3.3)}] (13,3.3) ellipse (0.4 and 0.1);
\end{scope}

\begin{scope}[shift={(10,3)}]
\node at (-1.5,0.5) {{\Large $v_6$}};
\draw[line width = 2pt] (0,-1) -- (0,0);
\draw[line width = 2pt] (0,1) -- (0,0);
\draw[line width = 2pt] (-1,0) -- (0,0);
\draw[line width = 2pt,dashed] (1,0) -- (0,0);
\node at (1.5,0) {\large $\bm{=}$};
\draw[line width = 2pt] (3,-1) -- (3,0);
\draw[line width = 2pt] (3,1) -- (3,0);
\draw[line width = 2pt] (2,0) -- (3,0);
\draw[line width = 2pt] (4,0) -- (3,0);
\draw[line width = 2pt] (2.4,0) -- (3,0.6) -- (3.6,0) -- (3,-0.6) -- (2.4,0);
\draw[line width = 1.3pt, color=blue] (2.2,0) ellipse (0.4 and 0.1);
\draw[line width = 1.3pt, color=blue] (3.3,0) ellipse (0.4 and 0.1);
\draw[line width = 1.3pt, color=blue,rotate around={90:(3,-0.8)}] (3,-0.8) ellipse (0.4 and 0.1);
\draw[line width = 1.3pt, color=blue,rotate around={90:(3,0.8)}] (3,0.8) ellipse (0.4 and 0.1);
\end{scope}
\begin{scope}[shift={(-7,3)}]
\node at (5.5,0.5) {{\Large $v_5$}};
\draw[line width = 2pt,dashed] (7,-1) -- (7,0);
\draw[line width = 2pt,dashed] (7,1) -- (7,0);
\draw[line width = 2pt,dashed] (6,0) -- (7,0);
\draw[line width = 2pt] (8,0) -- (7,0);
\node at (8.5,0) {\large $\bm{=}$};
\draw[line width = 2pt] (10,-1) -- (10,0);
\draw[line width = 2pt] (10,1) -- (10,0);
\draw[line width = 2pt] (9,0) -- (10,0);
\draw[line width = 2pt] (11,0) -- (10,0);
\draw[line width = 2pt] (9.4,0) -- (10,0.6) -- (10.6,0) -- (10,-0.6) -- (9.4,0);
\draw[line width = 1.3pt, color=blue,rotate around={135:(9.7,-0.3)}] (9.7,-0.3) ellipse (0.4 and 0.1);
\draw[line width = 1.3pt, color=blue] (10.8,0) ellipse (0.4 and 0.1);
\draw[line width = 1.3pt, color=blue,rotate around={90:(10,0.3)}] (10,0.3) ellipse (0.4 and 0.1);
\node at (11.5,0) {\large $\bm{+}$};
\draw[line width = 2pt] (13,-1) -- (13,0);
\draw[line width = 2pt] (13,1) -- (13,0);
\draw[line width = 2pt] (12,0) -- (13,0);
\draw[line width = 2pt] (14,0) -- (13,0);
\draw[line width = 2pt] (12.4,0) -- (13,0.6) -- (13.6,0) -- (13,-0.6) -- (12.4,0);
\draw[line width = 1.3pt, color=blue,rotate around={45:(12.7,0.3)}] (12.7,0.3) ellipse (0.4 and 0.1);
\draw[line width = 1.3pt, color=blue] (13.8,0) ellipse (0.4 and 0.1);
\draw[line width = 1.3pt, color=blue,rotate around={90:(13,-0.3)}] (13,-0.3) ellipse (0.4 and 0.1);
\end{scope}
\end{tikzpicture}
}
\caption{The correspondence between the odd 8-vertex model weights and dimer coverings.\label{fig:weights}}
\end{center}
\end{figure}

We consider a square lattice with toroidal boundary conditions of size $M\times N$ and use the dimer construction of~\cite{assis2017temp}, show in figure~\ref{fig:weights}. Each of the vertex weights can be written in terms of the lattice bond weights $z_i$, defined in figure~\ref{fig:bonddefs1}. 

\begin{figure}[htpb]
\begin{center}
\scalebox{1.}{
\begin{tikzpicture}
\draw[line width = 1pt] (-2.25,0) -- (2.25,0);
\draw[line width = 1pt] (0,-2.25) -- (0,2.25); 
\draw[line width = 1pt] (0,-1) -- (1,0) -- (0,1) -- (-1,0) -- (0,-1);
\node[above=4.5pt, right=-2pt] at (0,1) {\large\textbf{U}};
\node[below=5.5pt, right=-2pt] at (0,-1) {\large\textbf{D}};
\node[above=5.5pt, left=-2pt] at (-1,0) {\large\textbf{L}};
\node[above=5.5pt, right=-2.5pt] at (1,0) {\large\textbf{R}};
\node[above=5.5pt, right=-2.5pt] at (0,0) {\large\textbf{C}};
\node[above=3.5pt, left=-3.5pt] at (-0.5,0.5) {$z_1$};
\node[below=3.5pt, right=-3.5pt] at (0.5,-0.5) {$z_2$};
\node[above=3.5pt, right=-3.5pt] at (0.5,0.5) {$z_3$};
\node[below=3.5pt, left=-3.5pt] at (-0.5,-0.5) {$z_4$};
\node[below=3.5pt, left=-3.5pt] at (-0.5,-0.5) {$z_4$};
\node[below=1pt,left=-3.5pt] at (0,0.5) {$z_5$};
\node[above=1pt,left=-3pt] at (0,-0.5) {$z_6$};
\node[above=-2.5pt] at (-0.4,0) {$z_7$};
\node[below=-1.5pt] at (0.4,0) {$z_8$};
\node[above=4.5pt, right=-2pt] at (0,1.6) {1};
\node[below=5.5pt, right=-2pt] at (0,-1.6) {1};
\node[above=5.5pt, left=-2pt] at (-1.6,0) {1};
\node[above=5.5pt, right=-2.5pt] at (1.6,0) {1};
\end{tikzpicture}
}
\caption{Site and bond weight definitions around a cluster for the odd 8-vertex model.\label{fig:bonddefs1}}
\end{center}
\end{figure}
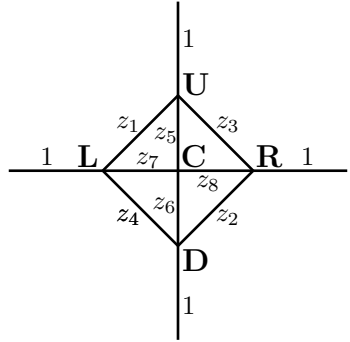

\begin{figure}[htpb]
\begin{center}
\scalebox{.55}{
\begin{tikzpicture}
\foreach \position in {(0,0),(6,0),(0,6),(6,6)}
{\begin{scope}[shift={\position}]
\draw[line width = 1pt] (-2,0) -- (5,0);
\draw[line width = 1pt] (-2,3) -- (5,3);
\draw[line width = 1pt] (0,-2) -- (0,5); 
\draw[line width = 1pt] (3,-2) -- (3,5); 
\draw[line width = 1.5pt, dashed] (-1.15,-1.15) -- (-1.15,4.15) -- (4.15,4.15) -- (4.15,-1.15) -- (-1.15,-1.15);
\draw[-{Latex[length=10pt]}] (1,0) -- (1.7,0);
\draw[-{Latex[length=10pt]}] (3,0) -- (4.7,0);
\draw[-{Latex[length=10pt]}] (1,3) -- (1.7,3);
\draw[-{Latex[length=10pt]}] (3,3) -- (4.7,3);
\draw[-{Latex[length=10pt]}] (0,1) -- (0,1.7);
\draw[-{Latex[length=10pt]}] (0,4) -- (0,4.7);
\draw[-{Latex[length=10pt]}] (3,2) -- (3,1.2);
\draw[-{Latex[length=10pt]}] (3,5) -- (3,4.2);

\draw[line width = 1pt] (0,-1) -- (1,0) -- (0,1) -- (-1,0) -- (0,-1);
\draw[-{Latex[length=10pt]}] (0,0) -- (-0.7,0);
\draw[-{Latex[length=10pt]}] (1,0) -- (0.2,0);
\draw[-{Latex[length=10pt]}] (0,0) -- (0,-0.7);
\draw[-{Latex[length=10pt]}] (0,1) -- (0,0.2);
\draw[-{Latex[length=10pt]}] (0,1) -- (-0.7,0.3);
\draw[-{Latex[length=10pt]}] (0,1) -- (0.7,0.3);
\draw[-{Latex[length=10pt]}] (0,-1) -- (0.7,-0.3);
\draw[-{Latex[length=10pt]}] (0,-1) -- (-0.7,-0.3);
\begin{scope}[shift={(3,0)}]
\draw[line width = 1pt] (0,-1) -- (1,0) -- (0,1) -- (-1,0) -- (0,-1);
\draw[-{Latex[length=10pt]}] (0,0) -- (-0.7,0);
\draw[-{Latex[length=10pt]}] (1,0) -- (0.2,0);
\draw[-{Latex[length=10pt]}] (0,0) -- (0,-0.7);
\draw[-{Latex[length=10pt]}] (0,1) -- (0,0.2);
\draw[-{Latex[length=10pt]}] (0,1) -- (-0.7,0.3);
\draw[-{Latex[length=10pt]}] (0,1) -- (0.7,0.3);
\draw[-{Latex[length=10pt]}] (0,-1) -- (0.7,-0.3);
\draw[-{Latex[length=10pt]}] (0,-1) -- (-0.7,-0.3);
\end{scope}
\begin{scope}[shift={(0,3)}]
\draw[line width = 1pt] (0,-1) -- (1,0) -- (0,1) -- (-1,0) -- (0,-1);
\draw[-{Latex[length=10pt]}] (0,0) -- (-0.7,0);
\draw[-{Latex[length=10pt]}] (1,0) -- (0.2,0);
\draw[-{Latex[length=10pt]}] (0,0) -- (0,-0.7);
\draw[-{Latex[length=10pt]}] (0,1) -- (0,0.2);
\draw[-{Latex[length=10pt]}] (0,1) -- (-0.7,0.3);
\draw[-{Latex[length=10pt]}] (0,1) -- (0.7,0.3);
\draw[-{Latex[length=10pt]}] (0,-1) -- (0.7,-0.3);
\draw[-{Latex[length=10pt]}] (0,-1) -- (-0.7,-0.3);
\end{scope}
\begin{scope}[shift={(3,3)}]
\draw[line width = 1pt] (0,-1) -- (1,0) -- (0,1) -- (-1,0) -- (0,-1);
\draw[-{Latex[length=10pt]}] (0,0) -- (-0.7,0);
\draw[-{Latex[length=10pt]}] (1,0) -- (0.2,0);
\draw[-{Latex[length=10pt]}] (0,0) -- (0,-0.7);
\draw[-{Latex[length=10pt]}] (0,1) -- (0,0.2);
\draw[-{Latex[length=10pt]}] (0,1) -- (-0.7,0.3);
\draw[-{Latex[length=10pt]}] (0,1) -- (0.7,0.3);
\draw[-{Latex[length=10pt]}] (0,-1) -- (0.7,-0.3);
\draw[-{Latex[length=10pt]}] (0,-1) -- (-0.7,-0.3);
\end{scope}
\end{scope}
}
\draw[-{Latex[length=10pt]}] (-2,0) -- (-1.3,0);
\draw[-{Latex[length=10pt]}] (-2,3) -- (-1.3,3);
\draw[-{Latex[length=10pt]}] (-2,6) -- (-1.3,6);
\draw[-{Latex[length=10pt]}] (-2,9) -- (-1.3,9);
\draw[-{Latex[length=10pt]}] (0,-2) -- (0,-1.3);
\draw[-{Latex[length=10pt]}] (3,-1) -- (3,-1.8);
\draw[-{Latex[length=10pt]}] (6,-2) -- (6,-1.3);
\draw[-{Latex[length=10pt]}] (9,-1) -- (9,-1.8);
\node at (0.85,-0.8) {\huge\textbf{1}};
\node at (3.85,-0.8) {\huge\textbf{2}};
\node at (0.85,2.2) {\huge\textbf{3}};
\node at (3.85,2.2) {\huge\textbf{4}};
\end{tikzpicture}
}
\caption{Orientation graph convention on a column staggered lattice with four dimer cluster units for the odd 8-vertex model. The numbering convention for each of the four cluster units is shown.\label{fig:orientgraph}}
\end{center}
\end{figure}
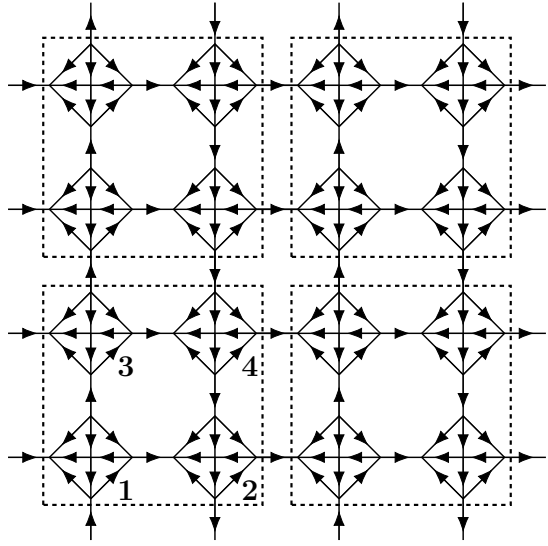

From figures~\ref{fig:weights} and~\ref{fig:bonddefs1}, the odd 8-vertex weights $v_i$ have the following expressions in terms of lattice bond weights $z_i$
\bat{4}
v_1 &= z_1z_8+z_3z_7,\qquad & v_2 &= z_6,\qquad & v_3  &= z_2z_7+z_4z_8,\qquad & v_4 &= z_5, \\
v_5 &= z_1z_6+z_4z_5,\qquad & v_6 &= z_8,\qquad & v_7  &= z_2z_5+z_3z_6,\qquad & v_8 &= z_7      
\eat
It can be seen from these relations that the vertex weights $v_i$ follow the free-fermion condition
\beq
v_1v_2+v_3v_4 = v_5v_6 + v_7v_8 
\eeq
One of the weights $z_i$ is superfluous and can be made arbitrary. We here take $z_2=1$.

Solving for the bond weights $z_i$ in terms of the vertex weights $v_i$, we have
\beqr
z_1 &=& \frac{v_4v_8 +v_5v_6 -v_3v_4}{v_2v_6} = \frac{v_1v_2+v_4v_8-v_7v_8}{v_2v_6}\\
z_2 &=& 1,\qquad z_3 = \frac{v_7-v_4}{v_2},\qquad z_4 = \frac{v_3-v_8}{v_6} \\
z_5 &=& v_4,\qquad z_6 = v_2,\qquad z_7 = v_8,\qquad z_8 = v_6 
\eeqr

For the four-unit staggered odd 8-vertex free-fermion model, we define independent vertex weights on each of the four cluster units. We show in figure~\ref{fig:orientgraph} the numbering convention for each cluster. From the first to the fourth we use the vertex weight notations $v_i$, $w_i$, $t_i$, $u_i$, respectively. Each independent set of vertex weights satisfies the free-fermion constraint.

We use a straightforward column staggering of the four-unit on the lattice. The model's partition function is given by a Pfaffian whose square is given by the determinant of the following matrix
\beq
M_{4U} = T\otimes I_N \otimes I_M + A_1\otimes H_N^\mathrm{T}\otimes I_M + A_2\otimes H_N\otimes I_M + B_1\otimes I_N\otimes H_M^\mathrm{T}+ B_2\otimes I_N\otimes H_M
\eeq
where the $I_n$ are $n\times n$ identity matrices, where the $n\times n$ matrix $H_n$ is defined as
\beq
H_n =
 \begin{pmatrix}
  0 & 1 & 0 & \cdots & 0 \\
  0 & 0 & 1 & \cdots & 0 \\
  \vdots  & \vdots & \vdots  & \ddots & \vdots  \\
  0 & 0 & 0 & \cdots & 1 \\
  1 & 0 & 0 & \cdots & 0
 \end{pmatrix}
\eeq
where the $20\times20$ matrix $T$ is defined as
\beq
T =
 \begin{pmatrix}
  T_1 & 0 & 0 & 0 \\
  0 & T_2 & 0 & 0 \\
  0 & 0 & T_3 & 0 \\
  0 & 0 & 0 & T_4
 \end{pmatrix}
\eeq
where $T_i$ are defined as 
\beq
T_i = \bordermatrix{
~ & U_i & D_i & L_i & R_i & C_i\cr
U_i & 0 & 0 & z_1 & z_3 & z_5  \cr
D_i & 0 & 0 & z_4 & z_2 & -z_6  \cr
L_i & -z_1 & -z_4 & 0 & 0 & -z_7  \cr
R_i & -z_3 & -z_2 & 0 & 0 & z_8  \cr
C_i & -z_5 & z_6 & z_7 & -z_8 & 0 \cr
}
\eeq
with the appropriate independent unit vertex weights substituted for the bond weights $z_i$ in each $T_i$ and with the further elements of $T$ given by
\bal
(T)_{U_1,D_3} &= -(T)_{D_3,U_1} = 1 \\
(T)_{R_1,L_2} &= -(T)_{L_2,R_1} = 1 \\
(T)_{U_2,D_4} &= -(T)_{D_4,U_2} = -1 \\
(T)_{R_3,L_4} &= -(T)_{L_4,R_3} = 1 
\eal
and where the non-zero elements of the $20\times20$ matrices $A_1,~A_2,~B_1,~B_2$ are given by 
\beqr
(A_1)_{D_1,U_3} &=& -(A_1)_{D_2,U_4} = -1 \\
(A_2)_{U_3,D_1} &=& -(A_2)_{U_4,D_2} = 1 \\
(B_1)_{L_1,R_2} &=& (B_1)_{L_3,R_4} = -1 \\
(B_2)_{R_2,L_1} &=& (B_2)_{R_4,L_3} = 1 
\eeqr

The matrices $H_n$ have eigenvalues $e^{2\pi i k/n}$, and since the $H_n$ are unitary
\beq
H_nH_n^{\mathrm{T}} = 1
\eeq
so that the eigenvalues of $H_n^{\mathrm{T}}$ are $e^{-2\pi i k/n}$.

The determinant of $M_{4U}$ is then given by the double product of the determinant $D(\theta_1,\theta_2)$, 
\beq
\mathrm{Det}(M_{4U}) = \prod_{\theta_1}\prod_{\theta_2}  D(\theta_1,\theta_2)
\eeq
where  $D(\theta_1,\theta_2)$ is the determinant of the matrix $T$ with additional entries coming from the diagonalization of the matrices $A_1,~A_2,~B_1,~B_2$,
\beqr
(T)_{D_1,U_3} &=& -(T)_{D_2,U_4} = -e^{-i\theta_1} \\
(T)_{U_3,D_1} &=& -(T)_{U_4,D_2} = e^{i\theta_1} \\
(T)_{L_1,R_2} &=& (T)_{L_3,R_4} = -e^{-i\theta_2} \\
(T)_{R_2,L_1} &=& (T)_{R_4,L_3} = e^{i\theta_2}
\eeqr
where
\beq
\theta_1 = \frac{2\pi n}{N},\qquad\theta_2 = \frac{2\pi m}{M}
\eeq
and where $n=1,\ldots,N$ and $m=1,\ldots,M$.

The partition function $Z_{4U}$ is the square root of the determinant of $M_{4U}$\footnote{On the finite lattice, for toroidal boundary conditions, four Pfaffians are actually needed, but they become degenerate in the thermodynamic limit~\cite{mccoy1973w,mccoy2014w}.}. In the thermodynamic limit, the free energy $f_{4U}$ is given by 
\beq
f_{4U} = -\frac{1}{4\beta}\lim_{M\to\infty}\lim_{N\to\infty}\frac{1}{MN}\ln Z_{4U} = -\frac{1}{4\beta}\lim_{M\to\infty}\lim_{N\to\infty}\frac{1}{MN}\ln \mathrm{Det}(M_{4U})^{1/2}
\eeq
where the leading factor of 4 is due to the fact that there are four dimer clusters in each unit. 

The logarithm of the products in the determinant can be expanded and written as integrals in the thermodynamic limit, giving the following free energy expression
\beq
-\beta f_{4U} =  \frac{1}{32\pi^2}\int_0^{2\pi}\int_0^{2\pi} \ln\big[D(\theta_1,\theta_2)\big] d\theta_1d\theta_2
\eeq
The expression for the free-energy is of the form
\bal
-\beta f_{4U} =  \frac{1}{32\pi^2}\int_0^{2\pi}\int_0^{2\pi} \ln\big[&A+2B\cos(\theta_1)+2C\cos(\theta_2)+2D\cos(\theta_1+\theta_2)+ 2E\cos(\theta_1-\theta_2)\nonumber\\
&+2F\cos(2\theta_1)+2G\cos(2\theta_2)+2H\cos(2\theta_1+\theta_2)+2I\cos(2\theta_1-\theta_2)\nonumber\\
&+2J\cos(\theta_1+2\theta_2)+2K\cos(\theta_1-2\theta_2)+2L\cos(2\theta_1+2\theta_2)\nonumber\\
&+2M\cos(2\theta_1-2\theta_2) \big] d\theta_1d\theta_2
\eal
where the $A,\ldots,M$ are large polynomials in the four sets of vertex weights $t_i$, $u_i$, $v_i$, $w_i$. 

Four phase transition conditions can be found by setting $\theta_1,\theta_2=0,\pi$ in $D(\theta_1,\theta_2)$, as explained in~\cite{green1964h,hurst1963}. A phase transition occurs when at least one of the four conditions are satisfied, and each of the conditions can define multiple phase transition points~\cite{hsue1975lw,lin1977w,sacco1975w,sacco1977w}. For the homogeneous lattice, that is, the unstaggered lattice, it can be proven that all of the physical phase transitions can be found from such conditions~\cite{green1964h,hurst1963}. For the staggered lattices we are considering, it appears that no such proof is known, however.

\section{Free-energies}\label{app:freeenergy}
Here we collect particular free-energy expressions. The expressions for the general four-unit staggered model are very large and we omit them. Using the mappings between the staggered even 8-vertex models and staggered odd 8-vertex models of~\cite{assis2017temp}, the expressions in~\cite{lin1977w} can alternatively be used for the free-energy of this most general case.

\subsection{Miura free-energy}
The expression for the Miura free-energy is of the form
\bal
-\beta f_{Mi} =  \frac{1}{32\pi^2}\int_0^{2\pi}\int_0^{2\pi} \ln\big[&A+2B\cos(\theta_1)+2C\cos(\theta_2)+2D\cos(\theta_1+\theta_2)+ 2E\cos(\theta_1-\theta_2)\nonumber\\
&+2F\cos(2\theta_1)+2G\cos(2\theta_2)+2H\cos(2\theta_1+\theta_2)+2I\cos(2\theta_1-\theta_2)\nonumber\\
&+2J\cos(\theta_1+2\theta_2)+2K\cos(\theta_1-2\theta_2)+2L\cos(2\theta_1+2\theta_2)\nonumber\\
&+2M\cos(2\theta_1-2\theta_2) \big] d\theta_1d\theta_2
\eal
where after simplifications of the general terms using the free-fermion conditions for the four sets of vertex weights we have the following expressions for each term
\bal
A &= 
u_3^2v_2^2w_4^2t_1^2+t_2^2u_4^2w_3^2v_1^2+(v_5^2w_6^2+v_7^2w_8^2)(t_5^2u_6^2+t_7^2u_8^2)+(v_5^2w_7^2+v_7^2w_5^2)(t_5^2u_7^2+t_7^2u_5^2) \nonumber\\
&\quad +(v_6^2w_5^2+v_8^2w_7^2)(t_6^2u_5^2+t_8^2u_7
^ 2) + ( v_6 ^ 2 w_8^ 2 + v_8^ 2 w_6 ^ 2 ) ( t_6^ 2  
u_8^2+t_8^2u_6^2)+2t_1t_2v_1v_2u_3u_4w_3w_4 \nonumber\\
&\quad +2t_5t_6u_5u_6v_7v_8w_7w_8+2t_7t_8u_7u_8v_5v_6w_5w_6 \nonumber\\
&\quad
+2(t_1u_3v_2w_4+t_2u_4v_1w_3)(t_5u_6v_7w_8+t_6u_5v_8w_7+t_7u_8v_5w_6+t_8u_7v_6w_5) \nonumber\\
&\quad +2(t_7u_5v_5w_7+t_8u_6v_6w_8)(t_5
u_7 v_7 w_5 + t_6 u_8  v_8  w_6) + 2( t_7  u_8 v_7 w_8+ t_8  u_7 v_8w_7)(t_5u_6v_5w_6+t_6u_5v_6w_5) \\
B &= 
(-v_5^2w_6w_7+v_7^2w_5w_8)(t_5^2u_6u_7-t_7^2u_5u_8)+(-v_5v_8w_6^2+v_6v_7w_8^2)(t_5t_8u_6^2-t_6t_7u_8^2) \nonumber\\
&\quad +(-v_5v_8w_7^2+v_6v_7w_5^2)(t_5t_8u_7^2-t_6t_7u_5^2)+(v_6^2w_5w_8-
v_8^2  w_6 w_7) ( - t_6 ^2 u_5 u_8 + t_8^2 u_6 u_7 ) \nonumber\\
&\quad +( u_5 u_7w_5w_7+u_6u_8w_6w_8)(t_5t_6v_7v_8+t_7t_8v_5v_6)+(u_5u_6w_7w_8+u_7u_8w_5w_6)(t_5 t_7 v_5  v_7 + t_6  t_8 v_6  v_8) \nonumber\\
&\quad +(t_1u_3v_2w_4+t_2u_4v_1w_3)(t_5u_7v_7w_5+t_6u_8v_8w_6+t_7u_5v_5w_7+t_8u_6v_6w_8) \\
C &= 
(-u_6^2w_6w_8+u_7^2w_5w_7)(-t_5^2v_5v_7+t_8^2v_6v_8)+(u_5^2w_5w_7-u_8^2w_6w_8)(t_6^2v_6v_8-t_7^2v_5v_7)  \nonumber\\
&\quad -u_5u_7[t_5t_7(v_5^2w_7^2+v_7^2w_5^2)-t_6t_8(v_6^2w_5^2+v_8^2w_7^2)]+
u_6 u_8 [t_5t_7 ( v_5^ 2 w_6^ 2+ v_7^2 w_8^ 2) - t_6t_8(v_6^2w_8^2+v_8^2w_6^2)]  \nonumber\\
&\quad +(v_5v_8w_6w_7+v_6v_7w_5w_8)(t_5t_6u_5u_6+t_7t_8u_7u_8)+(v_5v_6w_5w_6+v_7v_8w_7w_8)(t_5t_8u_6u_7+t_6t_7u_5u_8) \nonumber\\
&\quad +(t_1u_3v_2w_4+t_2u_4v_1w_3)
( t_5  u_6 v_5 w_6 + t_6 u_5  v_6 w_5+ t_7  u_8 v_7 w_8+t_8 u_7  v_8w_7) \\
D &= 
-[u_8(t_6^2v_6v_8-t_7^2v_5v_7)u_5+u_6u_7(t_5^2v_5v_7-t_8^2v_6v_8)]w_7w_8 \nonumber\\
&\quad +[t_7(-v_5^2w_6w_7+v_7^2w_5w_8)t_5-t_6t_8(v_6^2w_5w_8-v_8^2w_6w_7)]u_7u_8 \nonumber\\
&\quad -[t_8(-u_6^2w_6w_8+u_7^2w_5 w_7 ) t_5- t_6 t_7  (u_5 ^2 w_5 w_7- u_8^ 2 w_6 w_8)] 
v_6v_5 \nonumber\\
&\quad +[-u_7(v_5v_8w_7^2-v_6v_7w_5^2)u_5+u_6u_8(v_5v_8w_6^2-v_6v_7w_8^2)]t_6t_5 \nonumber\\
&\quad -(t_5u_7v_5w_7+t_6u_8v_6w_8)(t_1u_3v_2w_4+t_2u_4v_1w_3) \\
E &= 
[u_8(t_6^2v_6v_8-t_7^2v_5v_7)u_5+u_6u_7(t_5^2v_5v_7-t_8^2v_6v_8)]w_5w_6 \nonumber\\
&\quad -[t_7(-v_5^2w_6w_7+v_7^2w_5w_8)t_5-t_6t_8(v_6^2w_5w_8-v_8^2w_6w_7)]u_5u_6 \nonumber\\
&\quad -[- u_7 ( v_5 v_8 w_7^2-v_6v_7w_5^2)u_5+u_6u_8(v_5v_8w_6^2-v_6v_7w_8^2)]t_8t_7 \nonumber\\
&\quad +[t_8(-u_6^2w_6w_8+u_7^2w_5w_7)t_5-t_6t_7(u_5^2w_5w_7-u_8^2w_6w_8)]v_7v_8 \nonumber\\
&\quad -(t_1u_3v_2w_4+t_2u_4v_1 w_3) ( t_7  u_5 v_7  w_5+ t_8 u_6 v_8 w_6 ) \\
F &= t_5t_6u_7u_8v_7v_8w_5w_6+t_7t_8u_5u_6v_5v_6w_7w_8+(v_5v_8w_6w_7+v_6v_7w_5w_8)(t_5t_8u_6u_7+t_6t_7u_5u_8) \\
G &= t_5t_6u_5u_6v_5v_6w_5w_6+t_7t_8u_7u_8v_7v_8w_7w_8+(u_5u_7w_5w_7+u_6u_8w_6w_8)(t_5t_7v_5v_7+t_6t_8v_6v_8) \\
H &= -w_7w_8v_5v_6(t_5t_8u_6u_7+t_6t_7u_5u_8)-u_7u_8t_5t_6(v_5v_8w_6w_7+v_6v_7w_5w_8) \\
I &= -w_5w_6v_7v_8(t_5t_8u_6u_7+t_6t_7u_5u_8)-u_5u_6t_7t_8(v_5v_8w_6w_7+v_6v_7w_5w_8) \\
J &= -u_7u_8w_7w_8(t_5t_7v_5v_7+t_6t_8v_6v_8)-v_5v_6t_5t_6(u_5u_7w_5w_7+u_6u_8w_6w_8) \\
K &= -u_5u_6w_5w_6(t_5t_7v_5v_7+t_6t_8v_6v_8)-v_7v_8t_7t_8(u_5u_7w_5w_7+u_6u_8w_6w_8) \\
L &= t_5t_6u_7u_8v_5v_6w_7w_8 \\
M &= t_7t_8u_5u_6v_7v_8w_5w_6
\eal

\subsection{Trapezoid free-energy}
The trapezoid free-energy can be specialized from the bi-partite staggered odd 8-vertex model~\cite{assis2017temp} rather than the four-unit staggered odd 8-vertex model, and it is of the form
\bal
-\beta f_{C} =  \frac{1}{16\pi^2}\int_0^{2\pi}\int_0^{2\pi} \ln\big[&A+2B\cos(\theta_1)+2C\cos(\theta_2)+2D\cos(\theta_1+\theta_2)\nonumber\\
&+2G\cos(2\theta_2)+2J\cos(\theta_1+2\theta_2)+2L\cos(2\theta_1+2\theta_2)\big] d\theta_1d\theta_2
\eal
where
\bal
A &= v_1^2w_3^2+2v_1v_2w_3w_4+v_2^2w_4^2+v_5^2w_7^2+v_6^2w_8^2+v_7^2w_5^2+v_8^2w_6^2 \\
B &= -v_5v_7w_5w_7-v_6v_8w_6w_8 \\
C &= (v_7w_5+v_8w_6)(v_1w_3+v_2w_4) \\
D &= -(v_5w_7+v_6w_8)(v_1w_3+v_2w_4) \\
G &= v_7v_8w_6w_5 \\
J &= -v_5v_8w_6w_7-v_6v_7w_5w_8 \\
L &= v_5v_6w_7w_8
\eal

\subsection{Barreto's Mars free-energy}
The Barreto's Mars free-energy can be specialized from the full four-unit column staggered odd 8-vertex model. There is a vast simplification so that only the $A$ term in the integrand is non-zero, giving the simple free-energy expression
\bal
-\beta f_{Ma} = \frac{1}{8}\, \ln\big[(t_1u_3v_2w_4+t_2u_4v_1w_3+t_5u_6v_7w_8+t_6u_5v_8w_7)^2-2u_3u_4w_3w_4(t_1t_2v_1v_2-t_5t_6v_7v_8)\big]
\eal
Upon expanding the argument of the logarithm, we see that all of the terms are positive.

\subsection{Kite free-energy}
The kite model free-energy can be specialized from the trapezoid free-energy, giving
\bal
-\beta f_{H} =  \frac{1}{16\pi^2}\int_0^{2\pi}\int_0^{2\pi} \ln\big[&A+2C\cos(\theta_2)+2G\cos(2\theta_2)\big] d\theta_1d\theta_2
\eal
where
\bal
A &= v_1^2w_3^2+v_2^2w_4^2+v_7^2w_5^2+2v_7v_8w_5w_6+v_8^2w_6^2 \\
C &= (v_7w_5+v_8w_6)(v_1w_3+v_2w_4) \\
G &= w_3w_4v_7v_8
\eal
We give in the main text an alternative derivation which does not require the free-fermion condition on each set of vertex weights. 

\subsection{Simple square free-energies}
The free-energy of the general four unit column staggered odd 8-vertex free-fermion model is quite large. 
As noted above, using mappings in~\cite{assis2017temp}, the expressions of~\cite{lin1977w} can alternatively be used to find the free-energy of this most general case.

The four-unit free-energy can be specialized to the column and bi-partite staggered odd 8-vertex models with units of two vertices, but we instead use the derivation of~\cite{assis2017temp}.  The respective free-energies are given as~\cite{assis2017temp}
\beqr
-\beta f_{SC} &=&  \frac{1}{16\pi^2}\int_0^{2\pi}\int_0^{2\pi} \ln\big[A + 2B\cos(\theta_1) +2C\cos(\theta_2) +2D\cos(\theta_1-\theta_2) +2E\cos(\theta_1 + \theta_2)\nonumber\\
&&\qquad\qquad\qquad\qquad + 2G\cos(2\theta_2)  + 2H\cos(\theta_1-2\theta_2) + 2I\cos(\theta_1+2\theta_2) \big] d\theta_1d\theta_2\quad
\eeqr
where
\beqr
A &=& (v_5^2+v_8^2)(w_6^2+w_7^2)+(v_6^2+v_7^2)(w_5^2 +w_8^2)+2v_1v_3w_1w_3+2v_2v_4w_2w_4+2v_5v_8w_6w_7+2v_6v_7w_5w_8 
\nonumber\\&&\\
B &=& v_1v_2w_3w_4+v_3v_4w_1w_2-v_5v_6w_7w_8-v_7v_8w_5w_6-(v_5v_7-v_6v_8)(w_5w_7-w_6w_8) \\
C &=& w_5w_8(v_6^2+v_7^2)-w_6w_7(v_5^2+v_8^2)-v_5v_8(w_6^2+w_7^2)+v_6v_7(w_5^2+w_8^2) \\
D&=& (v_3v_4-v_5v_6)(w_5w_7-w_6w_8)-(v_5v_7-v_6v_8)(w_3w_4-w_5w_6) \\
E &=& (v_1v_2-v_5v_6)(w_5w_7-w_6w_8)-(v_5v_7-v_6v_8)(w_1w_2-w_5w_6) \\
G &=& v_8v_5w_7w_6+v_7v_6w_8w_5-v_3v_1w_3w_1-v_4v_2w_4w_2 \\
H &=& (v_1v_2-v_7v_8)(w_1w_2-w_7w_8) \\
I &=& (v_1v_2-v_5v_6)(w_1w_2-w_5w_6) 
\eeqr
and
\beqr
-\beta f_{SB} &=&  \frac{1}{16\pi^2}\int_0^{2\pi}\int_0^{2\pi} \ln\big[A+2B\cos(\theta_1) +2C\cos(\theta_2)+ 2D\cos(\theta_1-\theta_2)   \nonumber\\
&&\qquad\qquad\qquad\qquad + 2E\cos(\theta_1+\theta_2) + 2F\cos(2\theta_1)+ 2G\cos(2\theta_2) \big] d\theta_1d\theta_2\quad
\eeqr
where
\beqr
A &=& w_8^2v_6^2+w_7^2v_5^2+w_5^2v_7^2+w_6^2v_8^2+w_3^2v_1^2 +w_4^2v_2^2 +w_1^2v_3^2+w_2^2v_4^2+2(v_8v_7+v_5v_6)(w_8w_7 +w_5w_6) 
\\
B &=& (v_1w_3+v_2w_4)(v_7w_5+v_8w_6)-(v_3w_1+v_4w_2)(v_5w_7+v_6w_8) \\
C &=& (v_3w_1+v_4w_2)(v_7w_5+v_8w_6)-(v_1w_3+v_2w_4)(v_5w_7+v_6w_8) \\
D &=& v_1v_4w_2w_3+v_2v_3w_1w_4-v_6v_8w_6w_8-v_5v_7w_5w_7 \\
E &=& v_1v_3w_1w_3+v_2v_4w_2w_4-v_6v_7w_5w_8-v_5v_8w_6w_7 \\
F &=& -(v_1v_2-v_5v_6)(w_1w_2-w_5w_6) \\
G &=& -(v_1v_2-v_7v_8)(w_1w_2-w_7w_8) 
\eeqr

Both of these free-energies can be specialized to the homogeneous lattice, giving~\cite{wu2004k,assis2017temp}
\beq
-\beta\,f_{S} = \frac{1}{8\pi^2}\int_0^{2\pi}\int_0^{2\pi}d\theta_1 d\theta_2 \ln(A+2B\cos(\theta_1)+2C\cos(\theta_2)+2D\cos(\theta_1-\theta_2)+2E\cos(\theta_1+\theta_2) \label{odd8vfreeenergy}
\eeq
where
\beqr
A &=& (v_1v_2+v_3v_4)(v_5v_6+v_7v_8)+v_1^2v_4^2+v_2^2v_3^2+v_5^2v_7^2 +v_6^2v_8^2 \\
B &=& 2v_5v_6v_7v_8-v_1^2v_4^2-v_2^2v_3^2 \\
C &=& 2v_1v_2v_3v_4-v_5^2v_7^2-v_6^2v_8^2 \\
D &=& (v_1v_2-v_7v_8)(v_5v_6-v_3v_4) \\
E &=& (v_1v_2-v_5v_6)(v_7v_8-v_3v_4)
\eeqr



\begin{thebibliography}{10}

\bibitem{hawkes2010abtkdrw}
E.~Hawkes, B.~An, N.~M. Benbernou, H.~Tanaka, S.~Kim, E.~D. Demaine, D.~Rus,
  and R.~J. Wood.
\newblock Programmable matter by folding.
\newblock {\em Proceedings of the National Academy of Sciences of the United
  States of America}, 107:12441--12445, 2010.

\bibitem{an2011bdr}
Byoungkwon An, Nadia Benbernou, Erik~D. Demaine, and Daniela Rus.
\newblock Planning to fold multiple objects from a single self-folding sheet.
\newblock {\em Robotica}, 29(1):87--102, 2011.

\bibitem{silverberg2014emhhsc}
Jesse~L. Silverberg, Arthur~A. Evans, Lauren McLeod, Ryan~C. Hayward, Thomas
  Hull, Christian~D. Santangelo, and Itai Cohen.
\newblock Using origami design principles to fold reprogrammable mechanical
  metamaterials.
\newblock {\em Science}, 345(6197):647--650, 2014.

\bibitem{dudte2016vtm}
Levi~H. Dudte, Etienne Vouga, Tomohiro Tachi, and L.~Mahadevan.
\newblock Programming curvature using origami tessellations.
\newblock {\em Nature Materials}, 15:583--588, 2016.

\bibitem{felton2013tsodrw}
Samuel~M. Felton, Michael~T. Tolley, ByungHyun Shin, Cagdas~D. Onal, Erik~D.
  Demaine, Daniela Rus, and Robert~J. Wood.
\newblock Self-folding with shape memory composites.
\newblock {\em Soft Matter}, 9:7688--7694, 2013.

\bibitem{tolley2014fmarw}
Michael~T Tolley, Samuel~M Felton, Shuhei Miyashita, Daniel Aukes, Daniela Rus,
  and Robert~J Wood.
\newblock Self-folding origami: shape memory composites activated by uniform
  heating.
\newblock {\em Smart Materials and Structures}, 23(9):094006, 2014.

\bibitem{tachi20104}
Tomohiro Tachi.
\newblock Geometric considerations for the design of rigid origami structures.
\newblock In {\em Proceedings of the International Association for Shell and
  Spatial Structures (IASS) Symposium 2010}, volume~12 of {\em Spatial
  Structures --- Permanent and Temporary}, pages 458--460, 2010.

\bibitem{shim2012khjy}
Tae~Soup Shim, Shin-Hyun Kim, Chul-Joon Heo, Hwan~Chul Jeon, and Seung-Man
  Yang.
\newblock Controlled origami folding of hydrogel bilayers with sustained
  reversibility for robust microcarriers.
\newblock {\em Angewandte Chemie International Edition}, 51(6):1420--1423,
  2012.

\bibitem{randall2012gg}
Christina~L. Randall, Evin Gultepe, and David~H. Gracias.
\newblock Self-folding devices and materials for biomedical applications.
\newblock {\em Trends in Biotechnology}, 30(3):138--146, 2012.

\bibitem{fernandes2012g}
Rohan Fernandes and David~H. Gracias.
\newblock Self-folding polymeric containers for encapsulation and delivery of
  drugs.
\newblock {\em Advanced Drug Delivery Reviews}, 64(14):1579--1589, 2012.

\bibitem{gracias2013}
David~H Gracias.
\newblock Stimuli responsive self-folding using thin polymer films.
\newblock {\em Current Opinion in Chemical Engineering}, 2(1):112--119, 2013.

\bibitem{silverberg2015nelhslhc}
Jesse~L. Silverberg, Jun-Hee Na, Arthur~A. Evans, Bin Liu, Thomas~C. Hull,
  Christian~D. Santangelo, Robert~J. Lang, Ryan~C. Hayward, and Itai Cohen.
\newblock Origami structures with a critical transition to bistability arising
  from hidden degrees of freedom.
\newblock {\em Nature Materials}, 14:389--393, 2015.

\bibitem{na2015ebcslhh}
Jun-Hee Na, Arthur~A. Evans, Jinhye Bae, Maria~C. Chiappelli, Christian~D.
  Santangelo, Robert~J. Lang, Thomas~C. Hull, and Ryan~C. Hayward.
\newblock Programming reversibly self-folding origami with micropatterned
  photo-crosslinkable polymer trilayers.
\newblock {\em Advanced Materials}, 27(1):79--85, 2015.

\bibitem{ocampo2003vfwkaosihn}
Jos\'{e} M.~Zanardi Ocampo, Pablo~O. Vaccaro, Thomas Fleischmann, Te-Sheng
  Wang, Kazuyoshi Kubota, Tahito Aida, Toshiaki Ohnishi, Akira Sugimura, Ryo
  Izumoto, Makoto Hosoda, and Shigeki Nashima.
\newblock Optical actuation of micromirrors fabricated by the micro-origami
  technique.
\newblock {\em Applied Physics Letters}, 83(18):3647--3649, 2003.

\bibitem{mulakkal2016swmt}
Manu~C Mulakkal, Annela~M Seddon, George Whittell, Ian Manners, and Richard~S
  Trask.
\newblock 4d fibrous materials: characterising the deployment of paper
  architectures.
\newblock {\em Smart Materials and Structures}, 25(9):095052, 2016.

\bibitem{perazahernandez2014hml}
Edwin~A Peraza-Hernandez, Darren~J Hartl, Richard J~Malak Jr, and Dimitris~C
  Lagoudas.
\newblock Origami-inspired active structures: a synthesis and review.
\newblock {\em Smart Materials and Structures}, 23(9):094001, 2014.

\bibitem{evans2017_temp}
Arthur~A. Evans, 2017.
\newblock Private communication.

\bibitem{ballinger2015defgh}
Brad Ballinger, Mirela Damian, David Eppstein, Robin Flatland, Jessica Ginepro,
  and Thomas Hull.
\newblock {\em Minimum Forcing Sets for Miura Folding Patterns}, pages
  136--147.

\bibitem{abel2016cdehklt}
Zachary Abel, Jason Cantarella, Erik~D. Demaine, David Eppstein, Thomas~C.
  Hull, Jason~S. Ku, Robert~J. Lang, and Tomohiro Tachi.
\newblock Rigid origami vertices: conditions and forcing sets.
\newblock {\em Journal of Computational Geometry}, 7(1):171--184, 2016.

\bibitem{waitukaitis2015mch}
Scott Waitukaitis, R\'{e}mi Menaut, Bryan Gin-ge Chen, and Martin van Hecke.
\newblock Origami multistability: From single vertices to metasheets.
\newblock {\em Phys. Rev. Lett.}, 114:055503, 2 2015.

\bibitem{stern2017pm_temp}
Menachem Stern, Matthew Pinson, and Arvind Murugan.
\newblock The difficulty of folding self-folding origami, 2017.
\newblock \url{https://arxiv.org/abs/1703.04161}.

\bibitem{demaine2007o}
Erik Demaine and Joseph O'Rourke.
\newblock {\em Geometric Folding Algorithms --- Linkages, Origami, Polyhedra}.
\newblock Cambridge University Press, 2007.

\bibitem{evans2015lmh}
Thomas~A. Evans, Robert~J. Lang, Spencer~P. Magleby, and Larry~L. Howell.
\newblock Rigidly foldable origami gadgets and tessellations.
\newblock {\em Royal Society Open Science}, 2(9):150067, 2015.

\bibitem{bern1996h}
Marshall Bern and Barry Hayes.
\newblock The complexity of flat origami.
\newblock In {\em Proceedings of the Seventh Annual ACM-SIAM Symposium on
  Discrete Algorithms}, SODA '96, pages 175--183, Philadelphia, PA, USA, 1996.
  Society for Industrial and Applied Mathematics.

\bibitem{mori1996k}
Shintaro Mori and Yasumasa Kajinaga.
\newblock Square lattice with attractive interactions.
\newblock {\em Phys. Rev. E}, 53:124--133, 1 1996.

\bibitem{francesco1994g2}
P.~Di~Francesco and E.~Guitter.
\newblock Folding transition of the triangular lattice.
\newblock {\em Phys. Rev. E}, 50:4418--4426, 12 1994.

\bibitem{bowick1995fgg}
M.~Bowick, P.~Di Francesco, O.~Golinelli, and E.~Guitter.
\newblock Three-dimensional folding of the triangular lattice.
\newblock {\em Nuclear Physics B}, 450(3):463--494, 1995.

\bibitem{munkel1995h}
Christian M\"{u}nkel and Dieter~W. Heermann.
\newblock Folding transitions of self-avoiding membranes.
\newblock {\em Phys. Rev. Lett.}, 75:1666--1669, 8 1995.

\bibitem{bowick1996fgg}
M~Bowick, P~Di~Francesco, O~Golinelli, and E~Guitter.
\newblock Discrete folding, 1996.
\newblock Talk given by M. Bowick at the 4th Chia meeting on Condensed Matter
  and High-Energy Physics.

\bibitem{cirillo1996gp}
Emilio N.~M. Cirillo, Giuseppe Gonnella, and Alessandro Pelizzola.
\newblock Folding transitions of the triangular lattice with defects.
\newblock {\em Phys. Rev. E}, 53:1479--1484, 2 1996.

\bibitem{cirillo1996gp2}
Emilio N.~M. Cirillo, Giuseppe Gonnella, and Alessandro Pelizzola.
\newblock Folding transition of the triangular lattice in a discrete
  three-dimensional space.
\newblock {\em Phys. Rev. E}, 53:3253--3256, Apr 1996.

\bibitem{mori1996k2}
Shintaro Mori and Shigeyuki Komura.
\newblock Monte carlo study of a self-avoiding polymerized membrane with
  negative bending rigidity.
\newblock {\em Journal of Physics A: Mathematical and General}, 29(23):7439,
  1996.

\bibitem{francesco1997gm}
P.~Di~Francesco, E.~Guitter, and S.~Mori.
\newblock Folding of the triangular lattice with quenched random bending
  rigidity.
\newblock {\em Phys. Rev. E}, 55:237--251, 1 1997.

\bibitem{bowick1997ggm}
M.~Bowick, O.~Golinelli, E.~Guitter, and S.~Mori.
\newblock Geometrical folding transitions of the triangular lattice in the
  face-centred cubic lattice.
\newblock {\em Nuclear Physics B}, 495(3):583--607, 1997.

\bibitem{mori1997g}
S~Mori and E~Guitter.
\newblock Folding of the triangular lattice in the face-centred cubic lattice
  with quenched random spontaneous curvature.
\newblock {\em Journal of Physics A: Mathematical and General}, 30(24):L829,
  1997.

\bibitem{popova2007m}
Hristina Popova and Andrey Milchev.
\newblock Structure, dynamics, and phase transitions of tethered membranes: A
  monte carlo simulation study.
\newblock {\em The Journal of Chemical Physics}, 127(19):194903, 2007.

\bibitem{popova2009m}
Hristina Popova and Andrey Milchev.
\newblock Structure, dynamic properties, and phase transitions of tethered
  membranes.
\newblock {\em Annals of the New York Academy of Sciences}, 1161(1):397--406,
  2009.

\bibitem{francesco1998}
P.~Di Francesco.
\newblock Folding transitions of the square-diagonal lattice.
\newblock {\em Nuclear Physics B}, 528(3):453--468, 1998.

\bibitem{francesco19982}
P.~Di Francesco.
\newblock Folding the square-diagonal lattice.
\newblock {\em Nuclear Physics B}, 525(3):507--548, 1998.

\bibitem{cirillo2000gp}
Emilio~N.M. Cirillo, Giuseppe Gonnella, and Alessandro Pelizzola.
\newblock Folding transitions of the square-diagonal two-dimensional lattice.
\newblock {\em Nuclear Physics B}, 583(3):584--596, 2000.

\bibitem{francesco1997gg}
P.~Di Francesco, O.~Golinelli, and E.~Guitter.
\newblock Meander, folding, and arch statistics.
\newblock {\em Mathematical and Computer Modelling}, 26(8--10):97--147, 1997.

\bibitem{francesco1997gg2}
P.~Di Francesco, O.~Golinelli, and E.~Guitter.
\newblock Meanders and the temperley-lieb algebra.
\newblock {\em Communications in Mathematical Physics}, 186(1):1--59, 1997.

\bibitem{francesco1998eg}
P.~Di Francesco, B.~Eynard, and E.~Guitter.
\newblock Coloring random triangulations.
\newblock {\em Nuclear Physics B}, 516(3):543--587, 1998.

\bibitem{francesco2001}
Philippe~Di Francesco.
\newblock Matrix model combinatorics: Applications to folding and coloring.
\newblock In Pavel Bleher and Alexander Its, editors, {\em Random Matrix Models
  and Their Applications}, volume~40 of {\em Mathematical Sciences Research
  Institute Publications}, pages 111--170. Cambridge University Press, 2001.
\newblock spring 1999 MSRI program lectures.

\bibitem{francesco2000}
P.~Di Francesco.
\newblock Folding and coloring problems in mathematics and physics.
\newblock {\em Bulletin of the American Mathematical Society}, 37(3):251--307,
  2000.

\bibitem{bowick2001t}
Mark~J. Bowick and Alex Travesset.
\newblock The statistical mechanics of membranes.
\newblock {\em Physics Reports}, 344(4--6):255--308, 2001.
\newblock Renormalization group theory in the new millennium.

\bibitem{francesco2005g}
P.~Di Francesco and E.~Guitter.
\newblock Geometrically constrained statistical systems on regular and random
  lattices: From folding to meanders.
\newblock {\em Physics Reports}, 415(1):1--88, 2005.

\bibitem{shender1993chb}
E.~F. Shender, V.~B. Cherepanov, P.~C.~W. Holdsworth, and A.~J. Berlinsky.
\newblock Kagom\'{e} antiferromagnet with defects: Satisfaction, frustration,
  and spin folding in a random spin system.
\newblock {\em Phys. Rev. Lett.}, 70:3812--3815, 6 1993.

\bibitem{francesco1994g}
P.~Di Francesco and E.~Guitter.
\newblock Entropy of folding of the triangular lattice.
\newblock {\em EPL (Europhysics Letters)}, 26(6):455, 1994.

\bibitem{ginepro2014h}
Jessica Ginepro and Thomas~C. Hull.
\newblock Counting miura-ori foldings.
\newblock {\em Journal of Integer Sequences}, 17(10):8, 2014.

\bibitem{lang2017_temp}
Robert~J. Lang, 2017.
\newblock Private communication.

\bibitem{wu2004k}
F.Y. Wu and H.~Kunz.
\newblock The odd eight-vertex model.
\newblock {\em Journal of Statistical Physics}, 116(1--4):67--78, 2004.

\bibitem{assis2017temp}
Michael Assis.
\newblock The 16-vertex model and its even and odd 8-vertex subcases on the
  square lattice, 2017.
\newblock https://arxiv.org/abs/1702.02110.

\bibitem{hsue1975lw}
C.~S. Hsue, K.~Y. Lin, and F.~Y. Wu.
\newblock Staggered eight-vertex model.
\newblock {\em Phys. Rev. B}, 12:429--437, 7 1975.

\bibitem{lin1977w}
K~Y Lin and I~P Wang.
\newblock Staggered eight-vertex model with four sublattices.
\newblock {\em Journal of Physics A: Mathematical and General}, 10(5):813,
  1977.

\bibitem{baxter19702}
R.~J. Baxter.
\newblock Three-colorings of the square lattice: A hard squares model.
\newblock {\em Journal of Mathematical Physics}, 11(10):3116--3124, 1970.

\bibitem{pegg1982}
N~E Pegg.
\newblock The generalised six-vertex model.
\newblock {\em Journal of Physics A: Mathematical and General}, 15(10):L549,
  1982.

\bibitem{pearce1989s}
Paul~A Pearce and Katherine~A Seaton.
\newblock Exact solution of cyclic solid-on-solid lattice models.
\newblock {\em Annals of Physics}, 193(2):326--366, 1989.

\bibitem{wu1972}
F.Y. Wu.
\newblock Exact results on a general lattice statistical model.
\newblock {\em Solid State Communications}, 10(1):115--117, 1972.

\bibitem{mccoy1973w}
Barry McCoy and Tai~Tsun Wu.
\newblock {\em The Two-Dimensional Ising Model}.
\newblock Harvard University Press, 1973.

\bibitem{mccoy2014w}
Barry McCoy and Tai~Tsun Wu.
\newblock {\em The Two-Dimensional Ising Model: Second Edition}.
\newblock Dover Publications, 2014.

\bibitem{fan1970w}
Chungpeng Fan and F.~Y. Wu.
\newblock General lattice model of phase transitions.
\newblock {\em Phys. Rev. B}, 2:723--733, 8 1970.

\bibitem{baxter1982}
Rodney Baxter.
\newblock {\em Exactly Solved Models in Statistical Mechanics}.
\newblock Academic Press, 1982.

\bibitem{baxter2007}
Rodney Baxter.
\newblock {\em Exactly Solved Models in Statistical Mechanics}.
\newblock Dover Books on Physics. Dover Publications, 2007.

\bibitem{gaunt1965f}
David~S. Gaunt and Michael~E. Fisher.
\newblock Hard-sphere lattice gases. i. plane-square lattice.
\newblock {\em The Journal of Chemical Physics}, 43(8):2840--2863, 1965.

\bibitem{runnels1966c}
L.~K. Runnels and L.~L. Combs.
\newblock Exact finite method of lattice statistics. i. square and triangular
  lattice gases of hard molecules.
\newblock {\em The Journal of Chemical Physics}, 45(7):2482--2492, 1966.

\bibitem{chan2012}
Yao ban Chan.
\newblock Series expansions from the corner transfer matrix renormalization
  group method: the hard-squares model.
\newblock {\em Journal of Physics A: Mathematical and Theoretical},
  45(8):085001, 2012.

\bibitem{chan2013}
Yao ban Chan.
\newblock Series expansions from the corner transfer matrix renormalization
  group method: Ii. asymmetry and high-density hard squares.
\newblock {\em Journal of Physics A: Mathematical and Theoretical},
  46(12):125009, 2013.

\bibitem{deguchi1991}
Tetsuo Deguchi.
\newblock Multivariable vertex models associated with the temperley-lieb
  algebra.
\newblock {\em Physics Letters A}, 159(3):163--169, 1991.

\bibitem{tachi2009}
Tomohiro Tachi.
\newblock {\em Generalization of rigid foldable quadrilateral mesh origami},
  pages 2287--2294.
\newblock Editorial Universitat Polit\`{e}cnica de Val\`{e}ncia, 2009.

\bibitem{tachi20103}
Tomohiro Tachi.
\newblock {\em One-DOF cylindrical deployable structures with rigid
  quadrilateral panels}, pages 2295--2305.
\newblock Editorial Universitat Polit\`{e}cnica de Val\`{e}ncia, 2010.

\bibitem{tachi2012}
Tomohiro Tachi.
\newblock Design of infinitesimally and finitely flexible origami based on
  reciprocal figures.
\newblock {\em Journal for Geometry and Graphics}, 16(2):223--234, 2012.

\bibitem{tachi2017h}
Tomohiro Tachi and Thomas~C. Hull.
\newblock Self-foldability of rigid origami.
\newblock {\em Journal of Mechanisms and Robotics}, 9:021008--021008--9, 2017.

\bibitem{tachi20092}
Tomohiro Tachi.
\newblock {\em Simulation of rigid origami}, pages 175--187.
\newblock A K Peters/CRC Press, 2009.

\bibitem{tachi2017}
Tomohiro Tachi.
\newblock Rigid origami simulator, version 0.09, 2017.
\newblock Software available at \url{http://www.tsg.ne.jp/TT/software/index.html}.

\bibitem{mccoy1967w}
Barry~M. McCoy and Tai~Tsun Wu.
\newblock Theory of toeplitz determinants and the spin correlations of the
  two-dimensional ising model. ii.
\newblock {\em Phys. Rev.}, 155:438--452, 3 1967.

\bibitem{izergin1981k}
A.~G. Izergin and V.~E. Korepin.
\newblock The inverse scattering method approach to the quantum
  shabat-mikhailov model.
\newblock {\em Communications in Mathematical Physics}, 79(3):303--316, 1981.

\bibitem{green1964h}
H.S. Green and C.A. Hurst.
\newblock {\em Order-disorder phenomena}, volume~5 of {\em Monographs in
  statistical physics and thermodynamics}.
\newblock Interscience Publishers, 1964.

\bibitem{hurst1963}
C.~A. Hurst.
\newblock Solution of plane ising lattices by the pfaffian method.
\newblock {\em The Journal of Chemical Physics}, 38(10):2558--2571, 1963.

\bibitem{sacco1975w}
J~E Sacco and F~Y Wu.
\newblock 32-vertex model on the triangular lattice.
\newblock {\em Journal of Physics A: Mathematical and General}, 8(11):1780,
  1975.

\bibitem{sacco1977w}
J~E Sacco and F~Y Wu.
\newblock 32-vertex model on the triangular lattice.
\newblock {\em Journal of Physics A: Mathematical and General}, 10(7):1259,
  1977.

\end{thebibliography}
\end{document}